\documentclass[12pt]{article}
\usepackage{graphicx}
\usepackage{amsmath, amsfonts, amssymb, amsthm}
\usepackage[english]{babel}
\usepackage{enumerate}
\usepackage{booktabs}
\usepackage{array}
\usepackage{caption}
\usepackage{subcaption}
\usepackage{natbib}
\usepackage{multicol}
\usepackage{multirow}
\usepackage{enumitem}
\usepackage{etoolbox}
\usepackage{relsize}
\usepackage{pifont}
\usepackage[flushleft]{threeparttable}

\numberwithin{equation}{section}
\theoremstyle{plain}
\newtheorem{theorem}{Theorem}[section]
\newtheorem{lemma}[theorem]{Lemma}
\newtheorem{prop}[theorem]{Proposition}
\newtheorem{corr}[theorem]{Corollary}
\theoremstyle{definition}

\newtheorem{defn}{Definition}[section]
\newenvironment{remark}[1][Remark.]{\begin{trivlist}
\item[\hskip \labelsep {\bfseries #1}]}{\end{trivlist}}

\newcommand{\bs}[1]{\boldsymbol{#1}}
\newcommand{\var}{\textnormal{Var}}

\newcommand{\te}[1]{\textnormal{#1}}
\newcommand{\nn}{\nonumber \\}
\newcommand{\tr}{\textnormal{Tr}}

\newcommand{\Mid}{\textnormal{Mid}}
\newcommand{\Log}{\textnormal{Log}}
\newcommand{\Exp}{\textnormal{Exp}}

\newcommand{\cmark}{\ding{51}}
\newcommand{\xmark}{\ding{55}}

\makeatletter
\newcommand*\rel@kern[1]{\kern#1\dimexpr\macc@kerna}
\newcommand*\widebar[1]{%
  \begingroup
  \def\mathaccent##1##2{%
    \rel@kern{0.8}%
    \overline{\rel@kern{-0.8}\macc@nucleus\rel@kern{0.2}}%
    \rel@kern{-0.2}%
  }%
  \macc@depth\@ne
  \let\math@bgroup\@empty \let\math@egroup\macc@set@skewchar
  \mathsurround\z@ \frozen@everymath{\mathgroup\macc@group\relax}%
  \macc@set@skewchar\relax
  \let\mathaccentV\macc@nested@a
  \macc@nested@a\relax111{#1}%
  \endgroup
}
\makeatother

\begin{document}

\def\spacingset#1{\renewcommand{\baselinestretch}%
{#1}\small\normalsize} \spacingset{1}



\newcommand{\blind}{0}

\if0\blind
{
  \title{\bf Intrinsic wavelet regression for curves of Hermitian positive definite matrices}
  \author{Joris Chau\footnote{
	Corresponding author, joris.chau@openanalytics.be, Institute of Statistics, Biostatistics, and Actuarial Sciences (ISBA), Universit\'e catholique de Louvain, Voie du Roman Pays 20, B-1348, Louvain-la-Neuve, Belgium. 
  }\quad and Rainer von Sachs
  }
  \date{}
  \maketitle
} \fi

\if1\blind
{
  \bigskip
  \bigskip
  \bigskip
  \begin{center}
    {\LARGE\bf Intrinsic wavelet regression for curves of Hermitian positive definite matrices}
\end{center}
  \medskip
} \fi

\bigskip

\begin{abstract} 
\noindent 
Intrinsic wavelet transforms and wavelet estimation methods are introduced for curves in the non-Euclidean space of Hermitian positive definite matrices, with in mind the application to Fourier spectral estimation of multivariate stationary time series. The main focus is on intrinsic average-interpolation wavelet transforms in the space of positive definite matrices equipped with an affine-invariant Riemannian metric, and convergence rates of linear wavelet thresholding are derived for intrinsically smooth curves of Hermitian positive definite matrices. In the context of multivariate Fourier spectral estimation, intrinsic wavelet thresholding is equivariant under a change of basis of the time series, and nonlinear wavelet thresholding is able to capture localized features in the spectral density matrix across frequency, always guaranteeing positive definite estimates. The finite-sample performance of intrinsic wavelet thresholding is assessed by means of simulated data and compared to several benchmark estimators in the Riemannian manifold. Further illustrations are provided by examining the multivariate spectra of trial-replicated brain signal time series recorded during a learning experiment.
\end{abstract}

\noindent%

\spacingset{1.45} 


\section{Introduction}\label{sec:1}

	In multivariate time series analysis, the second-order behavior of a multivariate time series is studied by means of its autocovariance matrices in the time domain, or its spectral density matrices in the frequency domain. Non-degenerate spectral density matrices are necessarily curves of Hermitian positive definite (HPD) matrices, and one generally constrains a spectral curve estimator to preserve these properties. This is important for several reasons: i) interpretation of the spectral estimator as the Fourier transform of symmetric positive definite (SPD) autocovariance matrices in the time domain or as HPD covariance matrices across frequency in the Fourier domain; ii) well-defined transfer functions in the Cram\'er representation of the time series for the purpose of e.g. simulation or bootstrapping; iii) sufficient regularity to avoid computational problems in subsequent inference procedures (requiring e.g., the inverse of the estimated spectrum). Our main contribution is the development of intrinsic wavelet transforms and nonparametric wavelet regression for curves in the non-Euclidean space of HPD matrices, exploiting the geometric structure of the space as a Riemannian manifold. The primary focus is on nonparametric spectral density matrix estimation of stationary multivariate time series, but we emphasize that the methodology applies equally to general matrix-valued curve estimation or denoising problems, where the target is a curve of symmetric or Hermitian positive definite matrices. Examples include curve denoising of SPD diffusion covariance matrices in diffusion tensor imaging as in e.g., \cite{Y12}, or estimation of time-varying autocovariance matrices of a locally stationary time series as in e.g., \cite{D12}. \\[3mm]
	A first important consideration to perform estimation in the space of HPD matrices is the associated metric in the space. The metric gives the space its curvature and induces a distance between HPD matrices. Standard nonparametric spectral matrix estimation commonly relies on smoothing the periodogram via e.g., kernel regression as in \cite[Chapter 5]{B81}, \cite[Chapter 11]{BD06}, or multitaper spectral estimation as in e.g, \cite{W00}. These approaches equip the space of HPD matrices with the Euclidean  (i.e., Frobenius) metric and view it as a \emph{flat} space. An important disadvantage is that this metric space is incomplete, as the boundary of singular matrices lies at a finite distance. For this reason, flexible nonparametric (e.g., wavelet- or spline-) periodogram smoothing embedded in a Euclidean space cannot guarantee a positive definite spectral estimate. Exceptions to this rule include inflexible kernel or multitaper periodogram smoothing, which rely on a sufficiently large equivalent smoothing bandwidth for each matrix component. To avoid this issue, \cite{DG04}, \cite{RS07} and \cite{KC13} among others construct an HPD spectral estimate as the square of an estimated curve of Cholesky square root matrices. This allows for more flexible estimation of the spectrum, such as individual smoothing of Cholesky matrix components, while at the same time guaranteeing an HPD spectral estimate. In this context, the space of HPD matrices is equipped with the Cholesky metric, where the distance between two matrices is given by the Euclidean distance between their Cholesky square roots. Unfortunately, the Cholesky metric and Cholesky-based smoothing are not equivariant to permutations of the components of the input time series. That is, if one reorders the time series components, the resulting spectral estimate is not necessarily a permuted version of the spectral estimate obtained under the original input time series.\\[3mm]
	In this work, we exploit the geometric structure of the space of HPD matrices equipped with the affine-invariant (\cite{PFA05}) --also natural invariant (\cite{S00}), canonical (\cite{H18}), trace (\cite{Y12}), Rao-Fisher (\cite{S17})-- Riemannian metric, or simply the Riemannian metric \cite[Chapter 6]{B09}, \cite{D09}). The affine-invariant Riemannian metric plays an important role in estimation problems in the space of symmetric or Hermitian positive definite matrices for several reasons: (i) the space of HPD matrices equipped with the Riemannian metric is a complete metric space, (ii) there is no swelling effect as with the Euclidean metric, where interpolating two HPD matrices may yield a matrix with a determinant larger than either of the original matrices (e.g., \cite{P10}), and (iii) the induced Riemannian distance is invariant to congruence transformation by any invertible matrix, see Section \ref{sec:2}. The first property guarantees an HPD spectral estimate, while allowing for flexible spectral matrix estimation as with Cholesky-based smoothing. The third property ensures that the spectral estimator is --not only-- permutation or unitary congruence equivariant, but also \emph{general linear congruence equivariant}, which essentially implies that the estimator does not nontrivially depend on the chosen coordinate system of the time series. In \cite{D09}, the authors list several additional suitable metrics to perform estimation in the space of HPD matrices, one of which is the Log-Euclidean metric, also discussed in e.g., \cite{Y12} or \cite{BA11}. The Log-Euclidean metric transforms the space of HPD matrices in a complete metric space and is unitary congruence invariant, but not general linear congruence invariant. \\[3mm]
	Several recent works on nonparametric curve regression in the space of SPD matrices equipped with the affine-invariant Riemannian metric include: intrinsic geodesic and linear regression in \cite{PFA05} and \cite{Z09} among others, intrinsic local polynomial regression in \cite{Y12} and intrinsic penalized spline-like regression in \cite{BA11}. In the context of frequency-specific spectral matrix estimation \cite{H18} recently introduced a Bayesian geodesic Lagrangian Monte Carlo (gLMC) approach based on the affine-invariant Riemannian metric. The latter may not be best-suited to estimation of the entire spectral curve, as this requires application of the gLMC to each individual Fourier frequency, which is computationally quite challenging. In this work, we develop fast intrinsic wavelet transforms in the manifold of HPD matrices equipped with the Riemannian metric. Wavelet-based estimation of spectral matrices allows us to capture potentially very localized features, such as local peaks or troughs in the spectral matrix at pointwise frequencies or frequency bands, in contrast to the approaches mentioned above, which rely on globally homogeneous smoothness in the frequency domain. This paper is accompanied by an R-package \texttt{pdSpecEst} (\textbf{p}ositive \textbf{d}efinite \textbf{Spec}tral \textbf{Est}imation), which contains implementations of the presented material and is available on CRAN (\cite{C17}). The technical proofs and additional descriptions of the geometric notions and tools used in this paper can be found in the supplemental materials. 


\section{Intrinsic AI wavelet transforms} \label{sec:2}

	We consider intrinsic wavelet transforms in the space of HPD matrices as generalizations of the average-interpolation (AI) wavelet transforms on the real line in \cite{D93}. In this sense, they are related to the \emph{midpoint-interpolation} (MI) wavelet transforms in \cite{R05} for general symmetric Riemannian manifolds with tractable exponential and logarithmic maps. The MI approach in \cite{R05} projects manifold-valued input data to a set of tangent spaces and applies a Euclidean refinement scheme to the projected data. Such an approach might introduce a certain degree of ambiguity as the base points of the projecting tangent spaces are specified by the user and different base points may lead to different wavelet coefficients. In contrast, the intrinsic AI transforms implement a refinement scheme --intrinsic to the considered geometry-- on the manifold itself, without first projecting the data to a set of Euclidean spaces. The primary advantage of such an intrinsic approach is that in contrast to the MI approach in \cite{R05}, the AI refinement scheme of order $k \geq 0$ reproduces \emph{intrinsic polynomial} curves up to order $k$ as defined in \cite{HFJ14}, whereas the MI refinement scheme reproduces only geodesic curves, i.e., intrinsic polynomials of order $k = 1$. This polynomial reproduction property is a necessary condition to derive wavelet coefficient decay and nonparametric estimation rates for (intrinsically) smooth curves of HPD matrices in Section \ref{sec:3}, which are not readily available in the same context for the MI wavelet transforms in \cite{R05}.

	\begin{table} 
	\begin{small}
	\begin{tabular}{llll}
  	\toprule
  	Manifold: & \hspace{.15cm} & \multicolumn{2}{l}{$\mathbb{P}_{d \times d} := \{ p \in \mathbb{C}^{d \times d}\, :\, p = p^* \te{ and } \vec{z}^* p \vec{z} > 0, \te{ for } \vec{z} \in \mathbb{C}^d, \vec{z} \neq \vec{0} \}$}\\
  	Tangent spaces: & \hspace{.15cm} & $T_p(\mathbb{P}_{d \times d}) \cong \mathbb{H}_{d \times d} := \{ h \in \mathbb{C}^{d \times d}\, :\, h = h^* \}$ & \\
  	Riemannian metric: & \hspace{.15cm} & $\langle h_1, h_2 \rangle_p = \tr((p^{-1/2} \ast h_1)(p^{-1/2} \ast h_2))$, & $h_1, h_2 \in T_p(\mathbb{P}_{d \times d})$\\
  	Distance: & \hspace{.15cm} & $\delta_R(p_1,p_2) = \Vert \Log(p_1^{-1/2} \ast p_2)\Vert_F$, & $p_1, p_2 \in \mathbb{P}_{d \times d}$\\
  	Geodesics: & \hspace{.15cm} & $\eta(p_1,p_2,t) = p_1^{1/2} \ast (p_1^{-1/2} \ast p_2)^t$, & $p_1, p_2 \in \mathbb{P}_{d \times d}, 0 \leq t \leq 1$\\
  	Exponential map: & \hspace{.15cm} & $\Exp_p(h) = p^{1/2} \ast \Exp(p^{-1/2} \ast h)$, & $p \in \mathbb{P}_{d \times d}, h \in T_p(\mathbb{P}_{d \times d})$ \\
  	Logarithmic map: & \hspace{.15cm} & $\Log_p(q) = p^{1/2} \ast \Log(p^{-1/2} \ast q)$, & $p, q \in \mathbb{P}_{d \times d}$ \\
  	Parallel transport: & \hspace{.15cm} & $\Gamma_p^q(h) = p^{1/2} \ast (p^{-1/2} \ast q)^{1/2} \ast p^{-1/2} \ast h$, & $p, q \in \mathbb{P}_{d \times d}, h \in T_p(\mathbb{P}_{d \times d})$\\
  	\bottomrule
	\end{tabular}
	\caption{Geometric tools for the Riemannian manifold of $(d \times d)$-dimensional HPD matrices $(\mathbb{P}_{d \times d}, g_R)$, equipped with the affine-invariant Riemannian metric. \label{tab:1}}
	\end{small}
	\end{table}

	\paragraph{Preliminaries and notations}

	The space of $(d \times d)$-dimensional Hermitian positive definite matrices $\mathbb{P}_{d \times d}$ is not a vector space due to its positive definite constraints, but it is an open subset of the vector space of Hermitian matrices $\mathbb{H}_{d \times d}$ and as such is also a smooth manifold, see e.g., \cite{D92}. For every $p \in \mathbb{P}_{d \times d}$, the tangent space $T_p(\mathbb{P}_{d \times d})$ is identified by $\mathbb{H}_{d \times d}$, and as detailed in \cite{PFA05}, the Frobenius inner product on $\mathbb{H}_{d \times d}$ induces the affine-invariant Riemannian metric $g_R$ on the manifold $\mathbb{P}_{d \times d}$. By \cite[Theorem 6.1.6 and Prop. 6.2.2]{B09}, the Riemannian manifold $(\mathbb{P}_{d \times d}, g_R)$, equipped with the affine-invariant metric, is \emph{geodesically complete}, and the geodesic segment joining any two points $p_1, p_2 \in \mathbb{P}_{d \times d}$ is uniquely existing. Further, for each $p \in \mathbb{P}_{d \times d}$ the exponential map $\te{Exp}_p$ and logarithmic map $\te{Log}_p$ are global diffeomorphisms with as domains $T_p(\mathbb{P}_{d \times d})$ and $\mathbb{P}_{d \times d}$ respectively. The parameterizations of the geometric notions in the Riemannian manifold $(\mathbb{P}_{d \times d}, g_R)$ used throughout this paper are summarized in Table \ref{tab:1}, and more detailed descriptions can be found in the supplementary material or \cite[Chapter 2]{C18}. Here and throughout this paper, $y^{1/2}$ always refers to the Hermitian square root matrix of $y \in \mathbb{P}_{d \times d}$, and we write $y \ast x$ for the matrix congruence transformation $y^* x y$, where $y^*$ is the conjugate transpose of $y$. The norm $\Vert \cdot \Vert_F$ refers to the matrix Frobenius norm, and $\Exp(\cdot)$ and $\Log(\cdot)$ denote the matrix exponential and the (principal) matrix logarithm. For convenience, the affine-invariant Riemannian metric is usually referred to simply as the Riemannian metric throughout this paper.  \\[3mm]
	A random variable $X: \Omega \to \mathbb{P}_{d \times d}$ is a measurable function from a probability space $(\Omega, \mathcal{A}, \nu)$ to the measurable space $(\mathbb{P}_{d \times d}, \mathcal{B}(\mathbb{P}_{d \times d}))$, with $\mathcal{B}(\mathbb{P}_{d \times d})$ the Borel algebra in the complete separable metric space $(\mathbb{P}_{d \times d}, \delta_R)$. By $P(\mathbb{P}_{d \times d})$, we denote the set of all probability measures on $(\mathbb{P}_{d \times d}, \mathcal{B}(\mathbb{P}_{d \times d}))$ and $P_m(\mathbb{P}_{d \times d})$ denotes the subset of probability measures in $P(\mathbb{P}_{d \times d})$ that have finite moments of order $m$ with respect to the Riemannian distance, i.e., the $L^m$-Wasserstein space, see \cite[Definition 6.4]{V09}. In the intrinsic AI refinement scheme described below, the center of a random variable $X \sim \nu$ is characterized by its intrinsic (also Karcher or Fr\'echet) mean. The set of intrinsic means is given by the points that minimize the second moment with respect to the Riemannian distance $\delta_R$,
	\begin{eqnarray*}
	\mu\ =\ \mathbb{E}_{\nu}[X] \ := \ \arg\min_{y \in \te{supp}(\nu)} \int_{\mathbb{P}_{d \times d}} \delta_R(y,x)^2\ \nu(dx).
	\end{eqnarray*}
	If $\nu \in P_2(\mathbb{P}_{d \times d})$, then at least one intrinsic mean exists and since $(\mathbb{P}_{d \times d}, g_R)$ is a geodesically complete manifold of non-positive curvature, the intrinsic mean $\mu$ is also unique. By \cite[Corollary 1]{P06}, the intrinsic mean is conveniently represented by $\mu \in \mathbb{P}_{d \times d}$ satisfying,
	\begin{eqnarray*} 
	\bs{E}_\nu[\Log_{\mu}(X)] &=& \bs{0}.
	\end{eqnarray*}
	Here, $\bs{0}$ is the zero matrix and $\bs{E}_\nu[\cdot]$ is the Euclidean mean in the space of Hermitian matrices. The sample intrinsic mean typically has no closed-form solution, but it can be computed efficiently through gradient descent as detailed in e.g., \cite{P06}.\\[3mm]
	In the remainder of this section, $\gamma: \mathcal{I} \to \mathbb{P}_{d \times d}$, with $\mathcal{I} \subset \mathbb{R}$, is assumed to be a square integrable matrix-valued curve, such that $\int_{\mathcal{I}} \delta_R(\gamma(u), y_0)^2\, du < \infty$ for some $y_0 \in \mathbb{P}_{d \times d}$. As input data observations we consider a finite sequence of intrinsic local averages $M_{J,k} = \te{Ave}_{I_{J,k}}(\gamma)$, across equally-sized non-overlapping intervals $(I_{J,k})_k$ with $0 \leq k \leq n-1$, such that $\bigcup_k I_{J,k} = \mathcal{I}$. Here, $\te{Ave}_{I_{J,k}}(\gamma)$ denotes the intrinsic mean of $\gamma$ over the interval $I_{J,k}$. Without loss of generality, it is assumed that $\mathcal{I} = [0,1]$ and that $n = 2^J$ is dyadic in order to allow for a straightforward construction of the scaling coefficient pyramid below. The latter is not an absolute limitation of the approach, as the intrinsic wavelet transforms can also be adapted to non-dyadic observation grids, as outlined in \cite[Chapter 5]{C18}. 

	\subsection{Intrinsic AI refinement scheme} \label{sec:2.1}
	
	\paragraph{Midpoint pyramid}
	
	The construction of the wavelet transforms is based on the idea of \emph{lifting} transforms. For an overview of first- and second-generation wavelet transforms using the lifting scheme, we refer to e.g., \cite{JO05} or \cite{KH00}. First, we build a redundant midpoint or scaling coefficient pyramid analogous to \cite{R05}, starting with the sequence of midpoint coefficients $(M_{J,k})_k$ at the finest scale $J$. At the next coarser scale $j=J-1$ set,
	\begin{eqnarray} \label{eq:2.1}
	M_{j,k} &:=& \eta(M_{j+1,2k}, M_{j+1,2k+1}, 1/2), \quad \te{for } k=0,\ldots,2^j-1, \quad \quad 
	\end{eqnarray}
    where $\eta(p_1, p_2, 1/2)$ is the halfway point or \emph{midpoint} on the geodesic segment connecting $p_1, p_2 \in \mathbb{P}_{d \times d}$ according to Table \ref{tab:1}, which coincides with the intrinsic sample mean of $p_1$ and $p_2$. This coarsening operation is continued up to scale $j = 0$, such that each scale $j$ contains a total of $2^j$ midpoints. We also use the notation $\te{Ave}(\cdot ; \cdot)$ to denote an intrinsic (weighted) sample mean. That is, if $X_1,\ldots,X_n \in \mathbb{P}_{d \times d}$, then $\widebar{X}_n = \te{Ave}( \{X_i \}_i ; \{ w_i \}_i )$ is the weighted intrinsic average of $X_1,\ldots,X_n$ with weights $w_1,\ldots,w_n$, such that $\widebar{X}_n$ solves:
	\begin{eqnarray} \label{eq:2.2}
	\widebar{X}_n &=& \Exp_{\widebar{X}_n} \left( \sum_{i=1}^n w_i \Log_{\widebar{X}_n}(X_i) \right).
	\end{eqnarray}
	If we write $\widebar{X}_n = \te{Ave}(\{ X_i \}_i)$, then $\widebar{X}_n$ is understood to be the unweighted intrinsic average of $X_1,\ldots,X_n$. In particular, we can write in a recursive fashion $M_{j,k} := \te{Ave}(\{M_{j+1,2k}, M_{j+1,2k+1} \})$.
	
	\paragraph{Intrinsic polynomials} 
	
	Intrinsic polynomials as defined in \cite{HFJ14} play a key role in the construction of the AI refinement scheme. Essentially, polynomial curves of degree $k \geq 0$ in the Riemannian manifold are defined as the curves with vanishing $k$-th and higher order covariant derivatives. Let $\gamma: \mathcal{I} \to \mathbb{P}_{d \times d}$ be a smooth curve on the manifold, with existing covariant derivatives of all orders, then it is said to be a polynomial curve of degree $k$ if, 
	\begin{eqnarray*}
	\nabla^\ell_{\gamma'} \gamma'(t) \ := \ (\nabla_{\gamma'})^{\ell} \gamma'(t) \ = \ \bs{0}, \quad \quad \forall\, \ell \geq k \te{ and } t \in \mathcal{I},
	\end{eqnarray*}
	where $\nabla_{\gamma'}^0 \gamma'(t) := \gamma'(t)$. A zero degree polynomial is a curve for which $\gamma'(t) = \bs{0}$, i.e., a constant curve. A first-degree polynomial is a curve for which $\nabla_{\gamma'} \gamma'(t) = \bs{0}$ corresponding to a geodesic curve, i.e., a \emph{straight} line in the manifold. In general, higher degree polynomials are difficult to represent in closed form, but discretized polynomial curves are straightforward to generate via numerical integration as described in \cite{HFJ14}.

	\begin{figure}[t]
	\centering
	\includegraphics[scale=0.15]{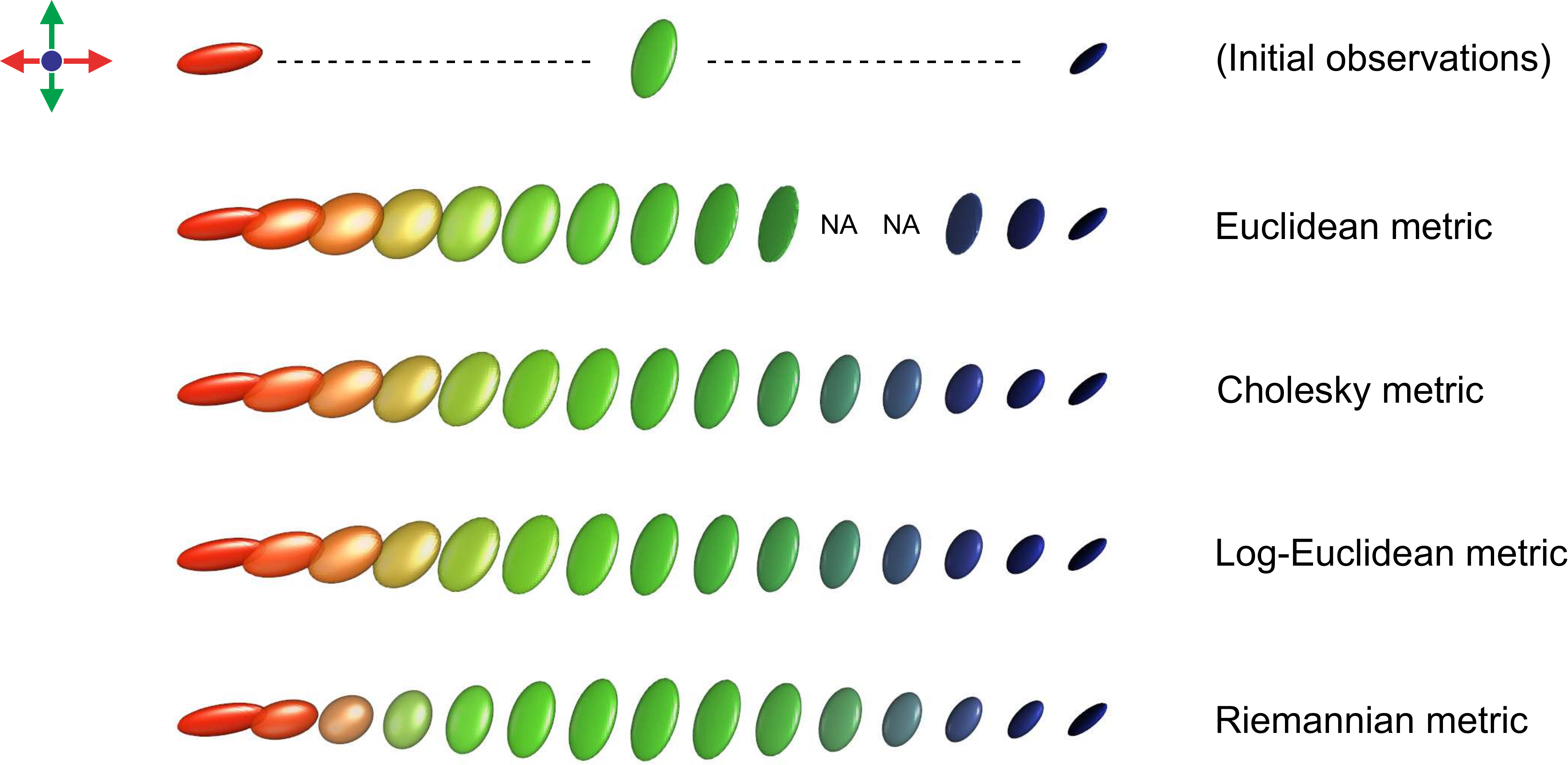}
	\caption{Illustration of intrinsic polynomial interpolation for $(3 \times 3)$-SPD matrices represented as 3D-ellipsoids. The colors indicate the direction of the principal eigenvectors. \label{fig:1}}
	\end{figure}

	\paragraph{Intrinsic polynomial interpolation}
	
	At scale $j \in \{0,\ldots,J-1\}$, the intrinsic AI refinement scheme takes as input coarse-scale midpoints $(M_{j,k})_{k}$ and outputs imputed or predicted finer-scale midpoints $(\widetilde{M}_{j+1,k'})_{k'}$. The predicted midpoints are computed as the $(j+1)$-scale midpoints of the unique intrinsic polynomial $\tilde{\gamma} : \mathcal{I} \to \mathbb{P}_{d \times d}$ with $j$-scale midpoints $(M_{j,k})_{k}$. In order to reconstruct intrinsic polynomials from a discrete set of points on the manifold, we consider a generalized intrinsic version of Neville's algorithm as in \cite[Chapter 9.2]{MF12}, replacing ordinary linear interpolation by \emph{geodesic interpolation}. \\[3mm]
	Given $P_0,\ldots,P_n \in \mathbb{P}_{d \times d}$ and $x_0 < \ldots < x_n \in \mathbb{R}$, set $p_{i,i}(x) := P_i$ for all $x$ and $i=0,\ldots,n$. The $p_{i,i}$ are zero-th order polynomials, since $p_{i,i}'(x) = \bs{0}$. Iteratively define,
	\begin{eqnarray*} 
	p_{i,j}(x) &:=& \Exp_{p_{i,j-1}(x)}\left( \frac{x-x_i}{x_j - x_i} \Log_{p_{i,j-1}(x)}(p_{i+1,j}(x)) \right), \quad \quad 0 \leq i < j \leq n,
	\end{eqnarray*}
	where $p_{i+1,j}(x)$ and $p_{i,j-1}(x)$ are the intrinsic polynomials of degree $j-i-1$ passing through $P_{i+1},\ldots,P_j$ at $x_{i+1},\ldots,x_j$ and through $P_i,\ldots,P_{j-1}$ at $x_i,\ldots,x_{j-1}$ respectively. Then $p_{i,j}(x)$ is the intrinsic polynomial of degree $j-i$ passing through $P_i, \ldots, P_j$ at $x_{i},\ldots,x_{j}$. Continuing the above iterative reconstruction, at the final iteration we obtain the intrinsic polynomial $p_{0,n}(x)$ of order $n$ passing through $P_0,\ldots,P_n$ at $x_0,\ldots,x_n$.\\[3mm]
	To illustrate, $p_{0,1}(x)$ is the geodesic, i.e., first-order intrinsic polynomial, passing through $P_0$ and $P_1$ at $x_0$ and $x_1$. In general, since $p_{i,j}(x)$ geodesically interpolates two polynomials of degree $j-i-1$, $p_{i,j}(x)$ is itself a polynomial of degree $j-i$ introducing one additional higher-degree non-vanishing covariant derivative. This is exactly analogous to the Euclidean setting, where linear interpolation of two polynomials of degree $r$ results in a polynomial of degree at most $r+1$. Intrinsic polynomial interpolation for a curve of HPD matrices by means of Neville's algorithm is demonstrated in Figure \ref{fig:1} by the interpolation of three $(3 \times 3)$-dimensional SPD matrices represented as 3D-ellipsoids using several different metric choices. Note that the interpolating second-order polynomial subject to the Euclidean metric is not everywhere positive definite, as indicated by the \texttt{NA} values.

	\subsubsection{Midpoint prediction via intrinsic average interpolation} 
	
	Reconstructing the intrinsic polynomial $\tilde{\gamma}(x)$ with $j$-scale midpoints $(M_{j,k})_{k}$ is not equivalent to reconstructing the intrinsic polynomial passing through the $j$-scale midpoints, which corresponds to an \emph{interpolating} refinement scheme instead of an \emph{average}-interpolating refinement scheme. Interpolating wavelet transforms are not well-suited to noise-removal applications, as noise would not get averaged out at coarser scales in the associated scaling coefficient pyramid. This is also discussed in more detail in \cite{R05} and \cite[Chapter 2]{C18}. To compute predicted midpoints via intrinsic average-interpolating refinement, instead consider the cumulative intrinsic mean of $\tilde{\gamma}(x)$, \@ $M_{y_0} : (y_0, 1] \to \mathbb{P}_{d \times d}$, given by:
	\begin{eqnarray} \label{eq:2.3}
	M_{y_0}(y) &=& \te{Ave}_{[y_0, y]}(\tilde{\gamma}).
	\end{eqnarray}
	If $\tilde{\gamma}(x)$ is the intrinsic polynomial with $j$-scale midpoints $(M_{j,k})_{k=0,\ldots,2^j-1}$, then $M_0((k+1)2^{-j})$ equals the cumulative intrinsic average of $\{M_{j,0},\ldots,M_{j,k-1}\}$. The main point is that the cumulative intrinsic mean of an intrinsic polynomial of order $r$ is again an intrinsic polynomial of order $\leq r$. For instance, given a geodesic segment, i.e., a first-order polynomial, its cumulative intrinsic mean is a geodesic segment moving at half the original speed. Again, this is analogous to the Euclidean setting, where an integrated polynomial is also a polynomial.\\[3mm]
	Fix a location $k \in \{L,\ldots, 2^{(j-1)} - (L + 1)\}$ at scale $j-1$ for some $L \geq 0$. Given the neighboring $(j-1)$-scale midpoints $\{ M_{j-1, k-L}, \ldots, M_{j-1,k}, \linebreak[1] \ldots, M_{j-1,k+L} \}$, we aim to predict the finer-scale midpoints $\{ M_{j,2k}, M_{j,2k+1} \}$. Here, $N := 2L + 1 \geq 0$ is referred to as the order or degree of the refinement scheme. First, to predict the midpoint $M_{j,2k+1}$, fit an intrinsic polynomial $\widehat{M}_{(k-L)2^{-(j-1)}}(y)$ of order $N - 1$ through the $N$ known points $\{ \widebar{M}_{j-1,0},\ldots, \widebar{M}_{j-1,N-1} \}$ by means of Neville's algorithm, where $\widebar{M}_{j-1,\ell}$ denotes the cumulative intrinsic average: 
	\begin{eqnarray} \label{eq:2.4}
	\widebar{M}_{j-1,\ell} &:=& M_{(k-L)2^{-(j-1)}}((k-L + \ell)2^{-(j-1)}) \ = \ \te{Ave}(\{M_{j-1, i} \}_{i=k-L}^{k-L+\ell}).
	\end{eqnarray}
	By construction of the cumulative intrinsic mean curve, $M_{(k-L)2^{-(j-1)}}((2k+1)2^{-j})$ lies on the geodesic segment connecting the known cumulative average $\widebar{M}_{j-1,L}$ and the midpoint $M_{j,2k+1}$. Replacing $M_{(k-L)2^{-(j-1)}}((2k+1)2^{-j})$ by its estimate $\widehat{M}_{(k-L)2^{-(j-1)}}((2k+1)2^{-j})$, the following expression for the predicted midpoint $\widetilde{M}_{j,2k+1}$ can be derived, see the proof of Proposition \ref{prop:3.2}:
	\begin{eqnarray*}
	\widetilde{M}_{j,2k+1} &=& \eta\left( \widebar{M}_{j-1,L}, \widehat{M}_{(k-L)2^{-(j-1)}}\big((2k+1)2^{-j}\big), -2L\right),
	\end{eqnarray*}
	using the notation $\eta(p_1, p_2, t)$ as in Table \ref{tab:1} for the geodesic passing through $p_1$ at $t = 0$ and  $p_2$ at $t = 1$. The value of $\widetilde{M}_{j,2k}$ directly follows from the midpoint relation $\te{Ave}(\{\widetilde{M}_{j,2k}, \widetilde{M}_{j,2k+1}\}) = M_{j-1,k}$ as,
	\begin{eqnarray*}
	\widetilde{M}_{j,2k} &=& M_{j-1,k} \ast \widetilde{M}_{j,2k+1}^{-1}.
	\end{eqnarray*}
	An important observation is that if the coarse-scale midpoints $\{ M_{j-1,k-L},\ldots,\linebreak[1] M_{j-1,k+L} \}$ are generated from an intrinsic polynomial $\gamma(x)$ of degree $\leq N-1$, then the midpoints $\{ M_{j,2k}, M_{j,2k+1} \}$ are reproduced without error. This is analogous to the scalar AI refinement scheme in \cite{D93} and is referred to as the \emph{intrinsic polynomial reproduction} property.\\[3mm]
	If $k \in \{0,\ldots,L-1\} \cup \{2^{(j-1)}-(L-1),\ldots,2^{(j-1)}-1\}$ is located near the boundary, not all symmetric neighbors around $M_{j-1,k}$ are available for prediction of $\{M_{j,2k}, M_{j,2k+1}\}$. Instead, collect the $N$ closest neighbors of $M_{j-1,k}$ either to the left or right and predict the $j$-scale midpoints as above through $(N-1)$-th order intrinsic polynomial interpolation based on the non-symmetric neighbors $(M_{j-1,k+\ell})_\ell$. This boundary modification preserves the intrinsic polynomial reproduction property. 

	\subsubsection{Faster midpoint prediction in practice}

	In the scalar AI refinement scheme on the real line in \cite{D93} or \cite[pg.\@ 95]{KH00}, the predicted $j$-scale scaling coefficients obtained via polynomial average-interpolation of the $(j-1)$-scale scaling coefficients are equivalent to weighted linear combinations of the input scaling coefficients, with weights depending on the average-interpolation order $N$. In the intrinsic version of Neville's algorithm, the only change with respect to its Euclidean counterpart is the nature of the interpolation, i.e., linear interpolation is substituted by geodesic interpolation. The predicted midpoints $\{ \widetilde{M}_{j,2k}, \widetilde{M}_{j,2k+1}\}$ remain weighted averages of the inputs $\{ M_{j-1,k-L},\ldots,M_{j-1,k},\ldots,M_{j-1,k+L} \}$, with the same weights as in the Euclidean case, but instead of weighted Euclidean averages the weighted averages are obtained as intrinsic weighted averages in the Riemannian manifold:
	\begin{eqnarray} \label{eq:2.5}
	\widetilde{M}_{j,2k} &=& \te{Ave}\left( \{M_{j-1,k+\ell}\}_{\ell=-L}^L ; \{C_{N,2\ell+N-1} \}_{\ell = -L}^L \right) \nn
	\widetilde{M}_{j,2k+1} &=& \te{Ave}\left( \{M_{j-1,k+\ell}\}_{\ell=-L}^L ; \{C_{N,2\ell+N} \}_{\ell = -L}^L \right),
	\end{eqnarray}
	where the weights $\bs{C}_N = (C_{N,i})_{i=0,\ldots,2N-1}$ depend on the refinement order $N \geq 1$ and sum up to 2. For instance, away from the boundary; for $N = 1$, $\bs{C}_1 = (1,1)$; for $N=3$, $\bs{C}_3 = (1,-1,8,8,-1,1)/8$; for $N=5$, $\bs{C}_5 = (-3,3,22,-22,128,128,-22,22,3,-3)/128$; and for $N=7$, $\bs{C}_7 = (5,-5,-44,44,201,-201,1024,1024,-201,201,44,-44,-5,5)/1024$. In the \texttt{pdSpecEst}-package, these prediction weights are pre-determined up to order $N \leq 9$ at all locations, allowing for faster computation of the predicted midpoints in practice. For higher refinement orders, the midpoints are predicted via the intrinsic version of Neville's algorithm.

	\subsection{Intrinsic forward and backward AI wavelet transform}
	
	\paragraph{Forward wavelet transform} 
	
\begin{figure}[t]
	\centering
	\includegraphics[scale=0.25]{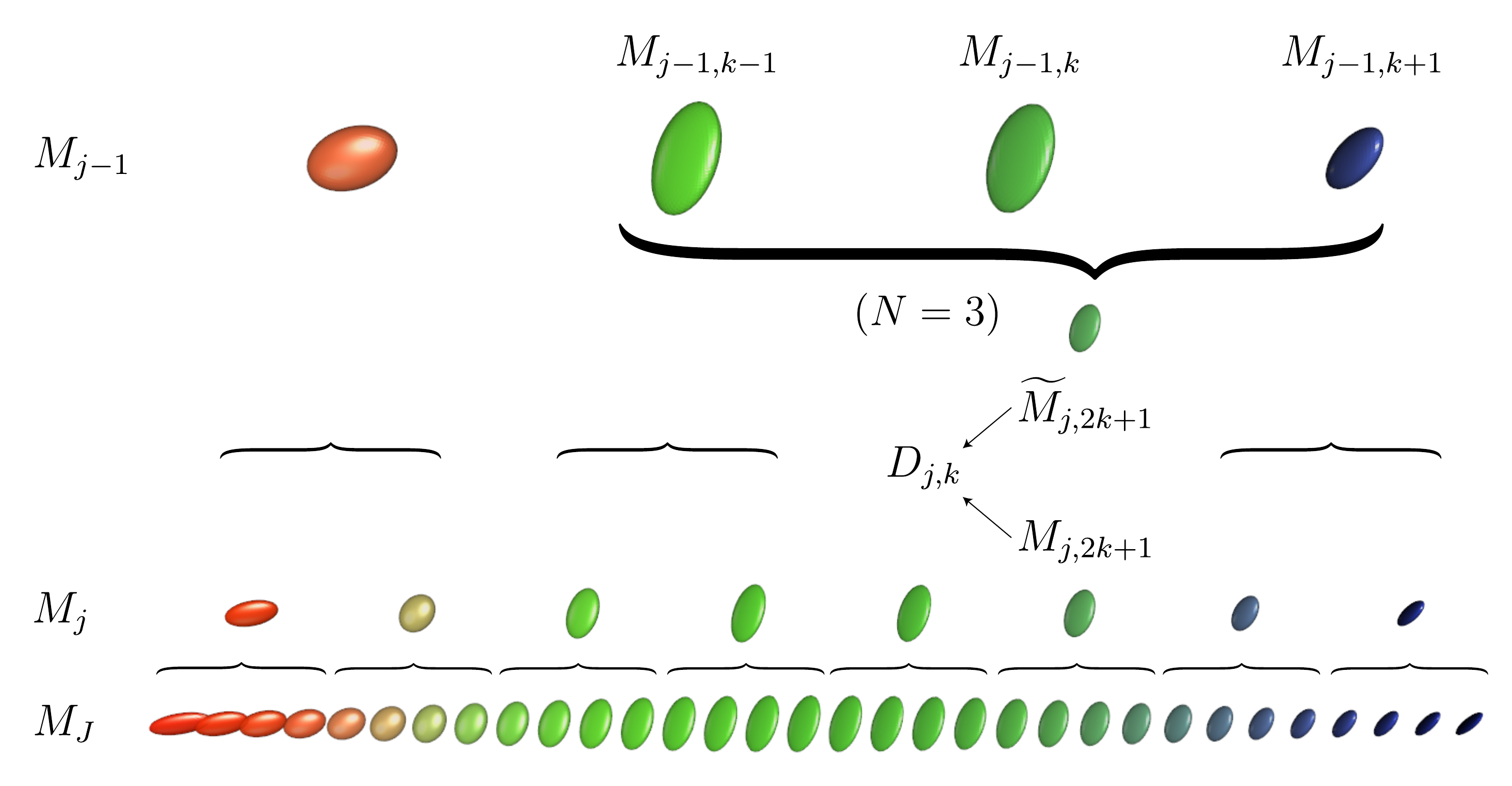}
	\caption{Illustration of intrinsic foward AI wavelet transform for a \emph{smooth} curve of three $(3 \times 3)$-dimensional SPD matrices represented as 3D-ellipsoids at different midpoint scales. \label{fig:2}}
	\end{figure}

	The intrinsic AI refinement scheme leads to an intrinsic AI wavelet transform passing from $j$-scale midpoints to $(j-1)$-scale midpoints plus $j$-scale wavelet coefficients. The steps in the intrinsic AI wavelet transform are also visualized in Figure \ref{fig:2} based on a sequence of $(3 \times 3)$-dimensional SPD matrices represented as 3D-ellipsoids.
	\begin{enumerate}
	\item \textbf{Coarsen/Predict:} given $j$-scale midpoints $(M_{j,k})_{k=0,\ldots,2^{j}-1}$, compute the $(j-1)$-scale midpoints $(M_{j-1,k})_{k=0,\ldots,2^{j-1}-1}$ via the midpoint relation in eq.(\ref{eq:2.1}). Select a refinement order $N \geq 1$ and generate the predicted midpoints $(\widetilde{M}_{j,k})_{k=0,\ldots,2^{j}-1}$ based on $(M_{j-1,k})_{k}$.
	\item \textbf{Difference:} given the true and predicted $j$-scale midpoints $M_{j,2k+1}, \widetilde{M}_{j,2k+1}$, define the wavelet coefficients as an \emph{intrinsic difference} according to,
	\begin{eqnarray} \label{eq:2.6}
	D_{j,k} &=& 2^{-j/2} \Log_{\widetilde{M}_{j,2k+1}}\big(M_{j,2k+1} \big)  \ \in \  T_{\widetilde{M}_{j,2k+1}}(\mathbb{P}_{d \times d}).
	\end{eqnarray}
	Note that $\Vert D_{j,k} \Vert_{\widetilde{M}_{j,2k+1}}^2 = 2^{-j}\delta_R(M_{j,2k+1}, \widetilde{M}_{j,2k+1})^2$ by definition of the Riemannian distance, giving the wavelet coefficients the interpretation of a (scaled) difference between $M_{j,2k+1}$ and $\widetilde{M}_{j,2k+1}$. In addition, we also keep track of the \emph{whitened} wavelet coefficients, 
	\begin{eqnarray} \label{eq:2.7}
	\mathfrak{D}_{j,k} &=& \widetilde{M}_{j,2k+1}^{-1/2} \ast D_{j,k} \ \in \ T_{\te{Id}}(\mathbb{P}_{d \times d}),
	\end{eqnarray}
	with $\te{Id} \in \mathbb{P}_{d \times d}$ the $(d \times d)$-dimensional identity matrix.
	The whitened coefficients correspond to the coefficients in eq.(\ref{eq:2.6}) transported to the same tangent space (at the identity). This allows for straightforward comparison of coefficients across scales and locations in Section \ref{sec:3} and \ref{sec:4}, since $\Vert \mathfrak{D}_{j,k} \Vert_F^2 = \Vert D_{j,k} \Vert_{\widetilde{M}_{j,2k+1}}^2$.
	\end{enumerate}

	\paragraph{Backward wavelet transform} 
	
	The backward wavelet transform passing from coarse $(j-1)$-scale midpoints plus $j$-scale wavelet coefficients to finer $j$-scale midpoints follows from reverting the above operations:
	\begin{enumerate}
	\item \textbf{Predict/Refine}: given $(j-1)$-scale midpoints $(M_{j-1,k})_{k=0,\ldots,2^{j-1}-1}$ and a refinement order $N \geq 1$, generate the predicted midpoints $(\widetilde{M}_{j,2k+1})_{k=0,\ldots,2^{j-1}-1}$ and compute the $j$-scale midpoints at the odd locations $2k+1$ for $k=0,\ldots,2^j-1$ through:
	\begin{eqnarray*}
	M_{j,2k+1} &=& \Exp_{\widetilde{M}_{j,2k+1}}(2^{j/2} D_{j,k}).
	\end{eqnarray*}
	\item \textbf{Complete}: the $j$-scale midpoints at the even locations $2k$ for $k=0,\ldots,2^j-1$ are retrieved from $M_{j-1,k}$ and $M_{j,2k+1}$ through the midpoint relation in eq.(\ref{eq:2.1}) as,
	\begin{eqnarray*}
	M_{j,2k} &=& M_{j-1,k} \ast M_{j,2k+1}^{-1}.
	\end{eqnarray*}  
	\end{enumerate}
	Given the coarsest midpoint $M_{0,0}$ at scale $j=0$ and the wavelet coefficient pyramid $(D_{j,k})_{j,k}$, for $j=1,\ldots,J$ and $k = 0,\ldots,2^{j-1}-1$, repeating the reconstruction procedure above up to scale $J$, we retrieve the original input sequence of local averages $(M_{J,k})_k$ for $k=0,\ldots,2^J-1$.


\section{Wavelet regression for smooth HPD curves} \label{sec:3}

	In this section, we derive the wavelet coefficient decay and linear wavelet thresholding convergence rates in the context of the intrinsic AI wavelet transforms for intrinsically smooth curves of HPD matrices. It turns out that the derived rates coincide with the usual scalar wavelet coefficient decay and linear thresholding convergence rates on the real line. Nonlinear thresholding will not improve the convergence rates in the case of a homogeneous smoothness space. However, nonlinear wavelet thresholding is expected to improve the convergence rates in the case of globally non-homogeneous smoothness spaces. This requires a well-defined intrinsic generalization to the Riemannian manifold of e.g., the family of Besov smoothness spaces, which is outside the scope of this paper.
		
	\paragraph{Repeated midpoint operator}
	
	The repeated midpoint operator in eq.(\ref{eq:2.1}) in the construction of the midpoint pyramid is a valid intrinsic averaging operator in the sense that it converges to the intrinsic mean in the metric space $(\mathbb{P}_{d \times d}, \delta_R)$ at the same rate of convergence as in the Euclidean setting. As in \cite{R05}, recursively define,
	\begin{eqnarray} \label{eq:3.1}
	\mu_n \ :=\ \mu_n(X_1,\ldots,X_n) \ =\ \te{Ave}\big(\{\mu_{n/2}(X_1,\ldots,X_{n/2}), \mu_{n/2}(X_{n/2+1},\ldots,X_n) \} \big).
	\end{eqnarray}
	
	\begin{prop} \label{prop:3.1}
	(Convergence midpoint operator) Let $X_1,\ldots,X_n \overset{\te{iid}}{\sim} \nu$, such that $\nu \in P_2(\mathbb{P}_{d \times d})$ with intrinsic mean $\mathbb{E}_\nu[X] = \mu$, and $n = 2^J$ for some $J>0$. Then, 
	\begin{eqnarray*}
	\bs{E}[\delta_R(\mu_n, \mu)^2] & \lesssim & n^{-1},
	\end{eqnarray*} 
	with $\lesssim$ smaller or equal up to a constant. Moreover, $\mu_n \overset{p}{\to} \mu$ as $n \to \infty$, where the convergence holds with respect to the Riemannian distance, i.e., for every $\epsilon > 0$, $P(\delta_R(\mu_n, \mu) > \epsilon ) \to 0$. 
	\end{prop}
	
	\paragraph{Wavelet coefficient decay of smooth curves} 
	
	The derivation of the wavelet coefficient decay of intrinsically smooth curves in the Riemannian manifold relies on the fact that the derivative $\gamma'(t) \in T_{\gamma(t)}(\mathbb{P}_{d \times d})$ of a smooth curve $\gamma: \mathcal{I} \to \mathbb{P}_{d \times d}$ can be Taylor expanded in terms of the parallel transport and covariant derivatives according to \cite[Chapter 9, Proposition 5.1]{L95} as,
	\begin{eqnarray} \label{eq:3.2}
	\gamma'(t) &=& \sum_{k=0}^m \Gamma(\gamma)_{t_0}^{t}\left( \nabla_{\gamma'}^k \gamma'(t_0) \right) \frac{(t-t_0)^k}{k!} + O((t-t_0)^{m+1}), \quad \quad \te{as } t \to t_0,
	\end{eqnarray}
	where the parallel transport $\Gamma(\gamma)_{t_0}^t(v)$ transports a vector $v \in T_{\gamma(t_0)}(\mathbb{P}_{d \times d})$ to the tangent space $T_{\gamma(t)}(\mathbb{P}_{d \times d})$ along the curve $\gamma$. If $\gamma(t)$ is an intrinsic polynomial curve of order $r > 0$, then, since $\Gamma(\gamma)_{t_0}^t(\bs{0}) = \bs{0}$, all terms of order higher or equal to $r$ vanish and $\gamma'(t)$ simplifies to,
	\begin{eqnarray*}
	\gamma'(t) &=& \sum_{k=0}^{r-1} \Gamma(\gamma)_{t_0}^t(\nabla_{\gamma'}^k \gamma'(t_0)) \frac{(t-t_0)^k}{k!}.
	\end{eqnarray*}
	In the specific case of a first-order polynomial, the above expression reduces to $\gamma'(t) = \Gamma(\gamma)_{t_0}^t(\gamma'(t_0))$, i.e., $\gamma'$ is parallel transported along the curve $\gamma$ itself, or in other words, $\gamma(t)$ is a geodesic curve.
	
	\begin{prop} \label{prop:3.2} 
	(Coefficient decay) Given a refinement order $N \geq 1$, suppose that $\gamma: [0,1] \to \mathbb{P}_{d \times d}$ is a smooth curve with existing covariant derivatives of order $N$ or higher. Then, for each scale $j > 0$ sufficiently large and location $k$, 
	\begin{eqnarray*}
	\Vert \mathfrak{D}_{j,k} \Vert_F & \lesssim & 2^{-j/2}2^{-jN},
	\end{eqnarray*}
	where $\mathfrak{D}_{j,k}$ denotes the whitened wavelet coefficient at scale-location $(j,k)$ as in eq.(\ref{eq:2.7}) obtained from the intrinsic AI wavelet transform with refinement order $N$. Here, the finest-scale midpoints are given by the local intrinsic averages $M_{J,k} = \te{Ave}_{I_{J,k}}(\gamma)$, with $I_{J,k} = [k/2^J, (k+1)/2^J]$ for $k=0,\ldots,2^J-1$.
	\end{prop}
	
	\begin{remark}
	Note that the above decay rates correspond to the usual wavelet coefficient decay rates of smooth real-valued curves in a Euclidean space based on wavelets with $N$ vanishing moments, see e.g., \cite[Theorem 9.5]{W02}.
	\end{remark}
	
	\paragraph{Consistency and convergence rates}
	
	The following results detail the convergence rates of linear thresholding of wavelet scales of intrinsically smooth curves $\gamma: [0,1] \to \mathbb{P}_{d \times d}$ subject to noise. Let $M_{J,k} = \te{Ave}_{I_{J,k}}(\gamma)$, with $I_{J,k} = [k/n, (k+1)/n]$ for $k=0,\ldots,n-1$ as before, and suppose that $X_0,\ldots,X_{n-1}$ is an independent sample, such that $X_k \sim \nu_k$ with $\nu_k \in P_2(\mathbb{P}_{d \times d})$ and $\mathbb{E}_{\nu_k}[X] = M_{J,k}$ for each $k=0,\ldots,n-1$. The proposition below gives the estimation error of the empirical wavelet coefficients based on $X_0,\ldots,X_{n-1}$ with respect to the true wavelet coefficients based on the sequence $M_{J,0},\ldots, M_{J,n-1}$. The proof relies on the convergence rate in Proposition \ref{prop:3.1} above.
	
	\begin{prop} \label{prop:3.3} 
	(Estimation error) Let $M_{J,0},\ldots,M_{J,n-1}$ and $X_0,\ldots, X_{n-1}$ be as defined above, with $n = 2^J$ for some $J > 0$. Then, for each scale $j > 0$ sufficiently small and each location $k$, it holds that, 
	\begin{eqnarray*}
	\bs{E}\Vert \widehat{\mathfrak{D}}_{j,k,n} - \mathfrak{D}_{j,k} \Vert_F^2 & \lesssim & n^{-1},
	\end{eqnarray*}
	where $\widehat{\mathfrak{D}}_{j,k,n} = 2^{-j/2}\, \textnormal{Log}(\widetilde{M}^{-1/2}_{j,2k+1,n} \ast M_{j,2k+1,n})$ is the empirical whitened wavelet coefficient at scale-location $(j,k)$, with $M_{j,2k+1,n}$ the estimated repeated midpoint at scale-location $(j, 2k+1)$ based on $X_0,\ldots,X_{n-1}$ and $\widetilde{M}_{j,2k+1,n}$ the predicted midpoint based on the estimated midpoints $(M_{j-1,k',n})_{k'}$ and some refinement order $N \geq 1$.
	\end{prop}
	
	\noindent Combining Proposition \ref{prop:3.2} and \ref{prop:3.3}, the main theorem below provides the averaged mean squared Riemannian error of a linear wavelet estimator of a smooth curve $\gamma(t)$ based on the sample of observations $X_0,\ldots,X_{n-1}$. Again, the convergence rates correspond to the usual nonparametric convergence rates of linear wavelet estimators of smooth real-valued curves in a Euclidean space based on wavelets with $N$ vanishing moments, see e.g., \cite{A97}.
	
	\begin{theorem} \label{thm:3.4}
	(Convergence rates linear thresholding) Given a refinement order $N \geq 1$, suppose that $\gamma : [0,1] \to \mathbb{P}_{d \times d}$ 	is a smooth curve with existing covariant derivatives of order $N$ or higher, and let $M_{J,0},\ldots,M_{J,n-1}$ and $X_0,\ldots,X_{n-1}$ be as defined above, with $n = 2^J$ for some $J \geq 0$. Consider the linear wavelet estimator based on the observations $X_0,\ldots,X_{n-1}$ that thresholds all wavelet coefficients at scales $j \geq J_0$, such that $J_0 = \log_2(n)/(2N+1)$, with $N$ the order of the intrinsic AI wavelet transform. For $n$ sufficiently large, 
	\begin{eqnarray} \label{eq:3.3}
	\sum_{j,k} \bs{E} \Vert \widehat{\mathfrak{D}}_{j,k,n} - \mathfrak{D}_{j,k} \Vert_F^2 & \lesssim & n^{-2N/(2N+1)},
	\end{eqnarray}
	where $\widehat{\mathfrak{D}}_{j,k,n}$ is the empirical whitened wavelet coefficient after linear thresholding of wavelet scales and the sum ranges over all scales $1 \leq j \leq J$ and locations $0 \leq k \leq 2^{j-1}-1$. Moreover, denote by $(\widehat{M}_{J,k,n})_k$ the finest-scale midpoints based on the linear thresholded wavelet estimator. Then, for $n$ sufficiently large, also,
	\begin{eqnarray} \label{eq:3.4}
	\frac{1}{n} \sum_{k=0}^{n-1} \bs{E}\left[ \delta_R\big( M_{J,k}, \widehat{M}_{J,k,n} \big)^2 \right] & \lesssim & n^{-2N/(2N+1)}.
	\end{eqnarray}
	\end{theorem}

	\begin{remark}
	Denoting $\hat{\gamma}_n(t) = \widehat{M}_{J,k,n}\bs{1}_{\{t \in I_{J,k}\}}$ and $\gamma_n(t) = M_{J,k}\bs{1}_{\{t \in I_{J,k}\}}$, with $\bs{1}$ the indicator function. If it is further assumed that $\gamma(t) - \gamma_n(t) = O(n^{-N/(2N+1)})$ for $t \in [0,1]$, then the linear wavelet estimator $\hat{\gamma}_n(t)$ converges to the continuous curve $\gamma(t)$ at the same rate as in Theorem \ref{thm:3.4} above, 
	\begin{eqnarray*}
	\int_0^1 \bs{E}\left[ \delta_R(\hat{\gamma}_n(t), \gamma(t))^2 \right] \, dt & \lesssim & n^{-2N/(2N+1)}. 
	\end{eqnarray*} 
	The derivation of this result follows directly from the application of a generalized triangle inequality, the details of which can be found in Appendix II in the supplementary material. 
	\end{remark}


\section{Wavelet-based spectral matrix estimation} \label{sec:4}

	In the context of multivariate spectral matrix estimation, consider data observations from a $d$-dimensional strictly stationary time series of length $T = 2n$ with HPD spectral density matrix $f(\omega) \in \mathbb{P}_{d \times d}$ and raw periodogram matrix $I_T(\omega_\ell)$ at the Fourier frequencies $\omega_\ell = \pi \ell /n \in (0, \pi]$ for $\ell = 1,\ldots,n$. The aim of this section is to estimate $f(\omega)$ by denoising the inconsistent spectral estimator $I_T(\omega_\ell)$ through shrinkage or thresholding of coefficients in the intrinsic wavelet domain. Given the setup in Section \ref{sec:2.1}, we can define the equally-sized intervals $I_{J,k} = (\pi k/n, \pi (k+1)/n]$, with $0 \leq k \leq n-1$ and $\bigcup_k I_{J,k} = (0,\pi]$, such that each interval $I_{J,k}$ contains a single Fourier frequency $\omega_{k+1}$. As we only consider estimating the spectrum at the Fourier frequencies, we set the finest-scale local averages to $M_{J,k} = \te{Ave}_{I_{J,k}}(f) = f(\omega_{k+1})$.

	\paragraph{Pre-smoothed periodogram}
	
	By construction, the raw periodogram matrix $I_T(\omega_\ell)$ is Hermitian, but only positive semidefinite as the rank of $I_T(\omega_\ell)$ is one. The intrinsic wavelet transform acts on curves of HPD matrices and for this reason we pre-smooth the periodogram to guarantee that it is HPD or full rank analogous to e.g., \cite{DG04}. By \cite[Lemma 1]{DG04}, for $\omega_\ell \not\equiv 0\ (\textrm{mod}\ \pi)$, a multitaper spectral estimate $\bar{I}_T(\omega_\ell)$ of a strictly stationary time series, with a fixed number of tapers $L$, is asymptotically independent at the Fourier frequencies, and its asymptotic distribution satisfies:
	\begin{eqnarray*}
	\bar{I}_T(\omega_\ell) & \overset{d}{\to} & W_d^C(L, L^{-1}f(\omega_\ell)), \quad \quad \te{as } n \to \infty.
	\end{eqnarray*}
	Here, $W_d^C(L, L^{-1}f(\omega_\ell))$ denotes a complex Wishart distribution of dimension $d$ with $L$ degrees of freedom and Euclidean mean $f(\omega_\ell)$. If $L \geq d$, then the spectral estimate $\bar{I}_T(\omega_\ell)$ is positive definite with probability one. In order to pre-smooth the raw periodogram matrix $I_T(\omega_\ell)$, we choose $L = d$ as small as possible, so that only the necessary small amount of pre-smoothing is performed to guarantee an HPD periodogram matrix $\bar{I}_T(\omega_\ell)$.
	
	\paragraph{Asymptotic bias-correction}
	
	Suppose that $X \sim W_d^C(L, L^{-1} f)$ exactly, then the Euclidean mean of $X$ equals $f$, and if $X_1,\ldots,X_n \overset{\te{iid}}{\sim} W_d^C(L, L^{-1} f)$, the arithmetic mean $\frac{1}{n} \sum_{\ell=1}^n X_\ell$ is an unbiased and consistent estimator of $f$ as $n \to \infty$. Intrinsic averaging in the midpoint pyramid is performed through repeated application of the midpoint operator. By Proposition \ref{prop:3.1}, it is understood that if the Euclidean mean $\bs{E}[X_\ell] = f$ and the intrinsic mean $\mathbb{E}[X_\ell] = \mu$ do not coincide, the repeated midpoint functional is not a consistent estimator of $f$, the object of interest. By defining the notion of intrinsic bias as in \cite{S00}, the repeated midpoint functional of a multitaper spectral estimate is seen to be asymptotically biased with respect to the spectrum $f$. 
	
	\begin{defn} \label{def:4.1}
	Given an estimator $\hat{\mu}$ of $\mu \in \mathbb{P}_{d \times d}$, define the bias $b(\hat{\mu}, \mu) \in T_\mu(\mathbb{P}_{d \times d})$ of $\hat{\mu}$ as,
	\begin{eqnarray*}
	b(\hat{\mu}, \mu) &=& \bs{E}[\Log_{\mu}(\hat{\mu})].
	\end{eqnarray*}
	Note that in a Euclidean space, the Exp- and Log-maps reduce to ordinary matrix addition and subtraction, in which case the above definition simplifies to the usual vector space definition of the bias.
	\end{defn}
	
	\begin{theorem} (Bias-correction) \label{thm:4.1}
	Let $X \sim W_d^C(L, L^{-1}f)$ and $c(d,L) = -\log(L) + \frac{1}{d}\sum_{i=1}^d \psi(L-(d-i))$, with $\psi(\cdot)$ the digamma function, then the intrinsic bias of $X$ with respect to $f$ is,
	\begin{eqnarray*}
	b(X, f) \ =\ \bs{E}[\textnormal{Log}_f(X)] \ = \ c(d,L) \cdot f.
	\end{eqnarray*} 
	If $(\widetilde{X}_\ell)_{\ell=1,\ldots,n} := (e^{-c(d,L)} X_\ell)_{\ell=1,\ldots,n}$, such that $X_1,\ldots,X_n \overset{\te{iid}}{\sim} W_d^C(L, L^{-1} f)$ with $n=2^J$, then,
	\begin{eqnarray*}
	\mu_n(\widetilde{X}_1,\ldots,\widetilde{X}_n) & \overset{p}{\to} f, \quad \quad \te{as } n \to \infty,
	\end{eqnarray*}
	where the convergence in probability holds with respect to the Riemannian distance.
	\end{theorem}
	
	\begin{remark} It is observed that if $d = L = 1$, the bias-correction simplifies to multiplication by the scalar $\exp(-c(d,L)) = \exp(-\psi(1))$, the exponential of the Euler-Mascheroni constant. This corresponds to the asymptotic bias-correction for the ordinary log-periodogram with respect to the log-spectrum in the context of a univariate time series, see e.g., \cite{W80}.
	\end{remark}
	
	\begin{remark}
	As the bias-corrected (and pre-smoothed) periodogram $\bar{I}_T(\omega_\ell)$ is asymptotically equivalent in distribution to a sequence of bias-corrected independent Wishart matrices, linear wavelet estimation of the bias-corrected periodogram approximately enjoys the same convergence properties as discussed at the end of Section \ref{sec:3}.
	\end{remark}

	\subsection{Nonlinear intrinsic wavelet thresholding} \label{sec:4.1}
	
	Given a sequence of $d$-dimensional time series observations, wavelet-based spectral estimation exploits the sparsity of representations of smooth curves in the intrinsic AI wavelet domain by proceeding along the usual steps:
	\begin{enumerate}[noitemsep]
	\item Apply the intrinsic AI wavelet transform to the bias-corrected HPD periodogram.
	\item Shrink or threshold the coefficients in the intrinsic wavelet domain.
	\item Apply the inverse intrinsic AI wavelet transform to the modified coefficients.
	\end{enumerate}
	There are various possibilities to nonlinearly shrink or threshold coefficients in the intrinsic manifold wavelet domain. In particular, expanding the matrix-valued coefficients in a basis of the vector space of Hermitian matrices, nonlinear thresholding or shrinkage of individual components allows to capture inhomogeneous smoothness behavior across components of the spectral matrix, similar to the Cholesky-based smoothing procedures in e.g., \cite{DG04} or \cite{KC13}. The wavelet-denoised estimator is guaranteed to be HPD, as the inverse wavelet transform always outputs a curve in the manifold of HPD matrices. From the perspective of wavelet coefficients being intrinsic local differences in the manifold, another sensible approach is to shrink or threshold all components of a matrix-valued wavelet coefficient simultaneously, e.g., a kink or cusp in a curve in the manifold likely affects all components of the matrix-valued wavelet coefficients at the corresponding scale-locations instead of a single or only a few components. Here, we pursue the latter approach and consider keep-or-kill thresholding of entire wavelet coefficients. 
	
	\paragraph{Congruence equivariance}

	In general, the only requirement that is imposed on the intrinsic wavelet thresholding or shrinkage procedure is that it is \emph{unitary congruence equivariant}. That is, if $D^X$ is a noisy matrix-valued wavelet coefficient and $\widehat{D}^X$ is its shrunken or thresholded equivalent, then $U \ast \widehat{D}^X$ should be the shrunken or thresholded equivalent of $U \ast D^X$ for each $U \in \mathcal{U}$, where $\mathcal{U}$ is the space of unitary matrices. In practice, this property virtually always holds. For instance, if one thresholds or shrinks components of coefficients data-adaptively, the component-specific threshold or shrinkage parameters rotate in the same fashion as the components of the coefficients.
	
	\begin{prop} (Unitary congruence equivariance) \label{prop:4.2}
	Let $(X_\ell)_\ell$ be a sequence of HPD matrices and $(\hat{f}_\ell)_\ell$ its wavelet-denoised estimate. If the wavelet thresholding or shrinkage procedure is unitary congruence equivariant, then the same is true for the wavelet estimator, i.e., the wavelet-denoised estimate of $(U \ast X_\ell)_\ell$ is $(U \ast \hat{f}_\ell)_\ell$ for each $U \in \mathcal{U}$.
	\end{prop}
	
	\noindent This is an important property in the context of multivariate spectral estimation. 
	Rotation of the observed time series data, e.g., permuting the time series components, results in a congruence transformation $U \ast f(\omega)$ of the generating spectral matrix, with $U \in \mathcal{U}$. Such rotations should not nontrivially affect the spectral estimator, as the observed rotation of the time series is essentially an arbitrary representation of the data. The spectral estimation methods based on smoothing the Cholesky decomposition of an initial noisy spectral estimator (\cite{DG04}, \cite{RS07} or \cite{KC13}) do not necessarily satisfy this condition. This is due to the fact that Cholesky square root matrices are generally not unitary congruence-equivariant, i.e., $\te{Chol}(U \ast f(\omega)) \neq U \ast \te{Chol}(f(\omega))$ for a non-trivial unitary matrix $U \in \mathcal{U}$. To circumvent this problem, in \cite{Z17}, the authors propose to average a large set of Cholesky-based estimates based on random rotations of the data. The main drawback of such an approach is the significant increase in computational effort. 

	\paragraph{Trace thresholding of coefficients}
	
	A method that is particularly \emph{traceable} is thresholding or shrinkage based on the trace of the whitened wavelet coefficients. For a sequence of independent complex Wishart matrices, the trace of the noisy whitened coefficients decomposes into an additive signal plus mean-zero noise sequence model. Moreover, the variance of the trace of the noisy whitened coefficients is constant across wavelet scales, and since the trace operator outputs a scalar, one can directly apply ordinary scalar thresholding or shrinkage methods to the matrix-valued coefficients. Thresholding or shrinkage of the trace of the whitened coefficients is equivariant under unitary congruence transformations as in Proposition \ref{prop:4.2}. Moreover, it is equivariant under congruence transformation by any invertible matrix, i.e., \emph{general linear} congruence equivariant. In the context of spectral estimation of multivariate time series, this means that the estimator does not nontrivially depend on the chosen basis or coordinate system of the time series, as the spectral estimator is equivariant under a change of basis of the time series.
	
	\begin{lemma} (General linear congruence equivariance) \label{lem:4.3}
	Let $(X_\ell)_\ell$ be a sequence of HPD matrices and $(\hat{f}_\ell)_\ell$ its wavelet-denoised estimate based on linear or nonlinear shrinkage of the trace of the whitened wavelet coefficients. The estimator is equivariant under general linear congruence transformation in the sense that the wavelet-denoised estimate $(\hat{f}_{A,\ell})_\ell$ of $(A \ast X_\ell)_\ell$ equals $(A \ast \hat{f}_\ell)_\ell$ for each $A \in \te{GL}(\mathbb{C})$, with $\te{GL}(\mathbb{C})$ the space of invertible complex matrices.
	\end{lemma} 
	
	\noindent In the following, $\widetilde{P}_f$ denotes the probability distribution associated to a bias-corrected complex Wishart distribution $e^{-c(d,L)}W_d^C(L, L^{-1} f)$ as in Theorem \ref{thm:4.1}, with $L \geq d$ to ensure positive-definiteness of the Wishart matrix. Here, $\widetilde{P}_f \in P_2(\mathbb{P}_{d \times d})$ is understood to be the distribution of a random variable $X = f^{1/2} \ast W$, where $W$ is an HPD complex Wishart matrix, with $L$ degrees of freedom, not depending on $f$, and with intrinsic mean equal to the identity matrix $\te{Id}$. Note that the latter directly implies that the intrinsic mean of $f^{1/2} \ast W$ is equal to $f$.
	
	\begin{prop} \label{prop:4.4} (Trace properties)
	Let $X_\ell \sim \widetilde{P}_{f_\ell}$, independently distributed for $\ell=1,\ldots,n$, with $n=2^J$. For each scale-location $(j,k)$, the whitened wavelet coefficients obtained from the intrinsic AI wavelet transform of order $N = 2L+1\geq 1$ satisfy:
	\begin{eqnarray*}
	\tr(\mathfrak{D}_{j,k}^X) &=& \tr(\mathfrak{D}_{j,k}^f) + \tr(\mathfrak{D}_{j,k}^W),
	\end{eqnarray*}
	where $\mathfrak{D}_{j,k}^X$ is the random whitened coefficient based on the sequence $(X_\ell)_{\ell=1}^n$, $\mathfrak{D}_{j,k}^f$ is the deterministic whitened coefficient based on the sequence of intrinsic means $(f_\ell)_{\ell=1}^n$, and $\mathfrak{D}_{j,k}^W$ is the random whitened coefficient based on an i.i.d. sequence of Wishart matrices $(W_\ell)_{\ell=1}^n$, with intrinsic mean equal to the identity, independent of $(f_\ell)_{\ell=1}^n$.\\[3mm]
	Moreover, $\bs{E}[\tr(\mathfrak{D}_{j,k}^X)] \ = \ \tr(\mathfrak{D}_{j,k}^f)$, and,
	\begin{eqnarray} \label{eq:4.1}
	\te{Var}(\tr(\mathfrak{D}_{j,k}^X)) &=& \left( 2^{-J}\sum_{i=0}^{2N-1} \bs{C}_{L,i}^2 \right) \left( \sum_{i=1}^d \psi'(L - (d-i)) \right),
	\end{eqnarray}
	where $\psi'(\cdot)$ denotes the trigamma function, and $(\bs{C}_{L,i})_i$ are the filter coefficients as in eq.(\ref{eq:2.5}). In particular, $\te{Var}(\tr(\mathfrak{D}_{j,k}^X))$ is independent of the scale-location $(j,k)$ and whenever $\tr(\mathfrak{D}_{j,k}^f)$ vanishes $\bs{E}[\tr(\mathfrak{D}_{j,k}^X)] = 0$, e.g., when $(f_\ell)_\ell$ is sampled from an intrinsic polynomial of order smaller than $N$.
	\end{prop}
	
	\begin{corr} (Centered noise) \label{cor:4.5}
	With the same notation as in Proposition \ref{prop:4.4}, the random whitened wavelet coefficients $\mathfrak{D}_{j,k}^W$ based on a sequence of i.i.d.\@ Wishart matrices $(W_\ell)_{\ell=1}^n$, with identity intrinsic mean satisfy,
	\begin{eqnarray*}
	\bs{E}[\mathfrak{D}_{j,k}^W] &=& \bs{0},
	\end{eqnarray*}
	where $\bs{E}[\cdot]$ denotes the (ordinary) Euclidean expectation.
	\end{corr}
	
	\noindent Based on the trace of the whitened coefficients, by Proposition \ref{prop:4.4}, in the context of a sequence of approximate complex random Wishart matrices, such as a curve of periodogram matrices, any preferred standard wavelet shrinkage procedure can be applied well-suited to scalar additive signal plus noise sequence models, with homogeneous variances across coefficient scales. 


\section{Illustrative data examples} \label{sec:5}

	\begin{figure}
	\centering
	\includegraphics[scale=0.55]{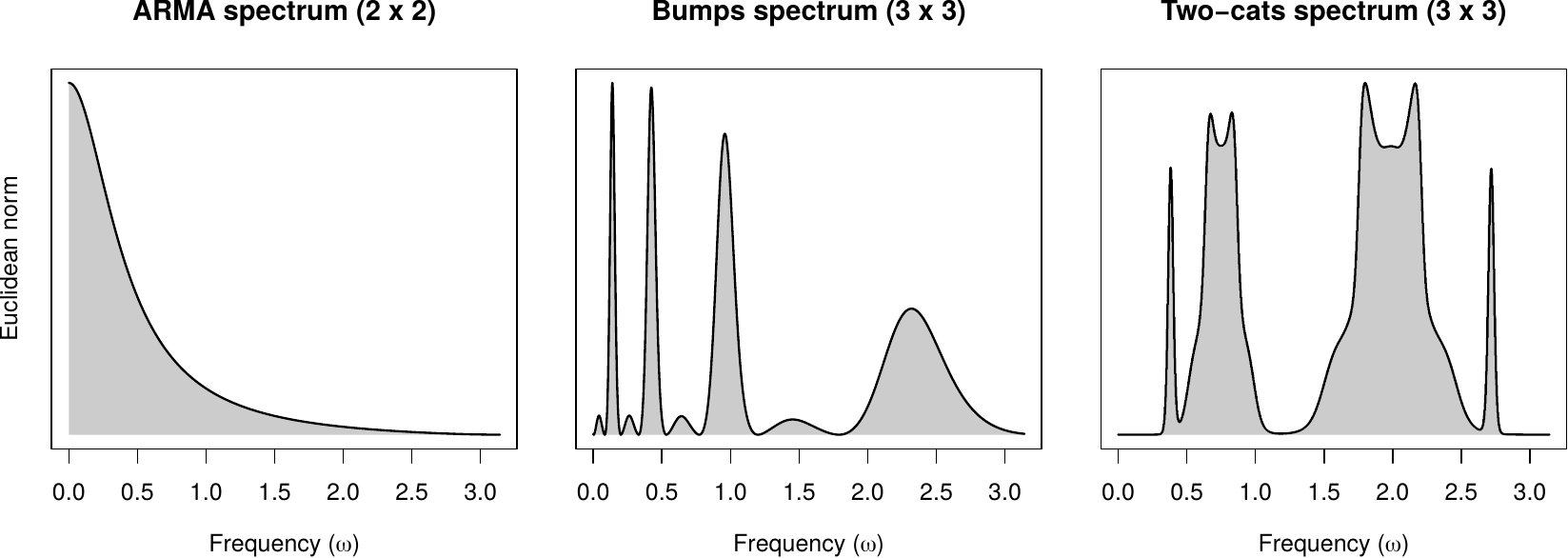}
	\caption{Euclidean norms of HPD test spectral matrices generated with \texttt{rExamples1D()}.} \label{fig:3}
	\end{figure}

	\subsection{Finite-sample performance} \label{sec:5.1}
	
	\paragraph{Simulation setup}
	
	In the figures below, we assess the finite-sample performance of intrinsic wavelet-based curve estimation in the space of HPD matrices and benchmark the performance against several alternative nonparametric smoothing procedures. In particular, we consider HPD test curves displaying both globally homogeneous and locally varying smoothness behavior, available through the function \texttt{rExamples1D()} in the \texttt{pdSpecEst}-package. The \texttt{arma} spectrum is a smooth ($2 \times 2$)-dimensional HPD spectral matrix generated by a stationary $\te{ARMA}(1,1)$ process based on \cite[Example 11.4.1]{BD06}. The \texttt{bumps} spectrum is a curve of ($3 \times 3$)-dimensional HPD matrices containing local bumps of various degrees of smoothness, and the \texttt{two-cats} spectrum visualizes the contours of two cats and consists of relatively smooth parts combined with local peaks and troughs. Figure \ref{fig:3} displays the Euclidean norm of the HPD matrix-valued curves as a function of frequency. Each test spectrum is normalized to have unit Euclidean norm over the integrated frequency range $[0, \pi]$. \\[3mm]
	Given the HPD test curves, random observations $(X_\ell)_{\ell = 1, \ldots, n} \in \mathbb{P}_{d \times d}$ are generated according to several different model distributions centered around the target curve $(f_\ell)_{\ell = 1,\ldots,n} \in \mathbb{P}_{d \times d}$. The data generating models and associated metrics used for estimation are summarized in Table \ref{tab:2}. In the periodogram noise scenario,  first a $d$-dimensional time series trace is generated from the target spectrum $f$ via its Cram\'er representation with complex normal random variates as in e.g., \cite[Section 4.6]{B81}, and second an initial HPD multitaper periodogram $(X_\ell)_\ell$ is computed based on $d$ discrete prolate spheroidal (DPSS) taper functions. The observations $(X_\ell)_\ell$ tend in distribution to the Wishart noise scenario as the length of the time series increases. The scale of the noise distributions in the Log-Gaussian and Riemannian Gaussian noise scenarios is chosen such that the signal-to-noise ratio is comparable to the  Wishart and periodogram noise scenarios. Additional details on the intrinsic signal-noise model in the Riemannian-Gaussian noise scenario are found in \cite[Section 2.2.6]{C18}.\\[3mm] 	
	
		\begin{table}
	\centering
	\begin{threeparttable}
	\caption{Simulation setup and signal-noise models. \label{tab:2}}
	\begin{footnotesize}
	\setlength{\tabcolsep}{1em}
	\begin{tabular}[t]{llllc}
	\toprule
	Simulation scenario & Signal-noise model & Noise distribution$^*$ & Metric & B-C$^\dagger$ \\ 
	\midrule
	\multirow{2}{*}{Wishart noise} & \multirow{2}{*}{$X_\ell = f_\ell^{1/2} \ast Z_\ell$} & \multirow{2}{*}{$Z_\ell \overset{\te{iid}}{\sim} \frac{1}{d}W_d^C(d, \te{Id})$} & Riemannian & \ding{51} \\ 
	&&& Cholesky & \ding{51} \\
	\multirow{2}{*}{Log-Gaussian noise} & \multirow{2}{*}{$X_\ell = \Exp(\Log(f_\ell) + Z_\ell)$} & $Z_\ell 	\overset{d}{=} \sum_{k=1}^{d^2} z_k e^k,$ & \multirow{2}{*}{Log-Euclidean} & \multirow{2}{*}{\ding{55}} \\
	&& $(z_k)_k \overset{\te{iid}}{\sim} N(0, 1/4)$ & \\
	\multirow{2}{*}{Riem.-Gaussian noise} & \multirow{2}{*}{$X_\ell = f_\ell^{1/2} \ast Z_\ell$} & $Z_\ell \overset{d}{=} \sum_{k=1}^{d^2} z_k e^k,$ & \multirow{2}{*}{Riemannian} & \multirow{2}{*}{\ding{55}} \\
	&& $(z_k)_k \overset{\te{iid}}{\sim} N(0, 1/4)$ & \\
	\multirow{2}{*}{Periodogram noise} & \multirow{2}{*}{$X_\ell = \bar{I}_T(\omega_\ell)$} & \scriptsize{Multitaper periodogram} & \multirow{2}{*}{Riemannian} & \multirow{2}{*}{\ding{51}} \\
	&& \scriptsize{with $d$ DPSS tapers} &&\\[1mm]
	\bottomrule
	\end{tabular}
 	\begin{tablenotes}
 	\begin{footnotesize}
    \item $*$: $\{e^1,\ldots,e^{d^2}\} \in \mathbb{H}_{d \times d}$ is an orthonormal basis of $(\mathbb{H}_{d \times d}, \langle \cdot, \cdot \rangle_F)$.
    \item $\dagger$: B-C denotes whether a bias-correction is required for the given data generating scenario and metric.
    \end{footnotesize}
    \end{tablenotes}
 	\end{footnotesize}
	\end{threeparttable}
	\end{table}	
	\vspace{-5mm}
	
	\noindent In each individual simulation experiment, the intrinsic integrated squared estimation error (IISE) is calculated as the integrated squared error based on the distance associated to the metric used for estimation. These are, respectively, the Riemannian distance $\delta_R$; the Log-Euclidean distance $\delta_L(x,y) = \Vert \Log(y) - \Log(x) \Vert_F$; and the Cholesky distance $\delta_C(x,y) = \Vert \te{Chol}(y) - \te{Chol}(x) \Vert_F$, with $x, y \in \mathbb{P}_{d \times d}$. For the Wishart noise scenario, HPD matrix curve estimation subject to the Riemannian or the Cholesky metric is biased with respect to the target HPD matrix curve. Under the Cholesky metric, this bias can be corrected by the bias-correction in \cite[Theorem 1]{DG04}. Under the Riemannian metric, we apply the bias-correction in Theorem \ref{thm:4.1}. For the periodogram noise scenario, we again make use of the bias-correction in Theorem \ref{thm:4.1}. For the Log-Gaussian noise scenario and estimation subject to the Log-Euclidean metric, the estimators are unbiased and no bias-correction is necessary. The same holds true for estimation in the Riemannian-Gaussian noise scenario and estimation with respect to the Riemannian metric.

	\paragraph{Estimation procedures} 

	The simulation experiments include linear thresholding of wavelet scales according to Section \ref{sec:3} and nonlinear trace thresholding of wavelet coefficients as in Section \ref{sec:4.1} in the space of HPD matrices equipped with the Riemannian, Log-Euclidean or Cholesky metric. As a straightforward nonlinear thresholding method, we consider scalar dyadic tree-structured thresholding based on the wavelet coefficient traces similar to \cite{D97}. More precisely, for each scale-location $(j,k)$, denote $d_{j,k} = \tr(\mathfrak{D}_{j,k}^X)$ for the trace of the observed whitened wavelet coefficient and let $w_{j,k} \in \{0, 1\}$ be a binary label. Given a regularization parameter $\lambda \geq 0$, we optimize the following complexity penalized loss criterion:
	\begin{eqnarray} \label{eq:5.1}
	\arg\!\min_{\bs{w}} L(\bs{w}) &=& \arg\!\min_{\bs{w}} \sum_{j,k} |d_{j,k}w_{j,k} - d_{j,k}|^2 + \lambda^2 \sum_{j,k} w_{j,k},
	\end{eqnarray} 
	under the constraint that the nonzero labels $\{ w_{j,k}\, |\, w_{j,k} = 1\}$ form a dyadic rooted tree, i.e., for each nonzero label $w_{j+1,2k+1}$ or $w_{j+1,2k}$, the label $w_{j,k}$ also has to be nonzero. This minimization problem can be solved in $O(n)$ computations via the tree-pruning algorithm in \cite{D97}, with $n$ the total number of coefficients, resulting in the estimated wavelet coefficients $\widehat{D}_{j,k} = w_{j,k} D_{j,k}^X$. Linear and nonlinear tree-structured wavelet thresholding are available in the \texttt{pdSpecEst}-package through the function \texttt{pdSpecEst1D()} and the argument \texttt{metric} set to the appropriate metric. The choice \texttt{metric = "Riemannian-Rahman"} replaces the forward and backward AI wavelet transforms by the MI wavelet transforms in \cite{R05} based on the affine-invariant Riemannian metric. The latter is slightly different to the metric suggested in \cite[Section 4.4]{R05}, which does not enjoy the same congruence invariance properties as the Riemannian metric. In addition to intrinsic wavelet-based curve estimation, we have implemented intrinsic versions of the following curve estimation procedures in the space of HPD matrices equipped with the Riemannian, Log-Euclidean and Cholesky metric: (i) Nearest-Neighbor (NN) regression, (ii) Cubic Spline (CS) regression (as in \cite{BA11}, \cite{BA11b}) and (iii) Local Polynomial (LP) regression (as in \cite{Y12}). In the periodogram noise scenario, a benchmark multitaper spectral estimator based on the generated time series has also been included. Details about the listed estimation procedures are found in Appendix III in the supplementary material.

	\begin{figure}
	\centering
	\includegraphics[scale=0.67]{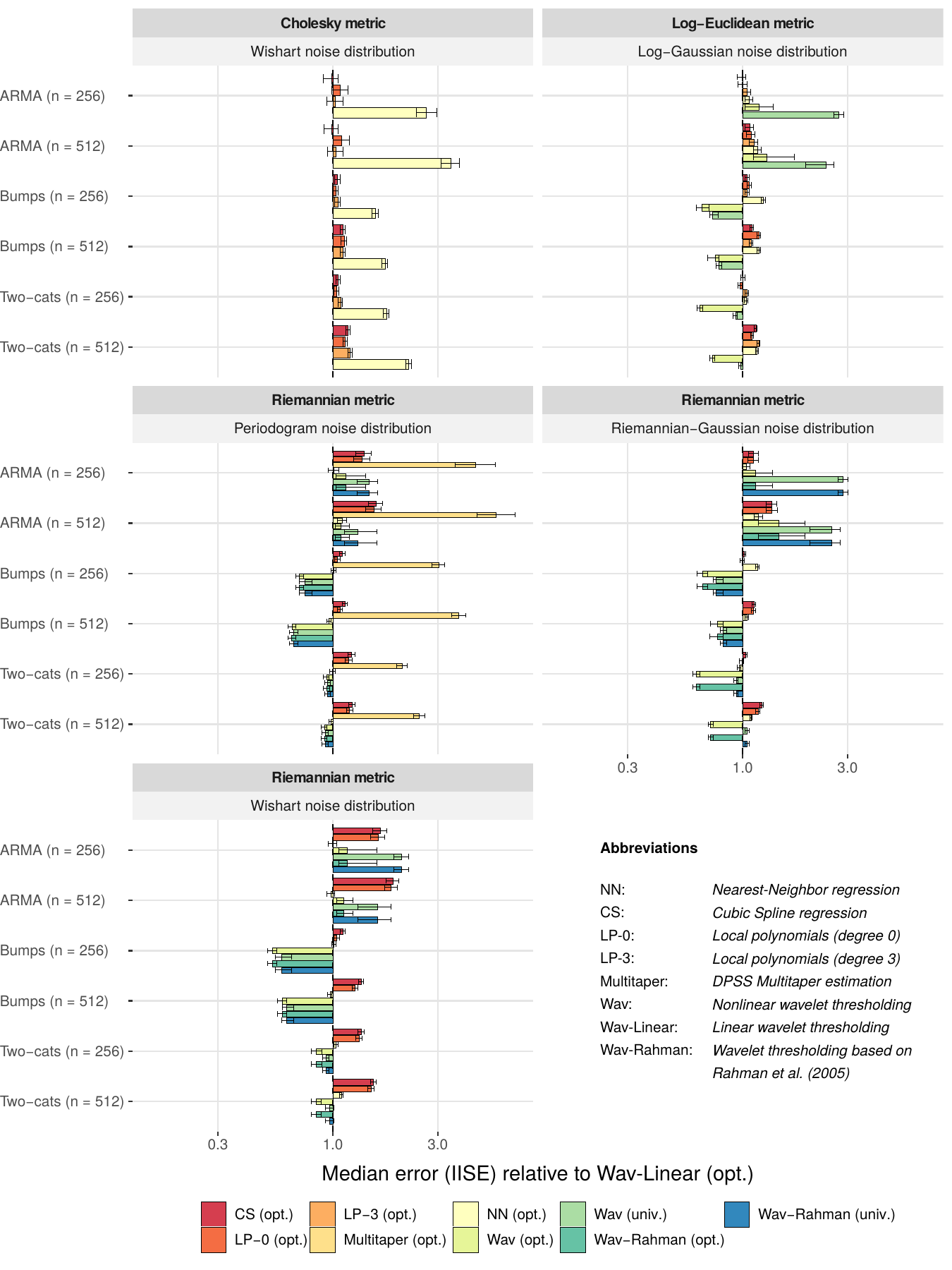}
	\caption{Relative median intrinsic integrated squared estimation errors (IISE) of the wavelet and benchmark estimation procedures for the \texttt{arma}, \texttt{bumps} and \texttt{two-cats} test spectra, relative to the IISE of intrinsic linear wavelet thresholding. \label{fig:4}}
	\end{figure}

	\paragraph{Simulation results}
	
	Figure \ref{fig:4} displays the relative median intrinsic integrated squared errors (IISEs), based on $M = 10\ 000$ replications per simulation scenario, with target spectral matrices sampled at $n = \{256, 512\}$ locations. For each simulation scenario replication, the IISEs are standardized with respect to the IISE of linear wavelet estimation, in order to allow for straightforward comparison of estimation performance across metrics and noise scenarios. More precisely, all error bars to the left of the vertical unit line outperform linear wavelet thresholding in terms of median IISE, and the opposite for error bars to the right. The included whiskers correspond to the first and third quartile of the relative error distributions. Each estimation procedure depends on a single main tuning parameter: for the linear wavelet estimator, this is the number of nonzero wavelet scales; for the tree-structured wavelet and cubic spline estimators, this is the regularization parameter; for the nearest neighbor regression, this is the number of nearest neighbors; for the local polynomial estimator, this is the bandwidth parameter; and for the multitaper estimator, this is the number of tapering functions. In each simulation experiment, an oracle tuning parameter, denoted by \texttt{(opt.)}, is determined by minimizing the IISE with respect to the true target HPD matrix curve. In addition, for the tree-structure wavelet estimators a choice of the regularization parameter based on a universal threshold is included, denoted by \texttt{(univ.)}. For the Wishart noise scenario subject to the Cholesky metric, nonlinear wavelet thresholding has been excluded from the simulations, as the traces of the wavelet coefficients cannot be shown to decompose into a scalar signal plus noise sequence model, which is the case for the other simulation scenarios subject to the Riemannian and Log-Euclidean metric.\\[3mm]
	The periodogram noise scenario mimics the distributional behavior of the periodogram in practice and its simulation results are therefore of primary interest. Subject to the Riemannian metric, the IISE of linear wavelet thresholding performs roughly similar in terms of the IISE to the nearest-neighbor, cubic spline and local polynomial benchmark procedures for the \texttt{bumps} and \texttt{two-cats} spectra and outperforms the benchmarks for the highly smooth \texttt{arma} spectrum. Replacing linear wavelet thresholding by nonlinear wavelet thresholding further reduces the IISE for the \texttt{bumps} and \texttt{two-cats} spectra. This is attributed to the fact that, in contrast to the other approaches, nonlinear wavelet thresholding is able to capture varying degrees of smoothness in the HPD matrix curve. On the other hand, linear wavelet thresholding does outperform nonlinear wavelet thresholding in estimating the \texttt{arma} spectrum, as a single global smoothing parameter is sufficient to capture the smooth behavior in the HPD spectral matrix. Furthermore, it is observed that the IISE of nonlinear tree-structured thresholding based on a standard universal threshold is relatively close to the optimal IISE for nonlinear tree-structured thresholding, thereby providing a fast heuristic choice of the main tuning parameter in practical applications. For the considered benchmark procedures, there is no simple heuristic choice for the main tuning parameter(s), and one needs to resort either to cross-validation procedures or manual smoothing parameter tuning. We point out that a drawback of the wavelet methods in their current form is the need for dyadic sample sizes, and additional data pre-processing or modifications to the wavelet transforms are required to handle non-dyadically sampled periodograms. The benchmark multitaper spectral estimator does not achieve the same level of performance as the other benchmark procedures in terms of the IISE. This is explained by the fact that, in contrast to the other approaches, the multitaper estimator considers the space of HPD matrices as a Euclidean space, but the estimation error is computed with respect to the Riemannian metric and not the Euclidean metric. Nonlinear wavelet thresholding based on the MI approach in \cite{R05} performs roughly similar to intrinsic nonlinear wavelet thresholding in the empirical setting, but lacks the same intrinsic consistency and convergence properties discussed in Sections \ref{sec:2} and \ref{sec:3}.\\[3mm]
	The simulation results for the Riemannian-Gaussian noise and the Wishart noise scenarios subject to the Riemannian metric display the same overall characteristics as observed for the periodogram noise scenario. In particular, the relative median errors under the Wishart noise scenario correspond roughly to a scaled version of the relative median errors under the periodogram noise scenario. This suggests that in the considered simulation scenarios, the distributional behavior of the periodogram and Wishart noise distributions are comparable, thereby providing further validation of the trace thresholding approach in Section \ref{sec:4.1} based on approximating the periodogram noise distribution by its asymptotically equivalent complex Wishart distribution. For the Wishart noise scenario subject to the Cholesky metric, the estimation performance of intrinsic cubic spline and local polynomial regression is similar to linear wavelet thresholding, whereas nearest-neighbor estimation performs somewhat worse in particular for the smooth \texttt{arma} spectrum. Similar observations can be made for the Log-Gaussian noise scenario subject to the Log-Euclidean metric. In addition, nonlinear wavelet thresholding is seen to outperform linear thresholding for the non-smooth \texttt{bumps} and \texttt{two-cats} spectrum, but performs worse than linear thresholding for the smooth \texttt{arma} spectrum, which is consistent with the observed results under the Riemannian metric.

	\subsection{Associative learning experiment LFP data} \label{sec:5.2}

	\begin{figure}[t]
	\centering
	\includegraphics[scale=0.7]{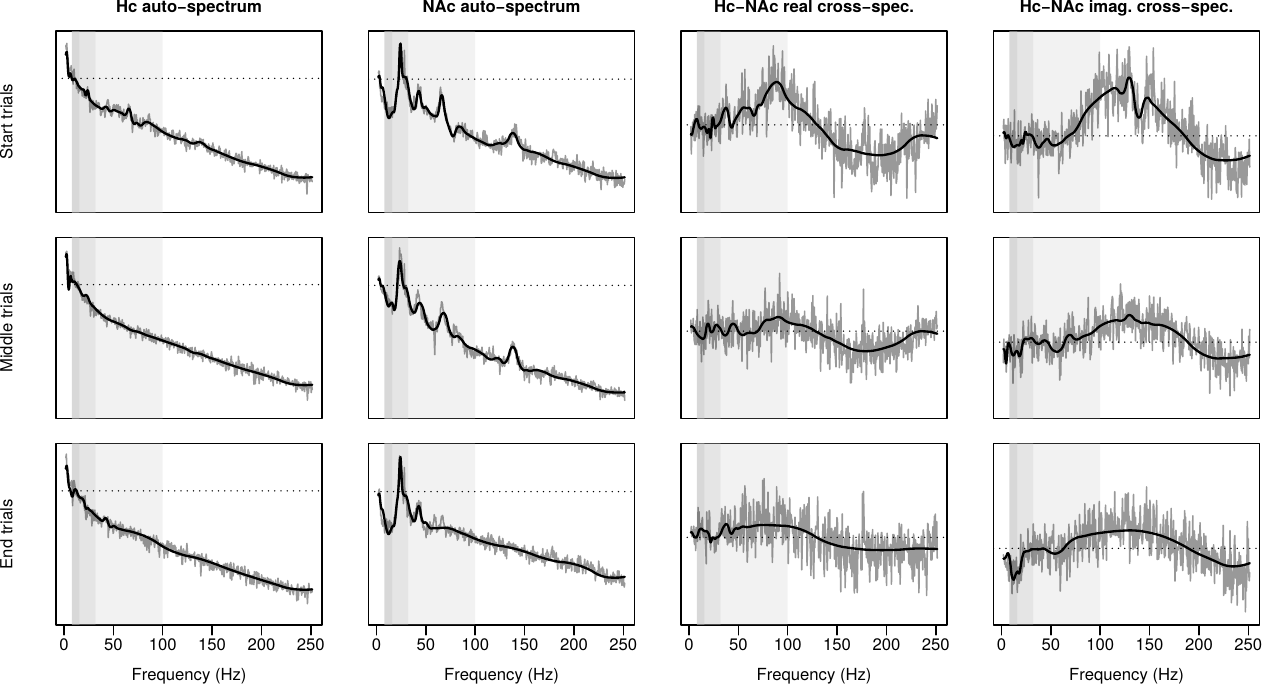}
	\caption{Row-expanded matrix logarithms of HPD periodograms averaged across LFP time series trials at the start, middle and end of the learning experiment across Fourier frequencies.}\label{fig:5}
	\end{figure}
	
\begin{figure}[t]
	\centering
	\includegraphics[scale = 0.7]{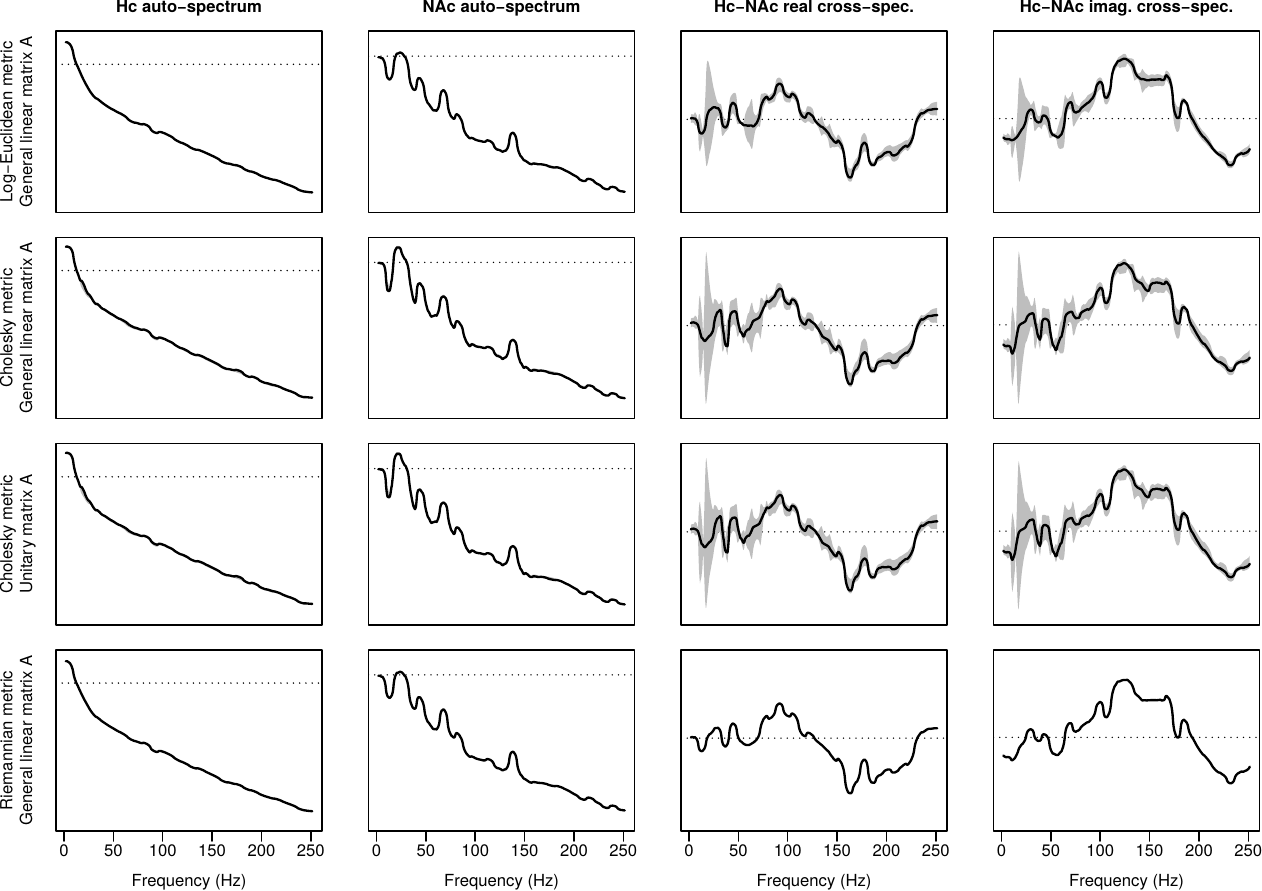}
	\caption{Row-expanded matrix logarithms of linear wavelet thresholded spectral estimates $\hat{f}(\omega)$ and $\hat{f}_m(\omega)$ subject to the Log-Euclidean, Cholesky and Riemannian metrics.}\label{fig:6}
	\end{figure}

	As an additional data example, we consider spectral matrix estimation for a subset of brain signal time series trials recorded over the course of an associative learning experiment with a macaque, see \cite{G12} or \cite{FO16} for additional details. During the learning experiment, the electrical activity in the brain of the macaque is measured by means of local field potentials (LFP). After preprocessing of the LFP time series, there remain a total of $S = 590$ trial-specific approximately stationary 2-dimensional time series traces of length $T = 2048$ sampled at 1\,000\,Hz. The two time series components correspond to LFP measurements in the hippocampus (Hc) and nucleus accumbens (NAc) regions of the macaque's brain, which have previously been implicated in cognitive processes involving memory and reward, as detailed in \cite{FO16} and the references therein. For demonstrational purposes, we extract trials from the start of the experiment ($s = 1,\ldots, 10$), the middle of the experiment ($s = 291,\ldots, 300$) and the end of the experiment ($s = 581,\ldots, 590$). For each of the trial subsets, an averaged HPD ($2 \times 2$)-periodogram matrix is computed by averaging the trial-specific raw ($2\times2$)-periodogram matrices across Fourier frequencies. Figure \ref{fig:5} displays the matrix logarithms of the initial noisy HPD periodograms up to $250$ Hz averaged across LFP trial subsets. The grey bands display respectively the $\alpha$-band (8-16 Hz), the $\beta$-band (16-32 Hz) and the $\gamma$-band (32-100 Hz). The overlayed black lines correspond to nonlinear wavelet denoised HPD periodograms subject to the Riemannian metric obtained with the function \texttt{pdSpecEst1D()}, with refinement order $N = 5$ and tree-structured trace thresholding based on a rescaled universal threshold.\\[3mm]
	Let $f(\omega)$ denote the theoretical ($2 \times 2$)-dimensional HPD spectral matrix of the stationary LFP time series process at frequency $\omega$. Among other steps, preprocessing of the raw LFP time series data includes standardizing the time series traces to have zero mean and unit variance. After standardization, the spectral matrix transforms as $A \ast f(\omega)$, with $A$ given by some diagonal matrix $A = ((\theta_1, 0)', (0, \theta_2)')$. If one also permutes the order of the time series traces, the matrix $A$ becomes $A = ((0, \theta_1)', (\theta_2, 0)')$. These are two straightforward examples of preprocessing steps that ideally should not have a nontrivial impact on the final spectral estimator as previously argued in Section \ref{sec:4.1}. In Figure \ref{fig:6}, we demonstrate the effects such transformations can have on the estimation of the LFP spectral matrices, focusing on the periodogram data associated to the middle of the experiment. Here, $M = 1\,000$ random $2 \times 2$-invertible matrices $A_m$ with standard complex Gaussian matrix entries are generated. First, the initial HPD periodograms $I_T(\omega)$ are transformed by $A_m \ast I_T(\omega)$, imitating a basis transformation of the LFP time series data. Second, a linear wavelet thresholded spectrum $\hat{f}_m(\omega)$ is calculated, discarding all coefficients above scale $J = 4$. Denoising by means of linear thresholding allows for straightforward visual comparisons between different metrics. Third, the spectral estimates are transformed back to the original basis of the LFP time series data, according to $A_m^{-1} \ast \hat{f}_m(\omega)$. Under the Cholesky metric, the same procedure is repeated with random $2 \times 2$-unitary matrices $A_m \in \mathcal{U}$, sampled with respect to the (additively invariant) Haar measure on $\mathcal{U}$. The black lines in Figure \ref{fig:6} display the spectral estimate $\hat{f}(\omega)$ obtained from the original periodogram data. The grey regions include all spectral estimates $\hat{f}_m(\omega)$ subject to congruence transformation by the random matrices $A_m$. For the Log-Euclidean and Cholesky metric, the estimated Hc and NAc auto-spectral components are nearly equivariant, but the estimated cross-spectral components potentially display a high degree of non-equivariance depending on the choice of $A_m$. Note that this observed non-equivariance directly extends to the estimated coherence, which are obtained as normalized versions of the estimated cross-spectra. 
	
\section{Concluding remarks}
	
	The primary contribution of this paper is the development of intrinsic average-interpolation (AI) wavelet transforms and intrinsic wavelet thresholding for curves in the space of HPD matrices equipped with the affine-invariant Riemannian metric. The intrinsic wavelet transforms are constructed independent of the chosen metric and although the wavelet coefficient decay and nonparametric convergence rates in Section \ref{sec:3} are derived exclusively for the affine-invariant Riemannian metric, similar arguments apply to other metrics as well. For instance, in a Euclidean space the intrinsic Taylor expansions reduce to ordinary Taylor expansions, as the parallel transport is the identity map and the covariant derivatives are standard matrix derivatives. In the context of high-dimensional time series, estimation of the spectral matrix with respect to the Riemannian metric may suffer from computational instability, as the estimation target may be located close to or at the boundary of the space of positive definite matrices. Alternative metrics, besides the Euclidean metric, that can handle rank deficient spectral matrices include e.g., the Procrustes shape-and-size metric or the Cholesky metric, see \cite{D09}. However, polynomial interpolation with respect to the Procrustes metric may lead to negative definite matrices, similar to the Euclidean metric, and the Cholesky square root matrix is not necessarily unique in the rank deficient case. The challenge of flexible estimation of nonnegative definite spectral matrices is currently a topic of interest for future research. Furthermore, Hermitian or symmetric positive definite matrices are encountered as autocovariance matrices or spectral density matrices in time series analysis, but also play an important role in the fields of medical imaging, computer vision or radar signal processing (e.g., \cite{PFA05}), and it is of interest to apply the intrinsic wavelet methods for the purpose of compression or denoising in other settings than spectral matrix estimation. For instance, applied to diffusion tensor imaging, intrinsic wavelet shrinkage or thresholding shows potential for fast denoising of large collections of non-smoothly varying diffusion tensors. \\[3mm]
	In \cite{COvS19}, the notion of intrinsic data depth in the space of HPD matrices equipped with the affine-invariant Riemannian metric is discussed, providing a center-to-outward ordering of a collection of HPD matrices. In the context of HPD spectral matrix estimation, the data depths are useful tools to construct confidence regions for the spectral matrix --intrinsic to the Riemannian geometry of the space-- based on for instance a parametric bootstrap using the data generating process of a stationary time series via its Cram\'er representation as detailed in \cite{DG04} and \cite{FO16} among others. In \cite[Chapter 5]{C18}, the intrinsic wavelet methods presented in this paper are extended to \emph{surfaces} of Hermitian positive definite matrices, with in mind the application to nonparametric estimation of the time-varying spectrum of a locally stationary time series. In addition to spectral matrix denoising, other potential applications of the intrinsic wavelet transforms include e.g., spectral matrix clustering, classification or peak detection based on the sparse representations in the intrinsic wavelet domain. 


\section*{Acknowledgments}

The authors gratefully acknowledge financial support from the following agencies and projects: the Belgian Fund for Scientific Research FRIA/FRS-FNRS (J. Chau), the contract ``Projet d'Actions de Recherche Concert\'ees'' No. 12/17-045 of the ``Communaut\'e fran\c{c}aise de Belgique'' (R. von Sachs), IAP research network P7/06 of the Belgian government (R. von Sachs). We thank the UC Irvine Space-Time Modeling Group and Dr. Emad Eskandar (Massachussetts General Hospital) for the local field potential data to illustrate the methodology and the anonymous referees for their suggestions that helped improving the presentation of this work.


\bibliographystyle{chicago}
\begin{small}
\bibliography{Main}

\begin{thebibliography}{}

\bibitem[\protect\citeauthoryear{Antoniadis}{Antoniadis}{1997}]{A97}
Antoniadis, A. (1997).
\newblock Wavelets in statistics: a review.
\newblock {\em Statistical methods \& applications\/}~{\em 6\/}(2), 97--130.

\bibitem[\protect\citeauthoryear{Bhatia}{Bhatia}{2009}]{B09}
Bhatia, R. (2009).
\newblock {\em Positive Definite Matrices}.
\newblock New Jersey: Princeton University Press.

\bibitem[\protect\citeauthoryear{Boumal and Absil}{Boumal and
  Absil}{2011a}]{BA11b}
Boumal, N. and P.-A. Absil (2011a).
\newblock A discrete regression method on manifolds and its application to data
  on {SO}(n).
\newblock {\em IFAC Proceedings Volumes\/}~{\em 44\/}(1), 2284--2289.

\bibitem[\protect\citeauthoryear{Boumal and Absil}{Boumal and
  Absil}{2011b}]{BA11}
Boumal, N. and P.-A. Absil (2011b).
\newblock Discrete regression methods on the cone of positive-definite
  matrices.
\newblock In {\em IEEE ICASSP, 2011}, pp.\  4232--4235.

\bibitem[\protect\citeauthoryear{Brillinger}{Brillinger}{1981}]{B81}
Brillinger, D. (1981).
\newblock {\em Time Series: Data Analysis and Theory}.
\newblock San Francisco: Holden-Day.

\bibitem[\protect\citeauthoryear{Brockwell and Davis}{Brockwell and
  Davis}{2006}]{BD06}
Brockwell, P. and R.~Davis (2006).
\newblock {\em Time Series: Theory and Methods}.
\newblock New York: Springer.

\bibitem[\protect\citeauthoryear{Chau}{Chau}{2017}]{C17}
Chau, J. (2017).
\newblock pd{S}pec{E}st: {A}n {A}nalysis {T}oolbox for {H}ermitian {P}ositive
  {D}efinite {M}atrices.
\newblock R Package version 1.2.3.
\newblock https://CRAN.R-project.org/package=pdSpecEst.

\bibitem[\protect\citeauthoryear{Chau}{Chau}{2018}]{C18}
Chau, J. (2018).
\newblock {\em Advances in Spectral Analysis for Multivariate, Nonstationary
  and Replicated Time Series}.
\newblock Ph.\ D. thesis, Universit{\'e} catholique de Louvain.

\bibitem[\protect\citeauthoryear{Chau, Ombao, and von Sachs}{Chau
  et~al.}{2019}]{COvS19}
Chau, J., H.~Ombao, and R.~von Sachs (2019).
\newblock Intrinsic data depth for {H}ermitian positive definite matrices.
\newblock {\em Journal of Computational and Graphical Statistics\/}~{\em
  28\/}(2), 427--439.

\bibitem[\protect\citeauthoryear{Dahlhaus}{Dahlhaus}{2012}]{D12}
Dahlhaus, R. (2012).
\newblock {\em Locally stationary processes}, Chapter in Time Series Analysis:
  Methods and Applications, Vol. 30, pp.\  351--413.
\newblock Amsterdam: Elsevier.

\bibitem[\protect\citeauthoryear{Dai and Guo}{Dai and Guo}{2004}]{DG04}
Dai, M. and W.~Guo (2004).
\newblock Multivariate spectral analysis using {C}holesky decomposition.
\newblock {\em Biometrika\/}~{\em 91\/}(3), 629--643.

\bibitem[\protect\citeauthoryear{do~Carmo}{do~Carmo}{1992}]{D92}
do~Carmo, M. (1992).
\newblock {\em Riemannian Geometry}.
\newblock Boston: Birkh\"auser.

\bibitem[\protect\citeauthoryear{Donoho}{Donoho}{1993}]{D93}
Donoho, D. (1993).
\newblock {\em Smooth wavelet decompositions with blocky coefficient kernels},
  Chapter in Recent Advances in Wavelet Analysis, pp.\  259--308.
\newblock New York: Academic Press.

\bibitem[\protect\citeauthoryear{Donoho}{Donoho}{1997}]{D97}
Donoho, D. (1997).
\newblock Cart and best-ortho-basis: a connection.
\newblock {\em The Annals of Statistics\/}~{\em 25\/}(5), 1870--1911.

\bibitem[\protect\citeauthoryear{Dryden, Koloydenko, and Zhou}{Dryden
  et~al.}{2009}]{D09}
Dryden, I., A.~Koloydenko, and D.~Zhou (2009).
\newblock Non-{E}uclidean statistics for covariance matrices, with applications
  to diffusion tensor imaging.
\newblock {\em The Annals of Applied Statistics\/}~{\em 3\/}(3), 1102--1123.

\bibitem[\protect\citeauthoryear{Fiecas and Ombao}{Fiecas and
  Ombao}{2016}]{FO16}
Fiecas, M. and H.~Ombao (2016).
\newblock Modeling the evolution of dynamic brain processes during an
  associative learning experiment.
\newblock {\em Journal of the American Statistical Association\/}~{\em
  111\/}(516), 1440--1453.

\bibitem[\protect\citeauthoryear{Gorrostieta, Ombao, Prado, Patel, and
  Eskandar}{Gorrostieta et~al.}{2012}]{G12}
Gorrostieta, C., H.~Ombao, R.~Prado, S.~Patel, and E.~Eskandar (2012).
\newblock Exploring dependence between brain signals in a monkey during
  learning.
\newblock {\em Journal of Time Series Analysis\/}~{\em 33\/}(5), 771--778.

\bibitem[\protect\citeauthoryear{Higham}{Higham}{2008}]{H08}
Higham, N.~J. (2008).
\newblock {\em Functions of Matrices: Theory and Computation}.
\newblock Philadelphia: Siam.

\bibitem[\protect\citeauthoryear{Hinkle, Fletcher, and Joshi}{Hinkle
  et~al.}{2014}]{HFJ14}
Hinkle, J., P.~Fletcher, and S.~Joshi (2014).
\newblock Intrinsic polynomials for regression on {R}iemannian manifolds.
\newblock {\em Journal of Mathematical Imaging and Vision\/}~{\em 50\/}(1-2),
  32--52.

\bibitem[\protect\citeauthoryear{Ho, Cheng, Salehian, and Vemuri}{Ho
  et~al.}{2013}]{H13}
Ho, J., G.~Cheng, H.~Salehian, and B.~Vemuri (2013).
\newblock Recursive {K}archer expectation estimators and recursive law of large
  numbers.
\newblock In {\em AISTATS, 2013}, pp.\  325--332.

\bibitem[\protect\citeauthoryear{Holbrook, Lan, Vandenberg-Rodes, and
  Shahbaba}{Holbrook et~al.}{2018}]{H18}
Holbrook, A., S.~Lan, A.~Vandenberg-Rodes, and B.~Shahbaba (2018).
\newblock Geodesic {L}agrangian {M}onte {C}arlo over the space of positive
  definite matrices: with application to {B}ayesian spectral density
  estimation.
\newblock {\em Journal of Statistical Computation and Simulation\/}~{\em
  88\/}(5), 982--1002.

\bibitem[\protect\citeauthoryear{Jansen and Oonincx}{Jansen and
  Oonincx}{2005}]{JO05}
Jansen, M. and P.~Oonincx (2005).
\newblock {\em Second Generation Wavelets and Applications}.
\newblock London: Springer-Verlag.

\bibitem[\protect\citeauthoryear{Jeuris, Vandebril, and Vandereycken}{Jeuris
  et~al.}{2012}]{J12}
Jeuris, B., R.~Vandebril, and B.~Vandereycken (2012).
\newblock A survey and comparison of contemporary algorithms for computing the
  matrix geometric mean.
\newblock {\em Electronic Transactions on Numerical Analysis\/}~{\em 39},
  379--402.

\bibitem[\protect\citeauthoryear{Klees and Haagmans}{Klees and
  Haagmans}{2000}]{KH00}
Klees, R. and R.~Haagmans (2000).
\newblock {\em Wavelets in the Geosciences}.
\newblock Berlin: Springer-Verlag.

\bibitem[\protect\citeauthoryear{Krafty and Collinge}{Krafty and
  Collinge}{2013}]{KC13}
Krafty, R. and W.~Collinge (2013).
\newblock Penalized multivariate {W}hittle likelihood for power spectrum
  estimation.
\newblock {\em Biometrika\/}~{\em 100\/}(2), 447--458.

\bibitem[\protect\citeauthoryear{Lang}{Lang}{1995}]{L95}
Lang, S. (1995).
\newblock {\em Differential and Riemannian Manifolds}.
\newblock New York: Springer-Verlag.

\bibitem[\protect\citeauthoryear{Le}{Le}{1995}]{Le95}
Le, H. (1995).
\newblock Mean size-and-shapes and mean shapes: a geometric point of view.
\newblock {\em Advances in Applied Probability\/}~{\em 27\/}(1), 44--55.

\bibitem[\protect\citeauthoryear{Ma and Fu}{Ma and Fu}{2012}]{MF12}
Ma, Y. and Y.~Fu (2012).
\newblock {\em Manifold Learning Theory and Applications}.
\newblock CRC Press, Taylor \& Francis.

\bibitem[\protect\citeauthoryear{Muirhead}{Muirhead}{1982}]{M82}
Muirhead, R. (1982).
\newblock {\em Aspects of Multivariate Statistical Theory}.
\newblock New Jersey: John Wiley \& Sons.

\bibitem[\protect\citeauthoryear{Pasternak, Sochen, and Basser}{Pasternak
  et~al.}{2010}]{P10}
Pasternak, O., N.~Sochen, and P.~Basser (2010).
\newblock The effect of metric selection on the analysis of diffusion tensor
  {MRI} data.
\newblock {\em NeuroImage\/}~{\em 49\/}(3), 2190--2204.

\bibitem[\protect\citeauthoryear{Pennec}{Pennec}{2006}]{P06}
Pennec, X. (2006).
\newblock Intrinsic statistics on {R}iemannian manifolds: Basic tools for
  geometric measurements.
\newblock {\em Journal of Mathematical Imaging and Vision\/}~{\em 25\/}(1),
  127--154.

\bibitem[\protect\citeauthoryear{Pennec, Fillard, and Ayache}{Pennec
  et~al.}{2006}]{PFA05}
Pennec, X., P.~Fillard, and N.~Ayache (2006).
\newblock A {R}iemannian framework for tensor computing.
\newblock {\em International Journal of Computer Vision\/}~{\em 66\/}(1),
  41--66.

\bibitem[\protect\citeauthoryear{Rahman, Drori, Stodden, Donoho, and
  Schr{\"o}der}{Rahman et~al.}{2005}]{R05}
Rahman, I., I.~Drori, V.~Stodden, D.~Donoho, and P.~Schr{\"o}der (2005).
\newblock Multiscale representations for manifold-valued data.
\newblock {\em Multiscale Modeling \& Simulation\/}~{\em 4\/}(4), 1201--1232.

\bibitem[\protect\citeauthoryear{Rosen and Stoffer}{Rosen and
  Stoffer}{2007}]{RS07}
Rosen, O. and D.~Stoffer (2007).
\newblock Automatic estimation of multivariate spectra via smoothing splines.
\newblock {\em Biometrika\/}~{\em 94\/}(2), 335--345.

\bibitem[\protect\citeauthoryear{Said, Bombrun, Berthoumieu, and Manton}{Said
  et~al.}{2017}]{S17}
Said, S., L.~Bombrun, Y.~Berthoumieu, and J.~Manton (2017).
\newblock Riemannian {G}aussian distributions on the space of symmetric
  positive definite matrices.
\newblock {\em IEEE Transactions on Information Theory\/}~{\em 63\/}(4),
  2153--2170.

\bibitem[\protect\citeauthoryear{Skovgaard}{Skovgaard}{1984}]{S84}
Skovgaard, L. (1984).
\newblock A {R}iemannian geometry of the multivariate normal model.
\newblock {\em Scandinavian Journal of Statistics\/}~{\em 11\/}(4), 211--223.

\bibitem[\protect\citeauthoryear{Smith}{Smith}{2000}]{S00}
Smith, S. (2000).
\newblock Intrinsic {C}ram{\'e}r-{R}ao bounds and subspace estimation accuracy.
\newblock In {\em Proceedings of the IEEE Sensor Array and Multichannel Signal
  Processing Workshop}, pp.\  489--493. IEEE.

\bibitem[\protect\citeauthoryear{Tulino and Verd{\'u}}{Tulino and
  Verd{\'u}}{2004}]{TV04}
Tulino, A. and S.~Verd{\'u} (2004).
\newblock {\em Random Matrix Theory and Wireless Communications}.
\newblock Hanover: Now Publishers Inc.

\bibitem[\protect\citeauthoryear{Villani}{Villani}{2009}]{V09}
Villani, C. (2009).
\newblock {\em Optimal Transport: Old and New}.
\newblock Berlin: Springer-Verlag.

\bibitem[\protect\citeauthoryear{Wahba}{Wahba}{1980}]{W80}
Wahba, G. (1980).
\newblock Automatic smoothing of the log periodogram.
\newblock {\em Journal of the American Statistical Association\/}~{\em
  75\/}(369), 122--132.

\bibitem[\protect\citeauthoryear{Walden}{Walden}{2000}]{W00}
Walden, A. (2000).
\newblock A unified view of multitaper multivariate spectral estimation.
\newblock {\em Biometrika\/}~{\em 87\/}(4), 767--788.

\bibitem[\protect\citeauthoryear{Walnut}{Walnut}{2002}]{W02}
Walnut, D. (2002).
\newblock {\em An Introduction to Wavelet Analysis}.
\newblock Boston: Birkh{\"a}user.

\bibitem[\protect\citeauthoryear{Yuan, Zhu, Lin, and Marron}{Yuan
  et~al.}{2012}]{Y12}
Yuan, Y., H.~Zhu, W.~Lin, and J.~Marron (2012).
\newblock Local polynomial regression for symmetric positive definite matrices.
\newblock {\em Journal of the Royal Statistical Society: Series B\/}~{\em
  74\/}(4), 697--719.

\bibitem[\protect\citeauthoryear{Zheng, Tsui, Kang, and Deng}{Zheng
  et~al.}{2017}]{Z17}
Zheng, H., K.-W. Tsui, X.~Kang, and X.~Deng (2017).
\newblock Cholesky-based model averaging for covariance matrix estimation.
\newblock {\em Statistical Theory and Related Fields\/}~{\em 1\/}(1), 48--58.

\bibitem[\protect\citeauthoryear{Zhu, Chen, Ibrahim, Li, Hall, and Lin}{Zhu
  et~al.}{2009}]{Z09}
Zhu, H., Y.~Chen, J.~Ibrahim, Y.~Li, C.~Hall, and W.~Lin (2009).
\newblock Intrinsic regression models for positive-definite matrices with
  applications to diffusion tensor imaging.
\newblock {\em Journal of the American Statistical Association\/}~{\em
  104\/}(487), 1203--1212.

\end{thebibliography}
\end{small}

\begin{small}

\section{Appendix I: Geometry of HPD matrices} 
The space of $(d \times d)$-dimensional Hermitian matrices together with matrix addition and scalar multiplication $(\mathbb{H}_{d \times d}, +, \cdot_S)$ is a real vector space and every finite-dimensional real vector space has a natural smooth manifold structure by considering a global coordinate chart induced by a basis of the real vector space. The space of $(d \times d)$-dimensional Hermitian positive definite (HPD) matrices is no longer a vector space due to the positive definite constraints, but it is an open subset of $\mathbb{H}_{d \times d}$ and as such it is also a smooth manifold, see e.g. \cite{D92}. 
\paragraph{Affine-invariant Riemannian metric} For notational convenience, in the remainder of the supplemental document, we denote $\mathcal{M} := \mathbb{P}_{d \times d}$ for the space of $(d \times d)$-dimensional HPD matrices, an $d^2$-dimensional smooth manifold. For every $p \in \mathcal{M}$, the tangent space $T_p(\mathcal{M})$ can be identified by $\mathcal{H} := \mathbb{H}_{d \times d}$, the space of $(d \times d)$-dimensional Hermitian matrices. As detailed in \cite{PFA05}, the Frobenius inner product on $\mathbb{H}_{d \times d}$ induces the affine-invariant Riemannian metric $g_R$ on the manifold $\mathcal{M}$ given by the smooth family of inner products:
\begin{eqnarray} \label{supp-eq:7.1}
\langle h_1, h_2 \rangle_p &=& \tr((p^{-1/2} \ast h_1)(p^{-1/2} \ast h_2)), \quad \quad \forall\: p \in \mathcal{M},
\end{eqnarray}
with notation as in the main document and $h_1,h_2 \in T_p(\mathcal{M})$. The Riemannian distance on $\mathcal{M}$ derived from the Riemannian metric is given by:
\begin{eqnarray}
\delta_R(p_1,p_2) &=& \Vert \Log(p_1^{-1/2} \ast p_2) \Vert_F,  \label{supp-eq:7.2}
\end{eqnarray}
The mapping $x \mapsto a \ast x$ is an isometry for each invertible matrix $a \in \te{GL}(d,\mathbb{C}) = \{ A \in \mathbb{C}^{d \times d} \ | \ \te{det}(A) \neq 0 \}$, i.e., it is distance-preserving:
\begin{eqnarray*}
\delta_R(p_1, p_2) &=& \delta_R(a \ast p_1, a \ast p_2),\quad \forall\: a \in \te{GL}(d, \mathbb{C}).
\end{eqnarray*}
\paragraph{Geodesics}
By \cite[Theorem 6.1.6 and Prop. 6.2.2]{B09}, the Riemannian manifold $(\mathcal{M}, g_R)$, with $g_R$ the affine-invariant metric, is geodesically complete, and the \emph{geodesic} segment joining any two points $p_1, p_2 \in \mathcal{M}$ is unique and can be parametrized as,
\begin{eqnarray} \label{supp-eq:7.3}
\eta(p_1, p_2, t) &=& p_1^{1/2} \ast \big(p_1^{-1/2} \ast p_2 \big)^t, \quad 0 \leq t \leq 1.
\end{eqnarray}
\paragraph{Exp- and Log-maps}
Since $(\mathcal{M}, g_R)$ is a geodesically complete manifold, the Hopf-Rinow Theorem says that for every $p \in \mathcal{M}$ the exponential map $\Exp_p$ and the logarithmic map $\Log_p$ are global diffeomorphisms with as domains $T_p(\mathcal{M})$ and $\mathcal{M}$ respectively. By (\cite{PFA05}), the exponential map $\Exp_p: T_p(\mathcal{M}) \to \mathcal{M}$ is given by,
\begin{eqnarray} \label{supp-eq:7.4}
\Exp_p(h) &=& p^{1/2} \ast \Exp\left(p^{-1/2} \ast h \right),\quad \forall\: h \in T_p(\mathcal{M}),
\end{eqnarray}
The logarithmic map $\Log_p: \mathcal{M} \to T_p(\mathcal{M})$ is given by the inverse exponential map:
\begin{eqnarray} \label{supp-eq:7.5}
\Log_p(q) &=& p^{1/2} \ast \Log\left(p^{-1/2} \ast q \right).
\end{eqnarray}
The Riemannian distance may now also be expressed in terms of the logarithmic map as:
\begin{eqnarray} \label{supp-eq:7.6}
\delta_R(p_1, p_2) \ \ = \ \ \Vert \Log_{p_1}(p_2) \Vert_{p_1} \ \ = \ \ \Vert \Log_{p_2}(p_1) \Vert_{p_2}, \quad \forall\: p_1, p_2 \in \mathcal{M},
\end{eqnarray}
where $\Vert h \Vert_p := \langle h, h \rangle_p$ denotes the norm of $h \in T_p(\mathcal{M})$ induced by the affine-invariant Riemannian metric.
\paragraph{Parallel transport}
As outlined in \cite{J12} among others, the covariant derivative at $p \in \mathcal{M}$ of a smooth vector field $Y \in \mathfrak{X}(\mathcal{M})$, with respect to a smooth vector field $X \in \mathfrak{X}(\mathcal{M})$ is given by:
\begin{eqnarray} \label{supp-eq:7.7}
(\nabla_{X_p}Y)_p &=& D(Y)(p)[X_p] - \frac{1}{2}(X_p p^{-1} Y_p + Y_p p^{-1} X_p).
\end{eqnarray}
Here, $X_p, Y_p \in T_p(\mathcal{M})$ denote the tangent vectors associated with the vector fields $X,Y$ at $p \in \mathcal{M}$ and $D(Y)(p)[X_p] := \lim_{h \to 0} (Y(p + h X_p) - Y(p))/h$ is the classical Fr\'echet derivative of $Y(p)$, where $Y: \mathcal{M} \to T\mathcal{M}$ maps $p \in \mathcal{M}$ to the tangent vector $Y_p \in T_p(\mathcal{M})$. This connection $\nabla$ is exactly the Levi-Civita connection on the Riemannian manifold $(\mathcal{M}, g_R)$, as it can be verified that it satisfies the Koszul formula, see \cite{J12}. \\[3mm]
The parallel transport can be derived from the covariant derivative, and it follows that the parallel transport of a vector $w \in T_p(\mathcal{M})$ from a point $p \in \mathcal{M}$ along a geodesic curve in the direction of $v \in T_p(\mathcal{M})$ for time $\Delta t$ is given by:
\begin{eqnarray} \label{supp-eq:7.8}
\mathfrak{T}(p, \Delta t v, w) & = & \Exp_{p}\left( \Delta t v / 2 \right) \ast p^{-1} \ast w.
\end{eqnarray}
Substituting $\Delta t v = \Log_p(q)$, we obtain the parallel transport $\Gamma_{p}^q: T_{p}(\mathcal{M}) \to T_q(\mathcal{M})$ that maps a vector in $T_p(\mathcal{M})$ to its parallel transported version along a geodesic curve in $T_q(\mathcal{M})$ given by:
\begin{eqnarray} \label{supp-eq:7.9}
\Gamma_p^q(w) &=& p^{1/2} \ast (p^{-1/2} \ast q)^{1/2} \ast p^{-1/2} \ast w.
\end{eqnarray}
\begin{remark}
If $q = \te{Id}$, where $\te{Id}$ denotes the identity matrix, we obtain the so-called \emph{whitening} transport as in e.g., \cite{Y12}, which parallel transports $w \in T_p(\mathcal{M})$ to $T_{\te{Id}}(\mathcal{M})$ along a geodesic curve,
\begin{eqnarray} \label{supp-eq:7.10}
\Gamma_p^{\te{Id}}(w) & = & p^{-1/2} \ast w \ \in T_{\te{Id}}(\mathcal{M}).
\end{eqnarray}
\end{remark}

\paragraph{Probability measures and random variables}
In order to perform statistics on the Riemannian manifold $(\mathcal{M}, g_R)$, we are concerned with the notions of probability distributions and random variables. A manifold-valued random variable $X: \Omega \to \mathcal{M}$ is a measurable function from some probability space $(\Omega, \mathcal{A}, \nu)$ to the measurable space $(\mathcal{M}, \mathcal{B}(\mathcal{M}))$, where $\mathcal{B}(\mathcal{M})$ is the Borel algebra, i.e., the smallest $\sigma$-algebra containing all open sets in the complete separable metric space $(\mathcal{M}, \delta_R)$. In the following, we always work directly with the induced probability on $\mathcal{M}$, $\nu(B) = \nu(\{ \omega \in \Omega: X(\omega) \in B \})$. By $P(\mathcal{M})$, we denote the set of all probability measures on $(\mathcal{M}, \mathcal{B}(\mathcal{M}))$ and $P_p(\mathcal{M})$ denotes the subset of probability measures in $P(\mathcal{M})$ that have finite moments of order $p$ with respect to the Riemannian distance $\delta_R$, i.e., the $L^p$-Wasserstein space, see \cite[Definition 6.4]{V09}. That is,
\begin{eqnarray} \label{supp-eq:7.11}
P_p(\mathcal{M}) &:=& \left\{ \nu \in P(\mathcal{M}) : \exists\: y_0 \in \mathcal{M},\ \te{s.t.} \int_{\mathcal{M}} \delta_R(y_0, x)^p\: \nu(dx) < \infty \right\}.
\end{eqnarray}
Note that if $\int_{\mathcal{M}} \delta_R(y_0, x)^p\: \nu(dx) < \infty$ for some $y_0 \in \mathcal{M}$ and $1 \leq p < \infty$, this is true for any $y \in \mathcal{M}$. This follows by the triangle inequality, \[\int_{\mathcal{M}} \delta_R(y, x)^p\: \nu(dx) \ \leq \ 2^p \left( \delta_R(y, y_0)^p + \int_{\mathcal{M}} \delta_R(y_0, x)^p \: \nu(dx) \right) \ <\ \infty,\] using that $\delta_R(p_1, p_2) < \infty$ for any $p_1,p_2 \in \mathcal{M}$ due to the Hopf-Rinow theorem for a geodesically complete manifold. For a sequence of probability measures $(\nu_n)_{n \in \mathbb{N}}$ in $P(\mathcal{M})$, $\nu_n \overset{w}{\to} \nu$ denotes weak convergence to the probability measure $\nu$ in the usual sense, i.e., $\int_{\mathcal{M}} \phi(x)\: \nu_n(dx) \to \int_{\mathcal{M}} \phi(x)\: \nu(dx)$ for every continuous and bounded function $\phi: \mathcal{M} \to \mathbb{R}$, and a sequence $(\nu_n)_{n \in \mathbb{N}}$ is said to be uniformly integrable if:\[\lim_{K \to \infty} \sup_{n \in \mathbb{N}} \int_{\mathcal{M}} \delta_R(y_0, x) \bs{1}_{\{\delta_R(y_0, x) > K \}}\: \nu_n(dx) = 0, \quad \te{for some } y_0 \in \mathcal{M}.\] Note that if $(\nu_n)_{n \in \mathbb{N}}$ is uniformly integrable for some $y_0 \in \mathcal{M}$, then the sequence is uniformly integrable for any $y \in \mathcal{M}$. 
\paragraph{Intrinsic means}
Equipped with the notions of probability distributions and random variables on the manifold, we can characterize the center of a manifold-valued random variable $X$ with probability measure $\nu$. One important measure of centrality of a probability distribution $\nu$ on the manifold is the \emph{intrinsic} mean, also \emph{Karcher} or \emph{Fr\'echet} mean, as its definition is intrinsic to the (Riemannian) distance on the space. The set of intrinsic  means is given by the points that minimize the second moment with respect to the Riemannian distance $\delta_R$, 
\begin{eqnarray} \label{supp-eq:7.12}
\mu\ =\ \mathbb{E}_{\nu}[X] \ := \ \arg\min_{y \in \te{supp}(\nu)} \int_{\mathcal{M}} \delta_R(y,x)^2\ \nu(dx).
\end{eqnarray}
If $\nu \in P_2(\mathcal{M})$, then at least one Karcher mean exists as the above expectation is finite for each $y \in \mathcal{M}$. Moreover, since the manifold $(\mathcal{M}, g_R)$ is a geodesically complete manifold of non-positive curvature, (see \cite{PFA05} or \cite{S84}), by \cite[Proposition 1]{Le95} the Karcher mean $\mu$ is unique for any distribution $\nu \in P_2(\mathcal{M})$. By \cite[Corollary 1]{P06}, the Karcher mean can also be represented by the unique point $\mu \in \mathcal{M}$ that satisfies,
\begin{eqnarray} \label{supp-eq:7.13}
\bs{E}_\nu[\Log_{\mu}(X)] &=& \bs{0}
\end{eqnarray}
where $\bs{0}$ is the zero matrix and $\bs{E}_\nu[\cdot]$ is the Euclidean mean in the space of Hermitian matrices. In general, the sample intrinsic mean of a set of observations $\{X_1,\ldots,X_n\} \in \mathcal{M}$ has no closed-form solution, but it can be computed efficiently through a gradient descent algorithm as described in e.g., \cite{P06}.
\begin{remark}
The representation of the intrinsic mean in eq.(\ref{supp-eq:7.13}) above has an intuitive interpretation if we view the logarithmic map as a generalized notion of subtraction on the Riemannian manifold. In particular, if we equip the Riemannian manifold of HPD matrices with the Euclidean metric, (instead of the affine-invariant Riemannian metric), the logarithmic map reduces to ordinary matrix subtraction $\Log_{x}(y) = y - x$ and the above representation becomes $\bs{E}_{\nu}[X - \mu] = \bs{0}$, or $\bs{E}_{\nu}[X] = \mu$.
\end{remark}

\section{Appendix II: Proofs}

\subsection{Proof of Proposition \ref{prop:3.1}}
\begin{proof}
Denote the distribution of $\mu_n := \mu_n(X_1,\ldots, X_n)$ by $\nu_n$, we show recursively that:
\begin{eqnarray*}
\bs{E}[\delta_R(\mu_n, \mu)^2] \ =\ \int_{\mathcal{M}} \delta_R(x, \mu)\, d\nu_n(x) \ \leq \ \frac{1}{n}  \bs{E}[\delta_R(X_1, \mu)^2].
\end{eqnarray*}
By \cite[Theorem 6.1.9]{B09}, if $X_1,X_2,X_3 \in \mathcal{M}$, then for $t \in [0,1]$,
\begin{eqnarray*}
\delta_R(\eta(X_1,X_2,t), X_3)^2 & \leq & (1-t)\delta_R(X_1,X_3)^2 + t\delta_R(X_2, X_3)^2 - t(1-t) \delta_R(X_1,X_2)^2.
\end{eqnarray*}
Substituting $X_3 = \mu$ and $t=1/2$, (note that $\mu_2 = \eta(X_1,X_2,1/2)$), and taking expectations on both sides yields:
\begin{eqnarray*}
\bs{E}_{X_1}\bs{E}_{X_2}[\delta_R(\mu_2, \mu)^2] & \leq & \frac{1}{2} \bs{E}_{X_1}[\delta_R(X_1,\mu)^2] + \frac{1}{2}\bs{E}_{X_2}[\delta_R(X_2,\mu)^2] - \frac{1}{4} \bs{E}_{X_1}\bs{E}_{X_2}[\delta_R(X_1,X_2)^2].
\end{eqnarray*}
Using that $X_1, X_2 \overset{\te{iid}}{\sim} \nu$ we obtain,
\begin{eqnarray}
\bs{E}[\delta_R(\mu_2, \mu)^2] & \leq & \bs{E}[\delta_R(X_1, \mu)^2] - \frac{1}{4}\bs{E}_{X_1}\bs{E}_{X_2}[\delta_R(X_1,X_2)^2]. \label{supp-eq:8.1}
\end{eqnarray}
From the semi-parallelogram law above, \cite[Proposition 1]{H13} derive:
\begin{eqnarray*}
\int_{\mathcal{M}} [\delta_R(x,y)^2 - \delta_R(x,\mu)^2]\ d\nu(x) & \geq & \delta_R(y, \mu)^2, \quad \quad \te{for any } y \in \mathcal{M}.
\end{eqnarray*}
By the above inequality (and independence of $X_1,X_2$),
\begin{eqnarray*}
\bs{E}_{X_2}[\delta_R(X_1,X_2)^2\ |\ X_1 = x_1] &=& \int_{\mathcal{M}} \delta_R(x_1,X_2)^2 \, d\nu(X_2) \nn
&\geq & \delta_R(x_1, \mu)^2 + \int_{\mathcal{M}} \delta_R(X_2, \mu)^2\, d\nu(X_2) \nn
&=& \delta_R(x_1, \mu)^2 + \bs{E}[\delta_R(X_2, \mu)^2],
\end{eqnarray*}
and consequently,
\begin{eqnarray*}
\bs{E}_{X_1}\bs{E}_{X_2}[\delta_R(X_1,X_2)^2] & \geq & \int_{\mathcal{M}} \delta_R(X_1, \mu)^2\, d\nu(X_1) + \bs{E}[\delta_R(X_2, \mu)^2] \nn
&=& 2 \bs{E}[\delta_R(X_1, \mu)^2].
\end{eqnarray*}
Returning to eq.(\ref{supp-eq:8.1}),
\begin{eqnarray*}
\bs{E}[\delta_R(\mu_2, \mu)^2] & \leq & \frac{1}{2} \bs{E}[\delta_R(X_1, \mu)^2].
\end{eqnarray*}
Repeating the same argument, using independence of $\eta(X_1,X_2,1/2)$ and $\eta(X_3,X_4,1/2)$,
\begin{eqnarray*}
\bs{E}[\delta_R(\mu_4, \mu)^2] \ \leq \ \frac{1}{2}\bs{E}[\delta_R(\mu_2, \mu)^2] \ \leq \ \frac{1}{4} \bs{E}[\delta_R(X_1, \mu)^2].
\end{eqnarray*}
Continuing this iteration up to $\mu_n$, we find the upper bound:
\begin{eqnarray*}
\bs{E}[\delta_R(\mu_n, \mu)^2] \ \leq \ \frac{1}{2}\bs{E}_{n/2}[\delta_R(\mu_{n/2}, \mu)^2] \ \leq \ \ldots \ \leq \ \frac{1}{n} \bs{E}[\delta_R(X_1, \mu)^2].
\end{eqnarray*}
By Markov's inequality, $P(\delta_R(\mu_n, \mu) > \epsilon) \to 0$ for each $\epsilon > 0$ as $n \to \infty$, since the distribution of $X_1$ is assumed to have finite second moment with respect to $\delta_R$, i.e., $\bs{E}[\delta_R(X_1, \mu)^2] < \infty$.
\end{proof}

\subsection{Proof of Proposition \ref{prop:3.2}}
\begin{proof}
Denote $L := (N-1)/2$, with $L \geq 0$, and fix $j \geq 1$ sufficiently large and $k \in [L, 2^{j-1}-(L+1)]$ away from the boundary, such that the neighboring $(j-1)$-midpoints $M_{j-1,k-L},\ldots,M_{j-1,k+L}$ exist. \\[2mm]
\textbf{Remark:} For $k < L$ or $k > 2^{j-1}-(L + 1)$ near the boundary, we collect the $N$ available closest neighbors of $M_{j-1,k}$ (either to the left or right). The remainder of the proof for the boundary case is exactly analogous to the non-boundary case and follows directly by mimicking the arguments outlined below.\\[3mm]
We predict $M_{j,2k+1}$ from $M_{j-1,k-L},\ldots,M_{j-1,k+L}$ via intrinsic polynomial interpolation of degree $N-1$ passing through the $N$ points $\widebar{M}_{j-1,0},\ldots,\widebar{M}_{j-1,N-1}$, where $\widebar{M}_{j-1,k}$ denotes the cumulative intrinsic average as in eq.(\ref{eq:2.4}) in the main document. The predicted midpoint $\widetilde{M}_{j,2k+1}$ is then a weighted intrinsic average of the estimated polynomial at $(2k+1)2^{-j}$, i.e., $\widehat{M}_{(k-L)2^{-(j-1)}}((2k+1)2^{-j})$, and the given midpoint $\widebar{M}_{j-1,L} = M_{(k-L)2^{-(j-1)}}(2k2^{-j})$, (with notation as in Section \ref{sec:2.1} in the main document).\\[3mm]
For notational simplicity, write $M(t) := M_{(k-L)2^{-(j-1)}}(t)$ and $\widehat{M}(t) := \widehat{M}_{(k-L)2^{-(j-1)}}(t)$ for the true and estimated intrinsic cumulative mean curves respectively, where the latter is an interpolating polynomial of order $N-1$ passing through $N$ equidistant points $x_0,\ldots,x_{N-1}$ on the curve $M(t)$. $M(t)$ itself is a smooth curve with existing covariant derivatives up to order $N$, and $|x_0 - x_{N-1}| \lesssim 2^{-j}$. The polynomial remainder of the interpolating polynomial in Newton form with respect to the smooth curve, for every $x \in [(k-L)2^{-(j-1)}, (k+L)2^{-(j-1)}]$, is upper bounded by:
\begin{eqnarray*}
\frac{d}{dt}\widehat{M}(t)|_{t = x} - \frac{d}{dt}M(t)|_{t = x} \ \lesssim \ \frac{(x-x_0)\cdots(x-x_{N-1})}{N!}\, \Gamma(M)_{\xi}^{x} \left( \nabla^N_{\frac{d}{dt}M}\frac{d}{dt}M\big|_{t = \xi}\right) \ = \ O(2^{-jN})
\end{eqnarray*}
for some $\xi \in [(k-L)2^{-(j-1)}, (k+L)2^{-(j-1)}]$ by the mean value theorem for divided differences. This is closely related to the Taylor expansion in eq.(\ref{eq:3.2}) in the main document. In particular, the limit of the Newton polynomial if all nodes coincide is the Taylor polynomial, as the divided differences become covariant derivatives, and the covariant derivatives in the Taylor expansions of the Taylor polynomial and the smooth curves match up to order $N-1$.\\[3mm]
By definition of the derivative $\widehat{M}'(t) := \frac{d}{dt} \widehat{M}(t) = \lim_{\Delta t \to 0} \frac{1}{\Delta t}\Log_{\widehat{M}(t)}(\widehat{M}(t + \Delta t))$ and the fundamental theorem of calculus, it is verified that:
\begin{eqnarray*}
\widehat{M}(t + \Delta t) &=& \Exp_{\widehat{M}(t)}\left( \int_{t}^{t + \Delta t} \widehat{M}'(u)\, du \right).
\end{eqnarray*}
Substituting $t = 2k2^{-j}$ and $\Delta t = 2^{-j}$ and using that $\widehat{M}(2k2^{-j}) = M(2k2^{j})$ by construction, we obtain:
\begin{eqnarray} \label{supp-eq:8.2}
\widehat{M}((2k + 1)2^{-j}) &=& \Exp_{M(2k2^{-j})}\left( \int_{2k2^{-j}}^{(2k+1)2^{-j}} \widehat{M}'(u)\, du \right) \nn
&=& \Exp_{M(2k2^{-j})} \left( \int_{2k2^{-j}}^{(2k+1)2^{-j}} M'(u)\, du + O(2^{-jN}) \right).
\end{eqnarray}
The second step in the above equation follows immediately if $L = 0$ (i.e., $N = 1$), since,
\begin{eqnarray*}
\int_{2k2^{-j}}^{(2k+1)2^{-j}} \widehat{M}'(u)\, du \ = \ \int_{2k2^{-j}}^{(2k+1)2^{-j}} [M'(u) + O(1)]\, du \ = \ \int_{2k2^{-j}}^{(2k+1)2^{-j}} M'(u)\, du + O(2^{-j}).
\end{eqnarray*}
If $L \geq 1$, the second step in eq.(\ref{supp-eq:8.2}) follows by the polynomial remainder error bound above, since $\widehat{M}'(u) = M'(u) + O(2^{-jN})$ for each $u \in [2k2^{-j}, (2k+1)2^{-j}] \subset [(k-L)2^{-(j-1)}, (k+L)2^{-(j-1)}]$.\\[3mm]
Application of the logarithmic map $\Log_{M(2k2^{-j})}(\cdot)$ to both sides in eq.(\ref{supp-eq:8.2}) and using that $\Log_{M(t)}(M(t + \Delta t)) = \int_{t}^{t + \Delta t} M'(u)\, du$ as above, we rewrite:
\begin{eqnarray} \label{supp-eq:8.3}
\Log_{M(2k2^{-j})}(\widehat{M}((2k+1)2^{-j}) &=& \Log_{M(2k2^{-j})}(M(2k+1)2^{-j}) + O(2^{-jN}).
\end{eqnarray}
For notational convenience, in the remainder of this proof, we write $\Lambda = \lambda E$ for some arbitrary (not necessarily fixed) deterministic matrix $E \in \mathbb{C}^{d \times d}$ and constant $\lambda \lesssim 2^{-jN}$, i.e., $\Lambda = O(2^{-jN})$.\\[3mm]
Let $M, M_1, M_2 \in \mathcal{M}$ be deterministic matrices, we verify the following implication:
\begin{description}
\item[\textbf{Claim.}] \emph{If $\Log_M(M_1) - \Log_M(M_2) = O(\lambda)$, then also $M_1 = M_2 + O(\lambda)$}. 
\begin{proof}
Starting from $\Log_M(M_1) - \Log_M(M_2) = O(\lambda)$, by the definition of the logarithmic map, we write out,
\begin{eqnarray*}
\begin{array}{rclr}
M^{1/2} \ast \Log(M^{-1/2} \ast M_1) & = & M^{1/2} \ast \Log(M^{-1/2} \ast M_2) + O(\lambda) & \quad \Rightarrow \nn
\Log(M^{-1/2} \ast M_1) &=& \Log(M^{-1/2} \ast M_2) + O(\lambda) & \quad \Rightarrow \nn
M^{-1/2} \ast M_1 &=& \Exp(\Log(M^{-1/2} \ast M_2) + O(\lambda)). &
\end{array}
\end{eqnarray*}
For $\lambda \to 0$ sufficiently small, $M_1 = \Exp(\Log(M_2) + O(\lambda))$ also implies $M_1 = M_2 + O(\lambda)$. This follows by Taylor expanding the matrix exponential, 
\begin{eqnarray*}
M_1 &=& \Exp(\Log(M_2) + O(\lambda)) \ = \ \sum_{k=0}^\infty \frac{(\Log(M_2) + O(\lambda))^k}{k!} \nn
&=& \sum_{k=0}^\infty \frac{(\Log(M_2))^k + O(\lambda)}{k!} \ =\ \sum_{k=0}^\infty \frac{(\Log(M_2))^k}{k!} + O(\lambda) \sum_{k=0}^\infty \frac{1}{k!} \ = \ M_2 + O(\lambda). 
\end{eqnarray*}
As a consequence, also, 
\begin{eqnarray*}
\begin{array}{rclr}
M^{-1/2} \ast M_1 &=& \Exp(\Log(M^{-1/2} \ast M_2) + O(\lambda)) &\quad \Rightarrow \nn
M^{-1/2} \ast M_1 &=& M^{-1/2} \ast M_2 + O(\lambda) &\quad \Rightarrow \nn
M^{-1/2} \ast ( M_1 - M_2) &=& O(\lambda) & \quad \Rightarrow \nn
M_1 &=& M_2 + O(\lambda). & 
\end{array}
\end{eqnarray*}
\end{proof}
\end{description} 
Applying the above implication to eq.(\ref{supp-eq:8.3}) yields,
\begin{eqnarray} \label{supp-eq:8.4}
\widehat{M}((2k+1)2^{-j}) &=& M((2k+1)2^{-j}) + O(2^{-jN}).
\end{eqnarray}
The predicted midpoint $\widetilde{M}_{j,2k+1}$ is reconstructed from $\widehat{M}((2k+1)2^{-j})$ and $M(2k2^{-j})$ as follows. By definition of $M(t)$ as the cumulative intrinsic mean curve, we can write $M((2k+1)2^{-j})$ as a weighted intrinsic average between $\widebar{M}_{j-1,L} = M(2k2^{-j})$ and $M_{j,2k+1}$ according to:
\begin{eqnarray*}
M((2k+1)2^{-j}) &=& \Exp_{M((2k+1)2^{-j})}\Bigg( \frac{(N-1)2^{-j}}{N2^{-j}}\Log_{M((2k+1)2^{-j})}(\widebar{M}_{j-1,L}) \nn
&& \hspace{4cm} + \ \frac{2^{-j}}{N2^{-j}} \Log_{M((2k+1)2^{-j})}(M_{j,2k+1}) \Bigg).
\end{eqnarray*}
Application of the logarithmic map $\Log_{M((2k+1)2^{-j})}(\cdot)$ to both sides and rearranging terms (substitute $N-1 = 2L$), gives,
\begin{eqnarray*}
\frac{-2L}{N}\Log_{M((2k+1)2^{-j})}(\widebar{M}_{j-1,L}) & = & \frac{1}{N} \Log_{M((2k+1)2^{-j})}(M_{j,2k+1}).
\end{eqnarray*} 
Or in terms of $M_{j,2k+1}$,
\begin{eqnarray*}
M_{j,2k+1} & = & \Exp_{M((2k+1)2^{-j})}\left(-2L \cdot \Log_{M((2k+1)2^{-j})}(\widebar{M}_{j-1,L}) \right) \nn
& = &  \eta\left(M((2k+1)2^{-j}), \widebar{M}_{j-1,L}, -2L \right).
\end{eqnarray*}
The predicted midpoint $\widetilde{M}_{j,2k+1}$ is given by replacing the true point $M((2k+1)2^{-j})$ by the estimated point $\widehat{M}((2k+1)2^{-j})$, ($\widebar{M}_{j-1,L}$ is known), i.e., 
\begin{eqnarray} \label{supp-eq:8.5}
\widetilde{M}_{j,2k+1} & = &  \eta\left(\widehat{M}((2k+1)2^{-j}), \widebar{M}_{j-1,L}, -2L \right).
\end{eqnarray}
Below, we use that $(M + \Lambda)^a = M^a + O(\lambda)$ for $a \in \mathbb{N}$, $(M + \Lambda)^{1/2} = M^{1/2} + O(\lambda)$ and $(M + \Lambda)^{-1} = M^{-1} + O(\lambda)$ for $M \in \mathcal{M}$ and $\lambda \to 0$ sufficiently small, as verified in the proof of Proposition \ref{prop:3.3}, (note that this is the deterministic version), combined with eq.(\ref{supp-eq:8.4}) and the definition of the geodesic in eq.(\ref{supp-eq:7.3}). Writing out eq.(\ref{supp-eq:8.5}) gives,
\begin{eqnarray} \label{supp-eq:8.6}
\widetilde{M}_{j,2k+1} &=& \left( M((2k+1)2^{-j})^{1/2} + \Lambda \right) \ast\ \Big( \left( M((2k+1)2^{-j})^{-1/2} + \Lambda \right) \ast \widebar{M}_{j-1,L} \Big)^{-2L} \nn
&=& \left( M((2k+1)2^{-j})^{1/2} + \Lambda \right) \ast\ \Big( \Big(M((2k+1)2^{-j})^{-1/2} \ast \widebar{M}_{j-1,L}\Big)^{-1} + \Lambda \Big)^{2L}\nn
&=& \left( M((2k+1)2^{-j})^{1/2} + \Lambda \right) \ast\ \Big( \Big(M((2k+1)2^{-j})^{-1/2} \ast \widebar{M}_{j-1,L}\Big)^{-2L} + \Lambda \Big) \nn
&=& M_{j,2k+1} + O(2^{-jN}).
\end{eqnarray}
Substituting the above result in the whitened wavelet coefficient $\mathfrak{D}_{j,k} = 2^{-j/2}\Log(\widetilde{M}_{j,2k+1}^{-1/2} \ast M_{j,2k+1})$, by the same identities as used above combined with $\Log(M + \Lambda) = \Log(M) + O(\lambda)$, (verified in the proof of Proposition \ref{prop:3.3}), it follows that for $j \geq 1$ sufficiently large,
\begin{eqnarray*}
\Vert \mathfrak{D}_{j,k} \Vert_F &=& \Big\Vert 2^{-j/2}\Log\big((M_{j,2k+1} + \Lambda)^{-1/2} \ast M_{j,2k+1} \big) \Big\Vert_F\nn 
&=& 2^{-j/2}\Big\Vert \Log\big((M_{j,2k+1}^{-1/2} + \Lambda) \ast M_{j,2k+1} \big) \Big\Vert_F\nn
&=& 2^{-j/2}\big\Vert \Log\big(\te{Id} + \Lambda) \big\Vert_F \ = \ O\left(2^{-j/2} 2^{-jN}\right),
\end{eqnarray*}
where in the final step we expanded $\Log(\te{Id} + \Lambda) = O(2^{-jN})$ via its Mercator series (see \cite[Section 11.3]{H08}), using that the spectral radius of $\Lambda$ is smaller than 1 for $j$ sufficiently large. 
\end{proof}

\subsection{Proof of Proposition \ref{prop:3.3}}
\begin{proof}
By the proof of Proposition \ref{prop:3.1}, $\bs{E}[\delta_R(M_{j,k,n}, M_{j,k})^2] = O(2^{-(J-j)})$ for each $j \geq 0$ and $0 \leq k \leq 2^j-1$. For notational convenience, in the remainder of this proof $\epsilon_{j,n}$ denotes a general (not necessarily the same) random error matrix that satisfies $\bs{E} \Vert \epsilon_{j,n} \Vert_F^2 = O(2^{-(J-j)})$. Furthermore, we can appropriately write $M_{j,k,n} = \Exp_{M_{j,k}}(\epsilon_{j,n})$, such that $M_{j,k,n} \overset{p}{\to} M_{j,k}$ as $J \to \infty$ at the correct rate since, 
\begin{eqnarray*}
\bs{E}[\delta_R(\Exp_{M_{j,k}}(\epsilon_{j,n}), M_{j,k})^2] &=& \bs{E}\Vert \Log(M^{-1/2}_{j,k} \ast \Exp_{M_{j,k}}(\epsilon_{j,n})) \Vert_F^2 \nn
&=& \bs{E}\Vert M^{-1/2}_{j,k} \ast \epsilon_{j,n} \Vert_F^2, \nn
&=& O(2^{-(J-j)})
\end{eqnarray*}
using the definitions of the Riemannian distance function and the logarithmic and exponential maps. In particular, by a first-order Taylor expansion of the matrix exponential, (abusing notation of $\epsilon_{j-1,n}$), $M_{j-1,k,n} = M^{1/2}_{j-1,k} \ast \Exp(\epsilon_{j-1,n}) = M^{1/2}_{j-1,k} \ast ( \te{Id} + \epsilon_{j-1,n} + \ldots) = M_{j-1,k} + \epsilon_{j-1,n}$.\\[3mm]
By eq.(\ref{eq:2.5}) in the main document, the predicted midpoint $\widetilde{M}_{j,2k+1,n}$ is a weighted intrinsic mean of $N$ coarse-scale midpoints $(M_{j-1,k,n})_k$ with weights summing up to 1. The rate of $\widetilde{M}_{j,2k+1,n}$ is therefore upper bounded by the (worst) convergence rate of the individual midpoints $(M_{j-1,k,n})_k$, and we can also write $\widetilde{M}_{j,2k+1,n} = \widetilde{M}_{j,2k+1} + \epsilon_{j-1,n}$. \\[3mm]
Below, we verify several implications that are needed to finish the proof. let $M \in \mathcal{M}$ be a deterministic matrix and $\lambda E = O_p(\lambda)$ a random error matrix, such that $\bs{E}\Vert \lambda E \Vert_F = O(\lambda)$.  
\begin{description}
\item[\textbf{Claim.}] If $\lambda \to 0$ sufficiently small, then $\textnormal{Log}(M + \lambda E) \ =\ \textnormal{Log}(M) + O_p(\lambda)$.
\begin{proof}
Rewrite $\Log(M + \lambda E) = \Log(M( \te{Id} + \lambda M^{-1}E))$. By the Baker-Campbell-Hausdorff formula (e.g., \cite[Theorem 10.4]{H08}), with $X = \Log(M)$ and $Y = \Log(\te{Id} + \lambda M^{-1} E))$,
\begin{eqnarray*}
\Log(M + \lambda E) &=& X + Y + \frac{1}{2}[X, Y] + \frac{1}{12}([X,[X,Y]] + [Y,[Y,X]]) + \frac{1}{24}[Y,[X,[X,Y]]] - \ldots,
\end{eqnarray*}
where $[X,Y] = XY - YX$ denotes the commutator of $X$ and $Y$. In particular, 
\begin{eqnarray*}
[X,Y] &=& [\Log(M), \Log(\te{Id} + \lambda M^{-1} E)] \nn
&=& \Log(M)\Log(\te{Id} + \lambda M^{-1}E) - \Log(\te{Id} + \lambda M^{-1}E)\Log(M) \nn
&=& \Log(M)(\lambda M^{-1}E + O_p(\lambda^2)) - (\lambda M^{-1}E + O_p(\lambda^2)) \Log(M) \nn
&=& O_p(\lambda).
\end{eqnarray*}
Here, we expanded $\Log(\te{Id} + \lambda M^{-1} E) = \lambda M^{-1} E + O_p(\lambda^2)$ via its Mercator series (e.g., \cite[Section 11.3]{H08}), using that the spectral radius $\rho(\lambda M^{-1} E) = \lambda \rho(M^{-1} E) < 1$ almost surely for $\lambda \to 0$ sufficiently small. \\[3mm]
Iterating the above argument, it follows that all the nested (higher-order) commutators are of the order $O_p(\lambda)$ as well, and we rewrite:
\begin{eqnarray*}
\Log(M + \lambda E) &=& \Log(M) + \Log(\te{Id} + \lambda M^{-1} E) + O_p(\lambda).
\end{eqnarray*}
Expanding again $\Log(\te{Id} + \lambda M^{-1} E) = \lambda M^{-1} E + O_p(\lambda^2) = O_p(\lambda)$, (for $\lambda$ sufficiently small), the claim follows.
\end{proof}
\item[\textbf{Claim.}] If $\lambda \to 0$ sufficiently small, then $(M + \lambda E)^{1/2} \ = \ M^{1/2} + O_p(\lambda)$ and $(M + \lambda E)^{-1} \ = \ M^{-1} + O_p(\lambda)$. 
\begin{proof}
For the first claim, Taylor expanding the matrix exponential,
\begin{eqnarray*}
(M + \lambda E)^{1/2} & = & \Exp\left(\frac{1}{2} \Log(M + \lambda E)\right) \ = \ \sum_{k=0}^\infty \frac{(\Log(M + \lambda E))^k}{2^k k!} \nn
&=& \sum_{k=0}^\infty \frac{(\Log(M) + O_p(\lambda))^k}{2^k k!} \ =\ \sum_{k=0}^\infty \frac{(\Log(M))^k}{2^k k!} + O_p(\lambda) \ =\ M^{1/2} + O_p(\lambda),
\end{eqnarray*}
using the previous claim $\textnormal{Log}(M + \lambda E) \ =\ \textnormal{Log}(M) + O_p(\lambda)$ for $\lambda \to 0$ sufficiently small.\\[3mm]
For the second claim, rewrite, (for $\lambda$ sufficiently small), 
\begin{eqnarray*}
(M + \lambda E)^{-1} &=& (M(\te{Id} + \lambda M^{-1}E))^{-1} \nn
&=& (\te{Id} + \lambda M^{-1} E))^{-1} M^{-1} \nn
&=& (\te{Id} - \lambda M^{-1} E + (\lambda M^{-1} E)^2 - \ldots ) M^{-1} \ = \ M^{-1} + O_p(\lambda),
\end{eqnarray*}
applying a binomial series expansion of the matrix inverse $(\te{Id} + \lambda M^{-1} E))^{-1}$, using that the spectral radius $\rho(\lambda M^{-1} E) = \lambda \rho(M^{-1} E) < 1$ almost surely for $\lambda \to 0$ sufficiently small. Combining the two claims, we find in particular also that $(M + \lambda E)^{-1/2} = M^{-1/2} + O_p(\lambda)$.
\end{proof}
\end{description}
Combining the above results, for $j < J$ sufficiently small such that the above claims hold, we write out for the empirical whitened wavelet coefficient $\widehat{\mathfrak{D}}_{j,k,n}$, (with some abuse of notation for $\epsilon_{j,n}$),
\begin{eqnarray*}
\widehat{\mathfrak{D}}_{j,k,n} &=& 2^{-j/2}\, \Log\left( (\widetilde{M}_{j,2k+1} + \epsilon_{j-1,n})^{-1/2} \ast (M_{j,2k+1} + \epsilon_{j,n}) \right) \nn
&=&  2^{-j/2}\, \Log\left( (\widetilde{M}_{j,2k+1}^{-1/2} + \epsilon_{j-1,n}) \ast (M_{j,2k+1} + \epsilon_{j,n}) \right) \nn
&=&  2^{-j/2}\, \Log\left( \widetilde{M}^{-1/2}_{j,2k+1} \ast M_{j,2k+1} + \epsilon_{j, n} + \ldots \right) \nn
&=&  2^{-j/2}\, \Log\left( \widetilde{M}^{-1/2}_{j,2k+1} \ast M_{j,2k+1}\right) + 2^{-j/2}\, O_p(2^{-(J-j)/2}) \nn
&=&  \mathfrak{D}_{j,k} + 2^{-j/2}\, O_p(2^{-(J-j)/2}).
\end{eqnarray*}
Plugging in the above result, it follows that for $j < J$ sufficiently small,
\begin{eqnarray*} 
\bs{E}\Vert \widehat{\mathfrak{D}}_{j,k,n} - \mathfrak{D}_{j,k} \Vert_F^2\ = \ O(2^{-j}\, 2^{-(J-j)})\ = \ O(n^{-1}).
\end{eqnarray*}
\end{proof}

\subsection{Proof of Theorem \ref{thm:3.4}}
\begin{proof} For the first part of the theorem, suppose that $J_0 = \log_2(n)/(2N+1) \gg 1$ is sufficiently large such that the rates in Propositions \ref{prop:3.2} and \ref{prop:3.3} hold. Then,
\begin{eqnarray} \label{supp-eq:8.7}
\sum_{j,k} \bs{E}\Vert \widehat{\mathfrak{D}}_{j,k} - \mathfrak{D}_{j,k} \Vert_F^2 &=& \sum_{j \geq J_0} \Vert \mathfrak{D}_{j,k}\Vert_F^2 + \sum_{j < J_0} \bs{E}\Vert \widehat{\mathfrak{D}}_{j,k} - \mathfrak{D}_{j,k} \Vert_F^2\nn
&\lesssim & \sum_{j \geq J_0} 2^j (2^{-j}2^{-2jN}) + \sum_{j < J_0} 2^j n^{-1} \nn
&=& \left(\sum_{j=0}^{J} (2^{-2N})^j - \sum_{j=0}^{J_0-1} (2^{-2N})^j \right) + n^{-1}\sum_{j=1}^{J_0-1} 2^j \nn
&= & \frac{(2^{-2N})^{J_0} - (2^{-2N})^{(J+1)}}{1-2^{-2N}} + n^{-1}(2^{J_0} - 2)  \nn
&\lesssim & (2^{-2N})^{J_0} + n^{-1}2^{J_0} + n^{-1} \nn
&\lesssim & n^{-2N/(2N+1)},
\end{eqnarray}
where the last step follows from substituting $J_0 = \log_2(n)/(2N+1)$ since,
\begin{eqnarray*}
\begin{array}{lclclcl}
(2^{-2N})^{J_0} &=& \exp(-2NJ_0 \log(2)) &=& \exp\left( \frac{-2N}{2N+1} \log(n) \right) &=& n^{-2N/(2N+1)} \nn
n^{-1}2^{J_0} &=& \exp(-\log(n) +  J_0 \log(2)) &=& \exp\left( \frac{-2N}{2N+1}\log(n) \right) &=& n^{-2N/(2N+1)}.
\end{array}
\end{eqnarray*}
For the second part of the theorem, 
if we can verify that $\bs{E}[\delta_R(M_{J,k}, \widehat{M}_{J,k,n})^2] \lesssim n^{-2N/(2N+1)}$ for each $k=0,\ldots,n-1$, the proof is finished.\\[3mm]
At scales $j=1,\ldots,J$, based on the estimated midpoints $(\widehat{M}_{j-1,k',n})_{k'}$ and the estimated wavelet coefficient $\widehat{D}_{j,k,n}$, in the inverse wavelet transform, the finer-scale midpoint $\widehat{M}_{j,k,n}$ is estimated through,
\begin{eqnarray*}
\widehat{M}_{j,k,n} &=& \Exp_{\widehat{\widetilde{M}}_{j,k,n}}\left(2^{j/2} \widehat{D}_{j,k,n} \right).
\end{eqnarray*}
where $\widehat{\widetilde{M}}_{j,k,n}$ is the predicted midpoint at scale-location $(j,k)$ based on $(\widehat{M}_{j-1,k',n})_{k'}$. In particular, at scale $j = 1$, $\widehat{\widetilde{M}}_{1,k,n} = \widetilde{M}_{1,k,n}$ as the estimated coarsest midpoints $(\widehat{M}_{0,k',n})_{k'}$ correspond to the empirical coarsest midpoints $(M_{0,k',n})_{k'}$.\\[3mm]
At scales $j = 1,\ldots, J_0 - 1$, we do not alter the wavelet coefficients. Assuming that $j \ll J$ is sufficiently small, such that the rate in Proposition \ref{prop:3.3} holds, we write $\widehat{\mathfrak{D}}_{j,k,n} = \mathfrak{D}_{j,k} + \eta_n$, with $\eta_n$ a general (not always the same) random error matrix satisfying $\bs{E}\Vert \eta_n \Vert_F = O(n^{-1/2})$. Also, by the proof of Proposition \ref{prop:3.3} (using the same notation), we can write $\widetilde{M}_{j,k,n} = \widetilde{M}_{j,k} + \epsilon_{j,n}$, where $\epsilon_{j,n}$ is a general (not always the same) random error matrix satisfying $\bs{E}\Vert \epsilon_{j,n} \Vert_F = O(2^{-(J-j)/2})$. \\[3mm]
In particular, at scale $j = 1$,
\begin{eqnarray} \label{supp-eq:8.8}
\widehat{M}_{1,k,n} &=& \Exp_{\widehat{\widetilde{M}}_{1,k,n}}\big( 2^{1/2} \widehat{D}_{1,k,n} \big) \nn
&=& \widetilde{M}_{1,k,n}^{1/2} \ast \Exp\big(2^{1/2} \widetilde{M}_{1,k,n}^{-1/2} \ast \widehat{D}_{1,k,n}\big) \nn
&=& \widetilde{M}_{1,k,n}^{1/2} \ast \Exp\big( 2^{1/2} \widehat{\mathfrak{D}}_{1,k,n} \big) \nn
&=&\left(\widetilde{M}_{1,k} + \epsilon_{1,n} \right)^{1/2} \ast \Exp\left( 2^{1/2} (\mathfrak{D}_{1,k} + \eta_n)  \right) \nn
&=& \left(\widetilde{M}_{1,k}^{1/2} + \epsilon_{1,n} \right) \ast \left(\Exp(2^{1/2} \mathfrak{D}_{1,k}) + 2^{1/2}\eta_n \right) \nn
&=& M_{1,k} + O_p(2^{1/2} n^{-1/2}) + O_p(2^{-(J-1)/2}) \nn
&=& M_{1,k} + O_p(2^{1/2} n^{-1/2}).
\end{eqnarray}
Here, we used that $(M + \lambda E)^{1/2} = M^{1/2} + O_p(\lambda)$ for $\lambda \to 0$ sufficiently small as in the proof of Proposition \ref{prop:3.3}, and a Taylor expansion of the matrix exponential:
\begin{eqnarray*}
\Exp(D + \eta_n)\ =\ \sum_{k=0}^\infty \frac{(D + \eta_n)^k}{k!} \ = \ \sum_{k=0}^\infty \frac{D^k}{k!} + O_p(n^{-1/2}) \ = \ \Exp(D) + O_p(n^{-1/2}).
\end{eqnarray*}
Iterating this same argument for each scale $j = 2,\ldots,J_0-1$, we find that:
\begin{eqnarray*}
\widehat{M}_{J_0-1,k,n} \ = \ M_{J_0-1,k} + \sum_{j=1}^{J_0-1} O_p(n^{-1/2} 2^{j/2}) \ =\  M_{J_0-1,k} + O_p(n^{-1/2} 2^{(J_0-1)/2}).
\end{eqnarray*}
As a consequence, (as in the proof of Proposition \ref{prop:3.3}), we can write $\widehat{\widetilde{M}}_{J_0,k,n} = \widetilde{M}_{J_0,k} + \epsilon_{J_0,n}$, where $\epsilon_{J_0,n} = O_p(n^{-1/2} 2^{J_0/2})$. At scales $j = J_0,\ldots,J$, we set $\widehat{D}_{j,k,n} = \bs{0}$ for each $k$. Assuming that $j \gg 1$ is sufficiently large, such that the rate in Proposition \ref{prop:3.2} holds, we can write $\widehat{D}_{j,k,n} = \bs{0} = \mathfrak{D}_{j,k} + \zeta_{j,N}$, with $\zeta_{j,N}$ a general (not always the same) deterministic error matrix satisfying $\Vert \zeta_{j,N} \Vert_F = O(2^{-j/2}2^{-jN})$. \\[3mm]
In particular, at scale $j = J_0$,
\begin{eqnarray*}
\widehat{M}_{J_0,k,n} &=& \Exp_{\widehat{\widetilde{M}}_{J_0,k,n}}\big(2^{J_0/2} \widehat{D}_{J_0,k,n} \big) \nn
&=& \big( \widetilde{M}_{J_0,k} + \epsilon_{J_0, n} \big)^{1/2} \ast \Exp\left( \big(\widetilde{M}_{J_0,k} + \epsilon_{J_0,n}\big)^{-1/2} \ast  2^{J_0/2}\big(\mathfrak{D}_{J_0,k} + \zeta_{J_0, n} \big)\right) \nn
&=& \big( \widetilde{M}_{J_0,k}^{1/2} + \epsilon_{J_0,n} \big) \ast \Exp\left( \big(\widetilde{M}_{J_0,k}^{-1/2} + \epsilon_{J_0,n}\big) \ast  \big(2^{J_0/2}\mathfrak{D}_{J_0,k} + 2^{J_0/2}\zeta_{J_0, n} \big)\right) \nn
&=& \left( \widetilde{M}_{J_0,k}^{1/2} + \epsilon_{J_0,n} \right) \ast \left( \Exp(2^{J_0/2}D_{J_0,k}) + 2^{J_0/2} \epsilon_{J_0,n} \mathfrak{D}_{J_0,k} + 2^{J_0/2} \zeta_{J_0,n} \right) \nn
&=& \left( \widetilde{M}_{J_0,k}^{1/2} + \epsilon_{J_0,n} \right) \ast \left( \Exp(2^{J_0/2}D_{J_0,k}) + O_p\big(2^{-J_0N}\big) \right) \nn
&=& M_{J_0,k} + O_p(n^{-1/2}2^{J_0/2}) + O_p\big(2^{-J_0N}\big),
\end{eqnarray*}
which follows in the same way as in eq.(\ref{supp-eq:8.8}) above, combined with the observation that $2^{J_0/2} \epsilon_{J_0,n} \mathfrak{D}_{J_0,k} = O_p(2^{-J_0 N})$, since $\Vert 2^{J_0/2} \epsilon_{J_0,n} \mathfrak{D}_{J_0,k}\Vert_F = O_p(2^{-(J-J_0)/2}2^{-J_0N}) = O_p(2^{-J_0N})$ by Proposition \ref{prop:3.2}. Iterating this same argument for each scale $j = J_0+1,\ldots,J$ yields, 
\begin{eqnarray*}
\widehat{M}_{J,k,n} \ =\ M_{J,k} + O_p(n^{-1/2}2^{J_0/2}) + \sum_{j=J_0}^{J} O_p\big(2^{-jN}\big) \ =\ M_{J,k} + O_p\big( 2^{-J_0N}\big) + O_p\big(n^{-1/2}2^{J_0/2}\big).
\end{eqnarray*}
Plugging in $J_0 = \log_2(n)/(2N+1)$, as previously demonstrated, the above expression reduces to: 
\begin{eqnarray*}
\widehat{M}_{J,k,n} &=& M_{J,k} + O_p\big(n^{-N/(2N+1)}\big), \quad \quad \te{for each } k = 0,\ldots,n-1. 
\end{eqnarray*}
For notational convenience, denote by $\xi_{n,N}$ a general (not always the same) random error matrix such that $\bs{E}\Vert \xi_{n,N} \Vert_F = O(n^{-N/(2N+1)})$. For each $k=0,\ldots,n-1$, by the previous result:
\begin{eqnarray*}
\bs{E}\left[\delta_R(M_{J,k}, \widehat{M}_{J,k,n})^2\right] &=& \bs{E}\left[\delta_R\big(M_{J,k}, M_{J,k} + \xi_{n,N} \big)^2\right] \nn
&=& \bs{E}\left\Vert \Log \left( M_{J,k}^{-1/2} \ast \big(M_{J,k} + \xi_{n,N} \big)\right) \right\Vert_F^2 \nn
&=& \bs{E}\left\Vert \Log \big( \te{Id} + \xi_{n,N} \big) \right\Vert_F^2 \ = \ O(n^{-2N/(2N+1)}),
\end{eqnarray*}
where in the final step we expanded $\Log(\te{Id} + \xi_{n,N}) = O_p(n^{-N/(2N+1})$ via its Mercator series, using that the spectral radius of $\xi_{n,N}$ is smaller than 1 almost surely for $n$ sufficiently large. 
\end{proof}

\subsubsection{Proof of remark Theorem \ref{thm:3.4}}
Let $\gamma_n(t) = \gamma(t) + \epsilon_{n,N}$ and $\hat{\gamma}(t)$ be as defined in the remark after Theorem 3.4, with $\epsilon_{n,N}$ a general error matrix, such that $\Vert \epsilon_{n,N} \Vert_F = O(n^{-N/(2N+1)})$. Then we can upper bound, 
\begin{eqnarray*}
\delta(\gamma(t), \gamma_n(t))^2 & = & \Vert \Log\big(\gamma(t)^{-1/2} \ast ( \gamma(t) + \epsilon_n) \big) \Vert_F^2 \nn
&=& \Vert \Log(\te{Id} + \epsilon_{n,N}) \Vert_F^2 \ = \ O(n^{-2N/(2N+1)}),
\end{eqnarray*}
where in the final step we again expand $\Log(\te{Id} + \epsilon_{n,N}) = O(n^{-N/(2N+1)})$ via its Mercator series, provided that $n$ is sufficiently large. \\[3mm]
By the triangle inequality, the integrated mean-squared error of the linear wavelet estimator with respect to the continuous curve $\gamma$ then also satisfies,
\begin{eqnarray*}
\int_0^1 \bs{E}\left[\delta_R(\hat{\gamma}_n(t), \gamma(t))^2\right]\, dt & \leq & 
2^2 \left( \int_0^1 \bs{E}\left[ \delta_R(\hat{\gamma}_n(t), \gamma_n(t))^2 \right]\, dt + \int_0^1 \delta_R(\gamma_n(t), \gamma(t))^2 \, dt \right) \nn
& = & 2^2 \left( \frac{1}{n} \sum_{k=0}^{n-1} \bs{E}\left[\delta_R(\widehat{M}_{J,k,n}, M_{J,k})^2\right] + \int_0^1 \delta_R(\gamma_n(t), \gamma(t))^2\, dt \right) \nn
& \lesssim & n^{-2N/(2N+1)},
\end{eqnarray*}
using the convergence rate for the linear wavelet estimator derived above.

\subsection{Proof of Theorem \ref{thm:4.1}}
\begin{proof}
First, we derive the bias $b(X, f) = c(d, L)\cdot f$. By linearity of the (ordinary) expectation:
\begin{eqnarray}
b(X, f) \ =\ \bs{E}[\Log_f(X)] \ =\ f^{1/2} \ast \bs{E}[\Log(f^{-1/2} \ast X)], \label{supp-eq:8.9}
\end{eqnarray}
using that $g \ast \Log_{X_1}(X_2) = \Log_{g \ast X_1}(g \ast X_2)$ for any $g \in \te{GL}(d, \mathbb{C})$. The transformed random variable $Y := f^{-1/2} \ast X$ is distributed as $Y \sim W_d^c(L, L^{-1} \te{Id})$, which is unitarily invariant (see e.g., \cite[Section 3.2]{M82}). By \cite[Section 2.1.5]{TV04}, taking the eigendecomposition of a unitarily invariant matrix $Y = Q \ast \Lambda$, the matrix of eigenvectors $Q$ is distributed according to the Haar measure, i.e., the uniform distribution on the set of unitary matrices $\mathcal{U}_d = \{ U \in \te{GL}(d, \mathbb{C})\ |\ U^*U = \te{Id} \}$, implying that the eigenvectors $(\vec{q}_i)_{i=1,\ldots,d}$ (the columns of $Q$) are identically distributed. Furthermore, $Q$ is independent of the diagonal eigenvalue-matrix $\Lambda$, therefore (see also \cite{S00}):
\begin{eqnarray}
\bs{E}[\Log(Y)] \ =\  \bs{E}\left[ \sum_{i=1}^d \log(\lambda_i) \vec{q}_i \vec{q}_i^* \right] \ =\ \bs{E}[\vec{q}_i \vec{q}_i^*] \bs{E}[\log(\det(\Lambda))]. \label{supp-eq:8.10}
\end{eqnarray}
Since $Y$ is Hermitian, $Q \in \mathcal{U}_d$, and therefore $\bs{E}[\log(\det(\Lambda))] \ =\ \bs{E}[\log(\det(Y))]$. By \cite[Theorem 3.2.15]{M82}, 
\begin{eqnarray*}
\log(\det(Y)) & \sim & -d \log(2L) + \sum_{i=1}^d \log\left(\chi^2_{2(L-(d-i))}\right),
\end{eqnarray*}
with $\chi^2_{2(L-(d-i))}$ mutually independent chi-squared distributions, with $2(L-(d-i))$ degrees of freedom. Using that $\bs{E}[\log(\chi_\nu^2)] = \log(2) + \psi(\nu/2)$, it follows that:
\begin{eqnarray*}
\bs{E}[\log(\det(\Lambda))] &=& -d\log(L) + \sum_{i=1}^d \psi(L-(d-i)).
\end{eqnarray*}
Following \cite{S00}, $\bs{E}[\vec{q}_i \vec{q}_i^*] = d^{-1} \te{Id}$, thus by eq.(\ref{supp-eq:8.10}):
\begin{eqnarray*}
\bs{E}[\Log(Y)] &=& \left( -\log(L) + \frac{1}{d}\sum_{i=1}^d \psi(L-(d-i)) \right) \cdot \te{Id} \ = \ c(d,L) \cdot \te{Id}.
\end{eqnarray*}
Plugging this back into eq.(\ref{supp-eq:8.9}) yields $b(X, f) = c(d,L) \cdot f$.\\[3mm]
For the second part of the theorem, observe that $\widetilde{X}_\ell$ ($1\leq \ell \leq n$) is unbiased with respect to $f$, since:
\begin{eqnarray*} 
b(\widetilde{X}_\ell, f) &=& f^{1/2} \ast \bs{E}[\Log( f^{-1/2} \ast \widetilde{X}_\ell)] \nn
&=& f^{1/2} \ast \bs{E}[\Log( e^{-c(d,L)}\te{Id}) + \Log(f^{-1/2} \ast X_\ell )] \nn
&=& f^{1/2} \ast (-c(d,L) \te{Id} + c(d,L) \te{Id}) \ = \ \bs{0},
\end{eqnarray*}
using that $\Log(AB) = \Log(A) + \Log(B)$ for commuting matrices $A,B$, and $\bs{E}[\Log(f^{-1/2} \ast X_\ell)] = c(d,L)\cdot \te{Id}$ as shown above. By eq.(\ref{supp-eq:7.13}), the unique intrinsic mean of $\widetilde{X}_\ell$ on $\mathcal{M}$ is characterized by $f$ such that $b(\widetilde{X}_\ell, f) = \bs{E}[\Log_f(\widetilde{X}_\ell)] = \bs{0}$, i.e., $f$ is the unique intrinsic mean of $\widetilde{X}_{\ell}$ for each $\ell=1,\ldots,n$. Since the distribution of $\widetilde{X}_\ell$ has finite second moment (rescaled complex Wishart distribution), the convergence in probability follows by Proposition \ref{prop:3.1}.
\end{proof}

\subsection{Proofs of Proposition \ref{prop:4.2} and Lemma \ref{lem:4.3}}
\begin{proof}
In this proof, we directly derive the stronger general linear congruence equivariance property in Lemma \ref{lem:4.3}. The weaker unitary congruence equivariance property in Proposition \ref{prop:4.2} then follows directly by substituting wavelet thresholding or shrinkage of coefficients that is only equivariant under unitary congruence transformation, (instead of trace thresholding as in Lemma \ref{lem:4.3}, which is equivariant under general linear congruence transformation of the coefficients).\\[3mm]
Let $M^X_{j,k}$, $M^{\hat{f}}_{j,k}$, $D^X_{j,k}$ and $D^{\hat{f}}_{j,k}$ be the midpoints and wavelet coefficients at scale-location $(j,k)$ based on the observations $(X_\ell)_\ell$ and the estimator $(\hat{f}_\ell)_\ell$ respectively. Analogously, let $M^{X,A}_{j,k}$, $M^{\hat{f}, A}_{j,k}$, $D^{X,A}_{j,k}$ and $D^{\hat{f}, A}_{j,k}$ be the midpoints and wavelet coefficients based on the observations $(A \ast X_\ell)_\ell$ and the estimator $(A \ast \hat{f}_\ell)_\ell$ respectively, where here and throughout this proof $A \in \te{GL}(d,\mathbb{C})$. Below, we repeatedly make use of the identities $A \ast \Exp_M(H) = \Exp_{A \ast M_1}(A \ast H)$ and $ A \ast \Log_{M_1}(M_2) = \Log_{A \ast M_1}(A \ast M_2)$ for $M_1, M_2 \in \mathcal{M}$ and $H \in \mathcal{H}$. In particular, denoting $\te{Mid}(M_1, M_2) := \eta(M_1, M_2, 1/2)$ for the geodesic midpoint, also,
\begin{eqnarray*}
A \ast \te{Mid}(M_1, M_2) \ = \ A \ast \Exp_{M_1}\left(\frac{1}{2} \Log_{M_1}(M_2) \right) & = & \nn
&& \hspace{-3cm} \Exp_{A \ast M_1}\left( \frac{1}{2} \Log_{A \ast M_1}(A \ast M_2) \right)\ =\ \te{Mid}(A \ast M_1, A \ast M_2).
\end{eqnarray*}
By construction, the finest-scale midpoints satisfy $M^{X, A}_{J,k} = A \ast M^X_{J,k}$. Repeated application of the above identity then implies, 
\begin{eqnarray} \label{supp-eq:8.11}
M_{j,k}^{X, A} &=& A \ast M_{j,k}^X \quad \quad \te{for all } j,k.
\end{eqnarray}
Furthermore, since the predicted midpoints $\widetilde{M}^{X,A}_{j,k}$ are weighted intrinsic means of $(M^{X,A}_{j-1,k'})_{k'}$ according to eq.(\ref{eq:2.5}) in the main document, the same relation holds for the predicted midpoints, i.e., $\widetilde{M}^{X,A}_{j,k} = A \ast \widetilde{M}^{X}_{j,k}$ for all $j,k$. Consequently, for the wavelet coefficients at each scale-location $(j,k)$, 
\begin{eqnarray}
D^{X,A}_{j,k} \ =\  2^{-j/2} \Log_{A \ast \widetilde{M}^X_{j,2k+1}}\big(A \ast M^X_{j,2k+1} \big) \ = \ A \ast D^X_{j,k}. \label{supp-eq:8.12}
\end{eqnarray}
In Lemma \ref{lem:4.3}, we threshold or shrink the wavelet coefficients based on the trace of the whitened coefficients, for which:
\begin{eqnarray}
\tr(\mathfrak{D}_{j,k}^{X,A}) &=& 2^{-j/2}\tr\left(\Log\big( (A \ast \widetilde{M}^X_{j,2k+1})^{-1/2} \ast (A \ast M^X_{j,2k+1}) \big) \right) \nn
&=& 2^{-j/2} \left( \tr\big( \Log(A \ast M_{j,2k+1}^X ) \big) - \tr\big( \Log(A \ast \widetilde{M}_{j,2k+1}^X)\big) \right) \nn
&=& 2^{-j/2} \left( \tr\big( \Log(M_{j,2k+1}^X) \big) - \tr\big(\Log(\widetilde{M}^X_{j,2k+1}) \big) \right) \nn
&=& \tr(\mathfrak{D}_{j,k}^X), \label{supp-eq:8.13}
\end{eqnarray}
using that $\tr(\Log(A \ast X)) = \tr(\Log(X)) + \tr(\Log(A A^*))$ and $\tr(\Log(X^t)) = t \tr(\Log(X))$ for $X \in \mathcal{M}$ and $t \in \mathbb{R}$, which follows from the fact that $\tr(\Log(X)) = \log(\det(X))$ and the properties of the determinant and ordinary logarithm.
Let $g(\tr(\mathfrak{D}^X_{j,k})) \in \mathbb{R}$ be a thresholding or shrinkage rule depending on $\tr(\mathfrak{D}^X_{j,k})$, such that $D_{j,k}^{\hat{f}} = g(\tr(\mathfrak{D}^X_{j,k})) D_{j,k}^X$. Due to the invariance in eq.(\ref{supp-eq:8.13}) combined with eq.(\ref{supp-eq:8.12}), it immediately follows that:
\begin{eqnarray*}
D_{j,k}^{\hat{f}, A} \ = \ g(\tr(\mathfrak{D}^{X,A}_{j,k})) D_{j,k}^{X,A} \ = \ A \ast \left(g(\tr(\mathfrak{D}^{X}_{j,k})) D_{j,k}^X\right) \ = \ A \ast D^{\hat{f}}_{j,k} \quad \quad \te{for all } j,k.
\end{eqnarray*}
The wavelet-thresholded estimator $(\hat{f}_\ell)_\ell$ is retrieved via the inverse wavelet transform applied to the set of thresholded wavelet coefficients (and coarse-scale midpoints). At scale $j = 0$, by eq.(\ref{supp-eq:8.11}), $M^{\hat{f},A}_{0,k} = M^{X,A}_{0,k} = A \ast M^X_{0,k} = A \ast M^{\hat{f}}_{0,k}$. At the odd locations $2k+1$ at the next coarser scale $j = 1$, 
\begin{eqnarray*}
M^{\hat{f},A}_{1,2k+1} &=& \Exp_{\widetilde{M}^{\hat{f},A}_{1,2k+1}}\left( 2^{1/2} D_{j,k}^{\hat{f},A} \right) \nn
&=& \Exp_{A \ast \widetilde{M}^{\hat{f}}_{1,2k+1}}\left( A \ast \big(2^{1/2} D_{j,k}^{\hat{f}}\big) \right) \nn
&=& A \ast \Exp_{\widetilde{M}^{\hat{f}}_{1,2k+1}}\left( 2^{1/2} D_{j,k}^{\hat{f}} \right) \nn
&=& A \ast M^{\hat{f}}_{1,2k+1},
\end{eqnarray*}
using that $\widetilde{M}^{\hat{f},A}_{1,2k+1} = A \ast \widetilde{M}^{\hat{f}}_{1,2k+1}$, since the same relation holds for $(M_{0,k'}^{\hat{f},A})_{k'}$ and the predicted midpoints are weighted intrinsic means of $(M_{0,k'}^{\hat{f},A})_{k'}$. Also, at the even locations $2k$, 
\begin{eqnarray*}
M^{\hat{f},A}_{1,2k} &=& M^{\hat{f},A}_{0,k} \ast \big( M^{\hat{f},A}_{1,2k+1} \big)^{-1} \nn
&=& (A \ast M^{\hat{f}}_{0,k}) \ast \big( A \ast M^{\hat{f}}_{1,2k+1} \big)^{-1} \nn
&=& A \ast \left( M^{\hat{f}}_{0,k} \ast \big(M^{\hat{f}}_{1,2k+1}\big)^{-1} \right) \nn
&=& A \ast M^{\hat{f}}_{1,2k}.
\end{eqnarray*}
Iterating the same argument up to the finest scale $j=J$ yields the desired result $\hat{f}_{A, \ell} = A \ast \hat{f}_\ell$ for each $\ell = 1,\ldots,2^J$.
\end{proof}

\subsection{Proof of Proposition \ref{prop:4.4}}
\begin{proof}
Let us write $M^X_{J,k-1} := X_{k} = f^{1/2}_k \ast W_k$ for $k=1,\ldots,n$, where the distribution of $W_k$ does not depend on $f_k$, and the intrinsic mean of $W_k$ is the identity $\te{Id}$. The latter follows from the fact that $X_k$ has intrinsic mean $f_k$, since: 
\begin{eqnarray*}
\bs{E}[\Log_{\te{Id}}(W_k)] &=& \bs{E}[ f_k^{-1/2} \ast \Log_{f_k}(f_k^{1/2} \ast W_k) ] \nn
&=& f_k^{-1/2} \ast \bs{E}[\Log_{f_k}(X_k)] \nn
&=& f_k^{-1/2} \ast \bs{0} \ =\ \bs{0},
\end{eqnarray*}
and the intrinsic mean $\mu$ of $W_k$ is uniquely characterized by $\bs{E}[\Log_\mu(W_k)] = \bs{0}$. First, we verify that:
\begin{eqnarray} \label{supp-eq:8.14}
\tr(\Log(M^X_{j,k})) &=& \tr(\Log(M^f_{j,k})) + \tr(\Log(M^W_{j,k})) \quad \quad \te{for all } j,k,
\end{eqnarray}
where $M^X_{j,k}$, $M^f_{j,k}$, and $M^W_{j,k}$ are the midpoints at scale-location $(j,k)$ based on the sequences $(X_\ell)_\ell$, $(f_\ell)_\ell$, and $(W_\ell)_\ell$ respectively. For convenience, as before, denote $\te{Mid}(X_1,X_2) := \eta(M_1,M_2,1/2)$ for the geodesic midpoint. Using that $\tr(\Log(AB)) = \tr(\Log(A)) + \tr(\Log(B))$ and $\Log(A^t) = t \Log(A)$ for any $A,B \in \mathcal{M}$, decompose:
\begin{eqnarray*}
\tr(\Log(M^X_{j,k})) &=& \tr(\Log(\Mid(M^X_{j+1,2k}, M^X_{j+1,2k+1})))\nn
&=& \tr\big(\Log\big((M^X_{j+1,2k})^{1/2} \ast \big((M^X_{j+1,2k})^{-1/2} \ast M^X_{j+1,2k+1}\big)^{1/2}\big)\big) \nn
&=&\frac{1}{2}\tr(\Log(M^X_{j+1,2k})) + \frac{1}{2} \tr(\Log(M^X_{j+1,2k+1})) \nn
& \vdots & \nn
&=& \frac{1}{2^{J-j}}\sum_{\ell=0}^{2^{J-j}-1} \tr(\Log(M^X_{J,(2k)^{J-j-1}+\ell}))\nn
&=& \frac{1}{2^{J-j}}\sum_{\ell=0}^{2^{J-j}-1} \tr(\Log(f_{(2k)^{J-j-1}+\ell+1})) \nn
&& + \quad \frac{1}{2^{J-j}}\sum_{\ell=0}^{2^{J-j}-1} \tr(\Log(W_{(2k)^{J-j-1}+\ell+1})) \nn
& \vdots & \nn
&=& \tr(\Log(\Mid(M^f_{j+1,2k},M^f_{j+1,2k+1}))) + \tr(\Log(\Mid(M^W_{j+1,2k}, M^W_{j+1,2k+1}))) \nn
&=& \tr(\Log(M^f_{j,k})) + \tr(\Log(M^W_{j,k})).
\end{eqnarray*}
Second, we also verify that for each scale $j$ and location $k$, 
\begin{eqnarray} \label{supp-eq:8.15}
\tr(\Log(\widetilde{M}^X_{j,2k+1})) &=& \tr(\Log(\widetilde{M}^f_{j,2k+1})) + \tr(\Log(\widetilde{M}^W_{j,2k+1})),
\end{eqnarray}
where $\widetilde{M}^X_{j,k'}, \widetilde{M}^f_{j,k'}$, and $\widetilde{M}^W_{j,k'}$ are the imputed midpoints at scale-location $(j,k')$ based on the sequences $(X_\ell)_\ell$, $(f_\ell)_\ell$, and $(W_\ell)_\ell$ respectively. By eq.(\ref{eq:2.5}) in the main document, the predicted midpoints at the odd locations $2k+1$ satisfy:
\begin{eqnarray*}
\widetilde{M}^X_{j,2k+1} &=& \Exp_{\widetilde{M}^X_{j,2k+1}}\left( \sum_{\ell=-L}^{L} C_{N, 2\ell + N} \Log_{\widetilde{M}^X_{j,2k+1}}(M^X_{j-1,k + \ell}) \right),
\end{eqnarray*}
with weights $\bs{C}_N = (C_{N,i})_{i=0,\ldots,2N-1}$ as in eq.(\ref{eq:2.5}). Here, without loss of generality we consider prediction away from the boundary, (at the boundary the sum runs over the $N = 2L + 1$ closest available neighbors to $M_{j,k}$). Using eq.(\ref{supp-eq:8.14}), we decompose,\\[1mm]
\begin{footnotesize}
\begin{eqnarray*}
\tr(\Log(\widetilde{M}^{X}_{j,2k+1})) &=& \tr\Big(\Log\Big(\Exp_{\widetilde{M}^X_{j,2k+1}}\Big( \sum_{\ell} C_{N, 2\ell + N} \Log_{\widetilde{M}^X_{j,2k+1}}(M^X_{j-1,k + \ell}) \Big)\Big)\Big)\nn
&=& \tr(\Log(\widetilde{M}^{X}_{j,2k+1})) + \tr\left((\widetilde{M}^{X}_{j,2k+1})^{-1/2} \ast \Big( \sum_{\ell} C_{N, 2\ell + N} \Log_{\widetilde{M}^X_{j,2k+1}}(M^X_{j-1,k + \ell}) \Big)\right) \nn
&=& \tr(\Log(\widetilde{M}^{X}_{j,2k+1})) + \tr\left( \sum_{\ell} C_{N, 2\ell + N} \Log\Big( (\widetilde{M}^{X}_{j,2k+1})^{-1/2} \ast M^{X}_{j-1,k+\ell} \Big) \right) \nn
&=& \tr(\Log(\widetilde{M}^{X}_{j,2k+1})) + \sum_{\ell} C_{N, 2\ell + N} \left( \tr(\Log(M^{X}_{j-1,k+\ell})) - \tr(\Log(\widetilde{M}^{X}_{j,2k+1})) \right) \nn
&=& \sum_{\ell} C_{N, 2\ell + N}\tr(\Log(M^{X}_{j-1,k+\ell})) \nn
&=& \sum_{\ell} C_{N, 2\ell + N} \tr(\Log(M^{f}_{j-1,k+\ell})) + \sum_{\ell} C_{N, 2\ell + N} \tr(\Log(M^{W}_{j-1,k+\ell})) \nn
&\vdots & \nn
&=& \tr(\Log(\widetilde{M}^f_{j,2k+1})) + \tr(\Log(\widetilde{M}^W_{j,2k+1})),\nn
\end{eqnarray*}
\end{footnotesize}
where we used in particular $g \ast \Log_{X_1}(X_2) = \Log_{g \ast X_1}(g \ast X_2)$ and $g \ast \Exp_{X_1}(X_2) = \Exp_{g \ast X_1}(g \ast X_2)$ for any $g \in \te{GL}(d, \mathbb{C})$, and the fact that $\sum_\ell C_{N,2\ell+N} = 1$.\\[3mm]
The first claim in the Proposition now follows from eq.(\ref{supp-eq:8.14}) and eq.(\ref{supp-eq:8.15}) through: 
\begin{eqnarray}
\tr(\mathfrak{D}^{X}_{j,k}) &=& 2^{-j/2}\tr\Big(\Log\Big((\widetilde{M}^{X}_{j,2k+1})^{-1/2} \ast M^{X}_{j,2k+1} \Big)\Big) \nn
&=& 2^{-j/2} \left( \tr(\Log( M^{X}_{j,2k+1} )) - \tr(\Log( \widetilde{M}^{X}_{j,2k+1} )) \right) \nn
&=& 2^{-j/2}\tr(\Log( M^f_{j,2k+1})) + 2^{-j/2}\tr(\Log(M^W_{j,2k+1})) \nn
&& - 2^{-j/2}\left( \tr(\Log( \widetilde{M}^{f}_{j,2k+1} )) + \tr(\Log( \widetilde{M}^{W}_{j,2k+1} )) \right) \nn
&=& \tr(\mathfrak{D}^f_{j,k}) + \tr(\mathfrak{D}^W_{j,k}). \label{supp-eq:8.16}
\end{eqnarray}
For the second claim in the Proposition, first observe:
\begin{eqnarray*}
\bs{E}[\tr(\Log(M^W_{j,k}))] \ =\ \frac{1}{2^{J-j}}\sum_{\ell=0}^{2^{J-j}-1} \bs{E}[\tr(\Log(W_{(2k)^{J-j-1}+\ell+1}))] \ = \ 0, \quad \quad \te{for each } j,k,
\end{eqnarray*}
using that $\bs{E}[\tr(\Log(W_\ell))] = 0$ for each $\ell=1,\ldots,n$, which is implied by $\bs{E}[\Log_{\te{Id}}(W_\ell)] = \bs{0}$. As a consequence, also,
\begin{eqnarray*}
\bs{E}[\tr(\Log(\widetilde{M}^W_{j,2k+1}))] \ = \ \sum_{\ell} C_{N, 2\ell + N}\tr(\Log(M^{W}_{j-1,k+\ell})) \ = \ 0, \quad \quad \te{for each } j,k,
\end{eqnarray*}
and therefore, 
\begin{eqnarray*}
\bs{E}[\tr(\mathfrak{D}^X_{j,k})] &=& \tr(\mathfrak{D}^f_{j,k}) + \bs{E}[\tr(\mathfrak{D}^W_{j,k})] \nn
&=& \tr(\mathfrak{D}^f_{j,k}) + 2^{-j/2}\bs{E}\left[ \tr(\Log(M^W_{j,2k+1})) - \tr(\Log(\widetilde{M}^W_{j,2k+1})) \right] \nn
&=& \tr(\mathfrak{D}^f_{j,k}).
\end{eqnarray*}
For the variance of $\tr(\mathfrak{D}_{j,k}^X)$, we first note that the random variables $(W_\ell)_{\ell=1,\ldots,n}$ are i.i.d., implying that the random variables $(\tr(\Log(M^W_{j,k}))_{k=0,\ldots,2^j-1}$ on scale $j$ are independent with equal variance. We write out:
\begin{eqnarray}
\var(\tr(\mathfrak{D}_{j,k}^{X})) &=& 2^{-j}\var\left(\tr(\Log(M^W_{j,2k+1})) - \tr(\Log(\widetilde{M}^W_{j,2k+1}))\right) \nn
&=& 2^{-j}\var\Big(\tr(\Log(M^W_{j,2k+1})) -  \sum_{\ell} C_{L, 2\ell + N}\tr(\Log(M^{W}_{j-1,k+\ell}))\Big) \nn
&=& 2^{-j}\var\Big(\tr(\Log(M^W_{j,2k+1})) - C_{N,N} \tr(\Log(M^W_{j-1,k}))\Big) \nn
&& +\quad 2^{-j}\sum_{-L \leq \ell \leq L; \ell \neq 0} C_{N, 2\ell+N}^2 \var(\tr(\Log(M^W_{j-1,k+\ell}))) \nn
&=& 2^{-(j+1)}\var(\tr(\Log(M^W_{j,2k}))) \nn
&& +\quad 2^{-j}\Big(\sum_{\ell} C_{N,2\ell+N}^2 - 1\Big) \var(\tr(\Log(M^W_{j-1,k+\ell}))) \nn
&=& 2^{-(j+1)}\sum_{\ell} C_{N,2\ell+N}^2 \var(\tr(\Log(M^W_{j,0}))), \label{supp-eq:8.17}
\end{eqnarray}
where in the final two steps we used that $C_{N,N} = 1$, and by the independence of the midpoints within each scale, for each $k$, 
\begin{eqnarray*}
\var(\tr(\Log(M^W_{j-1,k}))) &=& \var\left(\frac{1}{2}\tr(\Log(M^W_{j,2k})) + \frac{1}{2}\tr(\Log(M^W_{j,2k+1}))\right) \nn
&=& \frac{1}{2} \var(\tr(\Log(M^W_{j,0}))).
\end{eqnarray*}
It remains to derive an expression for $\var(\tr(\Log(M^W_{j,0})))$. By repeated application of the above argument, 
\begin{eqnarray}
\var(\tr(\Log(M^W_{j,0}))) &=& \frac{1}{2^{J-j}}\var(\tr(\Log(M^W_{J,0}))) \nn
&=& \frac{1}{2^{J-j}}\var(\tr(\Log(W_1))), \label{supp-eq:8.18}
\end{eqnarray}
with $W_1 \sim W_d^c(L, L^{-1}e^{-c(d,L)}\te{Id})$. As in the proof of Theorem \ref{thm:4.1}, 
\begin{eqnarray*}
\tr(\Log(W_1)) \sim - d\log(2e^{c(d,L)}L) + \sum_{i=1}^d \log\left(\chi^2_{2(L-(d-i))}\right).
\end{eqnarray*}
The variance of a $\log(\chi_\nu^2)$ distribution equals $\psi'(\nu/2)$, (with $\psi'(\cdot)$ the trigamma function), therefore:
\begin{eqnarray*}
\var(\tr(\Log(W_1))) &=& \sum_{i=1}^d \psi'(L-(d-i)).
\end{eqnarray*}
Combining the above result with eq.(\ref{supp-eq:8.17}) and eq.(\ref{supp-eq:8.18}) finishes the proof.  
\end{proof}

\subsection{Proof of Corollary \ref{cor:4.5}}
\begin{proof}
Analogous to the proof of Theorem \ref{thm:4.1}, $W_1,\ldots,W_n \overset{\te{iid}}{\sim} W_d^c(L, L^{-1}e^{-c(d,L)}\te{Id})$ are unitarily invariant, see \cite[Section 3.2]{M82}. By the same argument as in eq.(\ref{supp-eq:8.11}) the repeated midpoints based on unitarily invariant random variables satisfy $U \ast M^W_{j,k} \overset{d}{=} M^W_{j,k}$ for each $j, k$ and $U \in \mathcal{U}_d$. It follows that the predicted midpoints $\widetilde{M}_{j,2k+1}^W$ are unitarily invariant as well, as they can be expressed as weighted intrinsic averages of the midpoints $(M^W_{j-1,k})_k$, which are unitarily invariant themselves. That is, $U \ast \widetilde{M}^W_{j,2k+1} \overset{d}{=} \widetilde{M}^W_{j,2k+1}$ for each $j,k$ and $U \in \mathcal{U}_d$. Combining the above results, it follows that the random whitened coefficient $\mathfrak{D}_{j,k}^W$ is unitarily invariant, as for each $U \in \mathcal{U}_d$,
\begin{eqnarray*}
U \ast \mathfrak{D}_{j,k}^W &=& U \ast \Log\left( (\widetilde{M}^W_{j,2k+1})^{-1/2} \ast M_{j,2k+1} \right) \nn
&=& \Log\left((U \ast \widetilde{M}^W_{j,2k+1})^{-1/2} \ast (U \ast M_{j,2k+1})\right) \nn
&\overset{d}{=}& \Log\left((\widetilde{M}^{W}_{j,2k+1})^{-1/2} \ast M_{j,2k+1}\right) \nn
&=& \mathfrak{D}^W_{j,k},
\end{eqnarray*}
using that $U \ast \Log(X) = \Log(U \ast X)$ for $U \in \mathcal{U}_d$. By the same argument as in Theorem \ref{thm:4.1}, if we write the eigendecomposition $\mathfrak{D}^W_{j,k} = Q \ast \Lambda$, then for a unitarily invariant random matrix $\mathfrak{D}^W_{j,k}$, 
\begin{eqnarray*}
\bs{E}[\mathfrak{D}^W_{j,k}] &=& \bs{E}\left[ \sum_{i=1}^d \lambda_i \vec{q}_i \vec{q}_i^* \right] \nn
&=& \bs{E}[\vec{q}_i\vec{q}_i^*] \bs{E}[\tr(\Lambda)] \nn
&=& \bs{E}[\vec{q}_i\vec{q}_i^*] \bs{E}[\tr(\mathfrak{D}^W_{j,k})] \ = \ \bs{0}.
\end{eqnarray*}
Here we used that $\tr(Q \ast \Lambda) = \tr(\Lambda)$, since $Q$ is a unitary matrix ($\mathfrak{D}^W_{j,k}$ is Hermitian), combined with the result $\bs{E}[\tr(\mathfrak{D}_{j,k}^W)] = 0$ in Proposition \ref{prop:4.4}.
\end{proof}

\section{Appendix III: Additional details Section \ref{sec:5.1}}

\paragraph{Estimation procedures Section \ref{sec:5.1}}
This appendix section provides more details on the matrix curve estimation procedures considered in the simulated data scenarios in Section \ref{sec:5.1} in the main document. Each estimation procedure takes as input an initial dyadic sequence of random HPD matrix-valued observations $X_1,\ldots,X_n \in \mathcal{M}$ observed on an equidistant grid $t_1,\ldots, t_n \in \mathbb{R}$ and outputs a denoised sequence of HPD matrix-valued observations $\hat{f}(t_1),\ldots,\hat{f}(t_n) \in \mathcal{M}$. 

\begin{table}[t]
\centering
\begin{threeparttable}
\caption{Estimation procedure metrics and their properties.}
\begin{small}
\setlength{\tabcolsep}{1em}
\begin{tabular}[t]{lccccc}
\toprule
Metric & $U$-equiv.$^*$ & $A$-equiv.$^\dagger$ & PD Estimates & Wishart B-C$^{**}$ \\
\midrule\\[-4mm]
Riemannian & \cmark & \cmark & \cmark & \cmark \\
Log-Euclidean & \cmark & \xmark & \cmark & \xmark \\
Cholesky & \xmark & \xmark & \cmark & \cmark \\
Euclidean & \cmark & \xmark & \xmark & \cmark \\
\bottomrule
\end{tabular}
 \begin{tablenotes}
 \begin{footnotesize}
      \item $*,\dagger$: $U$-equiv. and $A$-equiv. respectively denote whether the estimator is equivariant under congruence transformation by a unitary matrix $U \in \mathcal{U}_d$ or a general linear matrix $A \in \te{GL}(\mathbb{C}, d)$, see Section \ref{sec:4.1}.
      \item $**$: Wishart B-C denotes whether a bias-correction (B-C) is available in the context of spectral matrix estimation, where the periodogram data is asymptotically Wishart distributed.
      \end{footnotesize}
    \end{tablenotes}
 \end{small}
\end{threeparttable}
\end{table}%

\begin{itemize}

\item{\textbf{Linear wavelet thresholding}}: the input data $X_1,\ldots,X_n$ is transformed to the intrinsic wavelet domain by means of the forward average-interpolating wavelet transform of Section \ref{sec:2} in the main document subject to respectively the Riemannian, Log-Euclidean or Cholesky metric, and all wavelet coefficients at scales $j > J_0$ are set to zero. The smoothed curve estimate $\hat{f}(t_1),\ldots,\hat{f}(t_n)$ is obtained by application of the intrinsic backward average-interpolating wavelet transform. The main tuning parameter in the case of linear wavelet thresholding is the maximum scale of nonzero wavelet coefficients $J_0$. The impact of the average-interpolation order of the wavelet transform is small in terms of the estimation error compared to the choice of the scale parameter $J_0$. For this reason the refinement order is fixed at $N = 5$ for all simulated scenarios in Section \ref{sec:5}. Linear wavelet thresholding is implemented in the \texttt{pdSpecEst}-package by the function \texttt{pdSpecEst1D()} with arguments \texttt{alpha = 0}, \texttt{jmax} set to the maximum scale of nonzero coefficients $J_0$, and \texttt{metric} set to metric considered for estimation.

\item{\textbf{Nonlinear wavelet thresholding}}: the input data $X_1,\ldots,X_n$ is transformed to the intrinsic wavelet domain the same way as for the linear wavelet thresholding procedure. The nonlinear wavelet thresholding procedure considers dyadic tree-structured thresholding based on the traces of the individual coefficients by minimizing the complexity penalized loss criterion given in eq.(\ref{eq:5.1}) and explained in more detail in the main document. The main tuning parameter is the regularization parameter $\lambda \geq 0$, and the refinement order of the wavelet transforms is fixed at $N = 5$ for all simulation scenarios equivalent to the linear thresholding procedure. For sufficiently large $n$, the scalar coefficients $d_{j,k}$ are approximately normally distributed at reasonably coarse scales $j$, as the scalar coefficients $d_{j,k}$ are essentially locally weighted averages of the observations. For normally distributed coefficients, a natural choice for the regularization parameter is the universal threshold $\lambda \sim \sigma_w \sqrt{2\log(n)}$, with $n$ the total number of wavelet coefficients and $\sigma_w^2$ the noise variance determined either via eq.(\ref{eq:4.1}) in the main document or from the data itself. Tree-structured trace thresholding is implemented in the \texttt{pdSpecEst}-package by the function \texttt{pdSpecEst1D()} with arguments \texttt{alpha = 1} to use a universal threshold multiplied by $\alpha = 1$, and \texttt{metric} set to the metric considered for estimation.

\item{\textbf{Nearest-Neighbor (NN) regression}}: intrinsic nearest-neighbor regression is implemented by replacing ordinary local Euclidean averages by their intrinsic counterparts based on the Riemannian, Log-Euclidean and Cholesky metric using the function \texttt{pdMean()} in the \texttt{pdSpecEst}-package. In the case of the Riemannian metric, the local intrinsic averages are calculated efficiently by the gradient descent algorithm in \cite{P06}. The main tuning parameter in this benchmark procedure is the number of nearest neighbors used in the local averages.

\item{\textbf{Cubic Spline (CS) regression}}: intrinsic cubic smoothing spline regression is implemented in the space of HPD matrices based on the Riemannian, Log-Euclidean and Cholesky metric. For the Riemannian metric, we implemented the penalized regression approach in \cite{BA11b} and \cite{BA11}, with penalty parameters $(\lambda = 0, \mu > 0)$, such that the minimizers of the objective function are approximate cubic splines in the manifold of HPD matrices. The Riemannian conjugate gradient descent method in \cite{BA11} to compute the estimator is available through the function \texttt{pdSplineReg()} in the \texttt{pdSpecEst}-package. Here, we use a backtracking line search based on the Armijo-Goldstein condition. The main tuning parameter in this benchmark procedure is the regularization parameter in the penalized loss criterion. 

\item{\textbf{Local polynomial (LP) regression}}: intrinsic local polynomial regression of degree $p = 0$ (LP-0) and degree $p = 3$ (LP-3) respectively is implemented in the space of HPD matrices based on the Riemannian metric, Log-Euclidean metric and Cholesky metric. For the Riemannian metric, we have only implemented the locally constant estimator, i.e.\@ degree $p = 0$, as local polynomial regression under the Riemannian metric for $p > 0$ requires the optimization of a non-convex objective function and is computationally quite challenging. We refer to \cite{Y12} for additional details. The main tuning parameter in this benchmark procedure is the bandwidth parameter of the local polynomials.

\item{\textbf{Multitaper spectral estimation}}: the multitaper benchmark estimator is only considered in the periodogram noise scenario given in Table \ref{tab:2}, as this is the only simulated scenario that provides input time series data in addition to the input (periodogram) observations $X_1,\ldots,X_n$. The multitaper spectral estimate takes as input the generated $d$-dimensional stationary time trace and is based on $L \geq d$ discrete prolate spheroidal (DPSS) taper functions using the function \texttt{pdPgram()}, thereby guaranteeing an HPD matrix curve estimate $\hat{f}(t_1),\ldots,\hat{f}(t_n) \in \mathcal{M}$. The main tuning parameter in this benchmark procedure is the number of DPSS tapers $L$.

\end{itemize}

\end{small}

\end{document}